\documentclass[letterpaper,twocolumn,10pt]{article}
\usepackage{usenix-2020-09}

\usepackage{tikz}
\usepackage{amsmath}
\usepackage{amsfonts}

\usepackage{caption}
\usepackage{subcaption}
\usepackage{booktabs}
\usepackage{multirow}
\usepackage{siunitx}
\usepackage{soul}
\usepackage{mathrsfs}
\usepackage{subcaption}
\usepackage{enumitem}
\setstcolor{red}
\setstcolor{blue}

\usepackage{breakurl}

\newlength{\bibitemsep}
\newlength{\bibparskip}
\let\oldthebibliography\thebibliography
\renewcommand\thebibliography[1]{%
  \oldthebibliography{#1}%
  \setlength{\parskip}{\bibitemsep}%
  \setlength{\itemsep}{\bibparskip}%
}

\newcommand{\cut}[1]{}

\newcommand{\ie}{i.e.}
\newcommand{\eg}{e.g.}
\newcommand{\op}{OpenPilot}

\newcommand{\circled}[1]{\raisebox{.5pt}{\textcircled{\raisebox{-.9pt} {{\small #1}}}}}

\newcommand{\diff}[1]{{#1}}
\newcommand{\diffst}[1]{}

\makeatletter
\renewcommand\p@subfigure{\thefigure\,}

\makeatother

\usepackage{mdframed}
\mdfdefinestyle{rebuttalstyle}{innerleftmargin=3pt, innerrightmargin=3pt, innertopmargin=3pt, innerbottommargin=3pt}

\newcounter{response}[section]

\newcommand{\nsection}[1]{\vspace{-0.4cm}\section{#1}\vspace{-0.25cm}}
\newcommand{\nsubsection}[1]{\vspace{-0.4cm}\subsection{#1}\vspace{-0.2cm}}
\newcommand{\nsubsubsection}[1]{\vspace{-0.4cm}\subsubsection{#1}\vspace{-0.15cm}}

\newcounter{revision}[section]

\newcounter{comment}[section]

\makeatletter
\def\thanks#1{\protected@xdef\@thanks{\@thanks
        \protect\footnotetext{#1}}}
\makeatother
\begin{document}

\date{}

\title{\Large \bf Dirty Road Can Attack: Security of Deep Learning based Automated Lane Centering under Physical-World Attack}

\author{
{\rm Takami Sato$^*$}\thanks{$^*$Co-first authors.}\\
UC Irvine\\
takamis@uci.edu
\and
{\rm Junjie Shen{{\footnotemark[1]}}}\\
UC Irvine\\
junjies1@uci.edu
\and
{\rm Ningfei Wang}\\
UC Irvine\\
ningfei.wang@uci.edu
\and
{\rm Yunhan Jia}\\
ByteDance\\
yunhan.jia@bytedance.com
\and
{\rm Xue Lin}\\
Northeastern University\\
xue.lin@northeastern.edu
\and
{\rm Qi Alfred Chen}\\
UC Irvine\\
alfchen@uci.edu
}

\maketitle

\begin{abstract}
Automated Lane Centering (ALC) systems are convenient and widely deployed today, but also highly security and safety critical. In this work, we are the first to systematically study the security of state-of-the-art deep learning based ALC systems in their designed operational domains under physical-world adversarial attacks. We formulate the problem with a safety-critical attack goal, and a novel and domain-specific attack vector: dirty road patches. To systematically generate the attack, we adopt an optimization-based approach and overcome domain-specific design challenges such as camera frame inter-dependencies due to attack-influenced vehicle control, and the lack of objective function design for lane detection models.

We evaluate our attack on a production ALC using 80 scenarios from real-world driving traces. The results show that our attack is highly effective with over 97.5\% success rates and less than 0.903 sec average success time, which is substantially lower than the average driver reaction time. This attack is also found (1) robust to various real-world factors such as lighting conditions and view angles, (2) general to different model designs, and (3) stealthy from the driver’s view. To understand the safety impacts, we conduct experiments using software-in-the-loop simulation and attack trace injection in a real vehicle. The results show that our attack can cause a 100\% collision rate in different scenarios, including when tested with common safety features such as automatic emergency braking. We also evaluate and discuss defenses.

\end{abstract}

\nsection{Introduction}
\label{sec:intro}

Automated Lane Centering (ALC) is a Level-2 driving automation technology that automatically steers a vehicle to keep it centered in the traffic lane~\cite{sae2018}. Due to its high convenience for human drivers, today it is widely available on various vehicle models such as Tesla, GM Cadillac, Honda Accord, Toyota RAV4, Volvo XC90, etc. While convenient, such system is highly security and safety critical: When the ALC system starts to make wrong steering decisions, the human driver may not have enough reaction time to prevent safety hazards such as driving off road or colliding into vehicles in adjacent lanes. Thus, it is imperative and urgent to understand the security property of ALC systems.

In an ALC system, the most critical step is \textit{lane detection}, which is generally performed using a front camera. So far, Deep Neural Network (DNN) based lane detection achieves the highest accuracy~\cite{tusimple} and is adopted in the most performant production ALC systems today such as Tesla Autopilot~\cite{tesla2020autopilot}. Recent works show that DNNs are vulnerable to physical-world adversarial attacks such as malicious stickers on traffic signs~\cite{eykholt2018robust, eykholt2018physical, chen2018shapeshifter, zhao2018seeing}. However, these methods cannot be directly applied to attack ALC systems due to two main design challenges. First, in ALC systems, the physical-world attack generation needs to handle \textit{inter-dependencies} among camera frames due to attack-influenced vehicle actuation. For example, if the attack deviates the detected lane to the right in a frame, the ALC system will steer the vehicle to the right accordingly. This causes the following frames to capture road areas more to the right, and thus directly affect their attack generation. Second, the optimization objective function designs in prior works are mainly for image classification or object detection models and thus aim at changing class or bounding box probabilities~\cite{eykholt2018robust, zhao2018seeing}. However, attacking lane detection requires to change the \textit{shape} of the detected traffic lane, making it difficult to directly apply prior designs.

The only prior effort that studied adversarial attacks on a production ALC is from Tencent~\cite{tencent2019}, which fooled Tesla Autopilot to follow fake lane lines created by white stickers on road regions \textit{without lane lines}. However, it is neither attacking the designed operational domain for ALC, i.e., roads \textit{with lane lines}, nor generating the perturbations systematically by addressing the design challenges above.

To fill this critical research gap, in this work we are the first to systematically study the security of DNN-based ALC systems in their designed operational domains (i.e., roads with lane lines) under physical-world adversarial attacks. Since ALC systems assume a fully-attentive human driver prepared to take over at any time~\cite{tesla2020support, sae2018}, we identify the attack goal as not only causing the victim to drive out of the current lane boundaries, but also achieving it shorter than the average driver reaction time to road hazard. This thus directly breaks the design goal of ALC systems and can cause various types of safety hazards such as driving off road and vehicle collisions.

Targeting this attack goal, we design a novel physical-world adversarial attack method on ALC systems, called \textit{DRP (Dirty Road Patch) attack}, which is the first to systematically address the design challenges above. First, we identify \textit{dirty road patches} as a novel and domain-specific attack vector for physical-world adversarial attacks on ALC systems. This design has 2 unique advantages: (1) Road patches can appear to be legitimately deployed on traffic lanes in the physical world, e.g., for fixing road cracks; and (2) Since it is common for real-world roads to have dirt or white stains, using similar dirty patterns as the input permutations can allow the malicious road patch to appear more normal and thus stealthier.

With this attack vector, we then design systematic malicious road patch generation following an optimization-based approach. To efficiently and effectively address the first design challenge without heavyweight road testing or simulations, we design a novel method that combines vehicle motion model and perspective transformation to dynamically synthesize camera frame updates according to attack-influenced vehicle control. Next, to address the second design challenge, one direct solution is to design the objective function to directly change the steering angle decisions. However, we find that the lateral control step in ALC that calculates steering angle decisions are generally not differentiable, which makes it difficult to effectively optimize. To address this, we design a novel lane-bending objective function as a differentiable surrogate function. We also have domain-specific designs for attack robustness, stealthiness, and physical-world realizability.

We evaluate our attack method on a production ALC system in \op{}~\cite{openpilot}, which is reported to have close performance to Tesla Autopilot and GM Super Cruise, and better than many others~\cite{vanderwerp2020opteslagm, openpilotreview, hall2020ophandson}. We perform experiments on 80 attack scenarios from real-world driving traces, and find that our attack is highly effective with over 97.5\% success rates for all scenarios, and less than 0.903 sec average success time, which is substantially lower than 2.5 sec, the average driver reaction time (\S\ref{sec:problem_formulation_goals}). This means that even for a fully-attentive driver who can take over as soon as the attack starts to take effect, the average reaction time is still not enough to prevent the damage. We further find this attack is (1) robust to real-world factors such as different lighting conditions, \diff{viewing angles}, printer color fidelity, and camera sensing capability, (2) general to different lane detection model designs, and (3) stealthy from the driver’s view based on a user study.

To understand the potential \diffst{end-to-end }safety impacts, we further conduct experiments using (1) \diff{software-in-the-loop simulation in} a production-grade simulator, and (2) \diff{attack trace injection} in a real vehicle. The simulation results show that our attack can successfully cause a victim running a production ALC to hit the highway concrete barrier or a truck in the opposite direction with 100\% success rates. The real-vehicle experiments show that it causes the vehicle to collide with (dummy) road obstacles in all 10 trials even with common safety features such as Automatic Emergency Braking (AEB) enabled. Demo videos are available at: \textbf{\url{https://sites.google.com/view/cav-sec/drp-attack/}}. We also explore and discuss possible defenses at DNN level and those based on sensor/data fusion.

In summary, this work makes the following contributions: %
\vspace{-\topsep}
\begin{itemize}[leftmargin=0.2in]
\setlength{\itemsep}{0pt}
\setlength{\parskip}{0pt}
\item We are the first to systematically study the security of DNN-based ALC in the designed operational domains under physical-world adversarial attacks. We formulate the problem with a safety-critical attack goal, and a novel and domain-specific attack vector, dirty road patches.

\item To systematically generate attack patches, we adopt an optimization-based approach with 2 major novel and domain specific designs: motion model based input generation, and lane-bending objective function. We also have domain-specific designs for improving the attack robustness, stealthiness, and physical-world realizability.

\item We perform evaluation on a production ALC using 80 attack scenarios from real-world driving traces. The results show that our attack is highly effective with $\ge$97.5\% success rates and $\le$0.903 sec average success time, which is substantially lower than the average driver reaction time. This attack is also found (1) robust to various real-world factors, (2) general to different lane detection model designs, and (3) stealthy from the driver’s view.

\item To understand the \diffst{end-to-end }safety impacts, %
we conduct experiments using (1) \diff{software-in-the-loop simulation}, and (2) \diff{attack trace injection in} a real vehicle. The results show that our attack can cause a 100\% collision rate in different scenarios, including when tested with safety features such as AEB. We also evaluate and discuss possible defenses.

\vspace{-\topsep}
\end{itemize}

\textbf{Code and data release.} 
Our code and data for the attack and evaluations are available at our project website~\cite{drp_attack_page}.

\vspace{-0.01in}
\nsection{Background} \label{sec:background}
\vspace{0.1in}
\nsubsection{Overview of DNN-based ALC Systems}
\vspace{-0.03in}
\label{sec:background_alc}

\begin{figure}[t]
\centering
\includegraphics[width=\columnwidth]{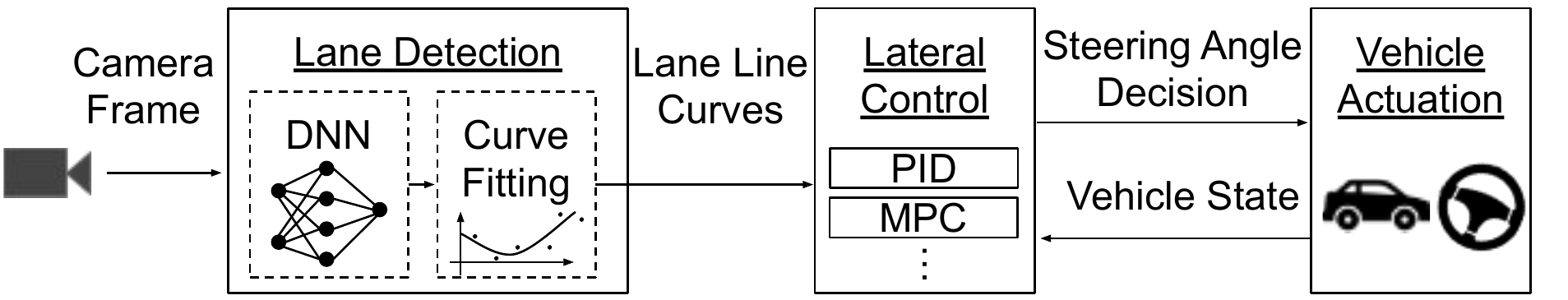}
\vspace{-0.22in}
\caption{Overview of the typical ALC system design.}
\label{fig:alc_overview}
\vspace{-0.15in}
\end{figure}

Fig.~\ref{fig:alc_overview} shows an overview of a typical ALC system design~\cite{simlink2020lane, lee2012unified, openpilot}, which operates in 3 steps:

\textbf{Lane Detection (LD).} Lane detection (LD) is the most critical step in an ALC system, since the driving decisions later are mainly made based on its output. Today, production ALC systems predominately use front cameras for this step~\cite{gmsupercruise, tesla2020autopilot, hondasensing, toyotasafetysense}.
On the camera frames, an LD model is used to detect lane lines. Recently, DNN-based LD models achieve the state-of-the-art accuracy~\cite{wang2018lanenet, pan2018spatial, ko2020key} and thus are adopted in the most performant production ALC systems today such as Tesla Autopilot~\cite{tesla2020autopilot}. Since lane line shapes do not change much across consecutive frames, recurrent DNN structure (e.g., RNN) is widely adopted in LD models to achieve more stable prediction~\cite{li2016deep, zou2019robust, openpilot}. LD models typically first predict the lane line points, and then post-process them to lane line curves using curve fitting algorithms~\cite{wang2018lanenet, smuda2006multiple, pan2018spatial, gackstatter2010stable}.

Before the LD model is applied, a Region of Interest (ROI) filtering is usually performed to the raw camera frame to crop the most important area out of it (\ie, the road surface with lane lines) as the model input. Such ROI area is typically around the center and much smaller than the original frame, to improve the model performance and accuracy~\cite{yenikaya2013keeping}. %
\\
\textbf{\hspace*{1em}Lateral control.}
This step calculates \textit{steering angle decisions} to keep the vehicle driving at the center of the detected lane. It first computes a desired driving path, typically at the center of the detected left and right lane lines~\cite{becker2018functional}. Next, a control loop mechanism, \eg, Proportional-Integral-Derivative (PID)~\cite{dorf2011modern} or Model Predictive Control (MPC) \cite{MPC}, is applied to calculate the optimal steering angle decisions that can follow the desired driving path as much as possible considering the vehicle state and physical constraints.\\
\textbf{\hspace*{1em}Vehicle actuation.} This step interprets the steering angle decision into actuation commands in the form of \textit{steering angle changes}. Here, such actuated changes are limited by a maximum value due to the physical constraints of the mechanical control units and also for driving stability and safety~\cite{becker2018functional}. For example, in our experiments with a production ALC with 100 Hz control frequency, such limit is 0.25$^\circ$ per control step (every 10 ms) for vehicle models~\cite{tinkla2018lane}.
As detailed later in~\S\ref{sec:problem_formulation_challenges}, such a steering limit prevents ALC systems from being affected too much from successful attack in one single LD frame, which introduces a unique challenge to our design.

\nsubsection{Physical-World Adversarial Attacks}
\label{sec:background_aml}

Recent works find that DNN models are generally vulnerable to adversarial examples, or adversarial attacks~\cite{Szegedy2014, goodfellow2014explaining}. Some works further explored such attacks in the physical world~\cite{kurakin2016adversarial_b, sharif2016accessorize, athalye2018synthesizing, brown2017advpatch, chen2018shapeshifter, eykholt2018physical, jack2018caraml, zhao2018seeing, pei2017deepxplore, tian2018deeptest, chernikova2019self, zhou2018deepbillboard}. While these prior works concentrate on DNN models for image classification and object detection tasks, we are the first to systematically study such attacks on production DNN-based ALC systems, which requires to address several new and unique design challenges as detailed later in~\S\ref{sec:problem_formulation_challenges}.

\nsection{Attack Formulation and Challenge} \label{sec:problem_formulation}

\vspace{0.15in}
\nsubsection{Attack Goal and Incentives}

\label{sec:problem_formulation_goals}
\label{sec:problem_formulation_incentive}

In this paper, we consider an attack goal that directly breaks the design goal of ALC systems: causing the victim vehicle a lateral deviation (i.e., deviating to the left or right) large enough to drive out of the current lane boundaries. Meanwhile, since ALC systems assume a fully-attentive human driver who is prepared to take over at any moment~\cite{tesla2020support, sae2018}, such deviation needs to be achieved fast enough so that the human driver cannot react in time to take over and steer back. Table~\ref{tbl:required_dev_time} shows concrete values of these two requirements for successful attacks on highway and local roads respectively, which will be used as evaluation metrics later in~\S\ref{sec:eval}. In the table, the required deviations are calculated based on representative vehicle and lane widths in the U.S., and the required success time is determined using commonly-used average driver reaction time to road hazards, which is detailed in Appendix~\ref{appendix:deviation_calculation}.

\textbf{Targeted scenario: Free-flow driving.} Our study targets the most common driving scenario for using ALC systems: \textit{free-flow} driving scenarios~\cite{zhao2017trafficnet}, in which a vehicle has at least 5--9 seconds clear headway~\cite{boora2017identification} and thus can drive freely without considering the front vehicle~\cite{zhao2017trafficnet}.

\textbf{Safety implications.} The attack goal above can directly cause various safety hazards in the real world: (1) \textit{Driving off road}, which is a direct violation of traffic rules~\cite{offroad_violate_law} and can cause various safety hazards such as hitting road curbs or falling down the highway cliff. (2) \textit{Vehicle collisions}, \eg, with vehicles parked on the road side, or driving in adjacent or opposite traffic lanes \diff{on a local road or a two-lane undivided highway}. Even with obstacle or collision avoidance, these collisions are still possible for two reasons. First, today's obstacle and collision avoidance systems are not perfect. For example, a recent study shows that the AEB (Automatic Emergency Braking) systems in popular vehicle models today fail to avoid crashes 60\% of the time~\cite{aeb-fail}. Second, even if they can successfully perform emergency stop, they cannot prevent the victim from being hit by other vehicles that fail to yield on time. Later in~\S\ref{sec:real_vehicle}, we evaluate the safety impacts of our attack with a simulator and a real vehicle.

\begin{table}[t]
\caption{Required deviations and success time for successful attacks on ALC systems on highway and local roads. Detailed calculations and explanations are in Appendix~\ref{appendix:deviation_calculation}.}
\vspace{-0.1in}
\label{tbl:required_dev_time}
\centering
\footnotesize
\renewcommand{\arraystretch}{0.8}
\begin{tabular}{@{}ccc@{}}
\toprule
Road Type & Required Lateral Deviation & Required Success Time                                                                                                   \\ \midrule
Highway   & 0.735 meters       &\multirow{2}{*}{\begin{tabular}[c]{@{}c@{}}\textless 2.5 seconds (average driver\\reaction time to road hazard)\end{tabular}} \\
Local road     & 0.285 meters       &                                                                                                                         \\ \bottomrule
\end{tabular}
\vspace{-0.2in}
\end{table}

\nsubsection{Threat Model}
\label{sec:problem_formulation_threat_model}
We assume that the attacker can obtain the same ALC system as the one used by the victim to get a full knowledge of its implementation details. This can be done through purchasing or renting the victim vehicle model and reverse engineering it, which has already been demonstrated possible on Tesla Autopilot~\cite{tencent2019}. Moreover, there exist production ALC systems that are open sourced~\cite{openpilot}. We also assume that the attacker can obtain a motion model~\cite{bicyclemodel} of the victim vehicle, which will be used in our attack generation process (\S\ref{sec:design_motion_model}). This is a realistic assumption since the most widely-used motion model (used by us in~\S\ref{sec:design_motion_model}) only needs vehicle parameters such as steering ratio and wheelbase as input~\cite{bicyclemodel}, which can be directly found from vehicle model specifications. We assume the victim drives at the speed limit of the target road, which is the most common case for free-flow driving. In the attack preparation time, we assume that the attacker can collect the ALC inputs (\eg, camera frames) of the target road by driving the victim vehicle model there with the ALC system on.

\vspace{-0.04in}
\nsubsection{Design Challenges}
\vspace{-0.04in}
\label{sec:problem_formulation_challenges}

Compared to prior works on physical-world adversarial attacks on DNNs, we face 3 unique design challenges:

\textbf{C1. Lack of legitimately-deployable attack vector in the physical world.} To affect the camera input of an ALC system, it is ideal if the malicious perturbations can appear legitimately around traffic lane regions in the physical world. To achieve high legitimacy, such perturbations also must not change the original human-perceived lane information. Prior works use small stickers or graffiti in physical-world adversarial attacks~\cite{eykholt2018robust, zhao2018seeing, tencent2019}. However, directly performing such activities to traffic lanes in public is illegal~\cite{graffiti_vandalism_law}. In our problem setting, the attacker needs to operate in the middle of the road when deploying the attack on traffic lanes. Thus, if the attack vector cannot be disguised as legitimate activities, it becomes highly difficult to deploy the attack in practice.

\textbf{C2. Camera frame inter-dependency due to attack-influenced vehicle actuation.} In real-world ALC systems, a successful attack on one single frame can barely cause any meaningful lateral deviations \diff{due to the steering angle change limit at the vehicle actuation step (\S\ref{sec:background_alc}). For example, for the vehicle models with 0.25$^\circ$ angle change limit per control loop (\S\ref{sec:background_alc}), even if a successful attack on a single frame causes a very large steering angle decision at MPC output (e.g., 90$^\circ$), it can only cause at most 1.25$^\circ$ actuated steering angle changes before the next frame comes, which can only cause up to \textit{0.3-millimeter} lateral deviations at 45 mph ($\sim$72 km/h). More detailed explanations are in Appendix~\ref{appendix:openpilot_details}.}

\diff{Thus, to achieve our attack goal in \S\ref{sec:problem_formulation_goals}, the attack must be \textit{continuously effective on sequential camera frames} to increasingly reach larger actuated steering angles and thus larger lateral deviations per frame.} In this process, due to the dynamic vehicle actuation applied by the ALC system, the attack effectiveness for later frames are directly dependent on that for earlier frames. For example, if the attack successfully deviates the detected lane to the right in a frame, the ALC system will steer the vehicle to the right accordingly. This causes the following frames to capture road areas more to the right, and thus directly affect their attack generation. There are prior works considering attack robustness across sequential frames, e.g., using EoT~\cite{athalye2018synthesizing, brown2017advpatch} and universal perturbation~\cite{shasha2019stealth}, but none of them consider frame inter-dependencies due to attack-influenced vehicle actuation in our problem setting.

\textbf{C3. Lack of differentiable objective function design for LD models.} To systematically generate adversarial inputs, prior works predominately adopt optimization-based approaches, which have shown both high efficiency and effectiveness~\cite{Szegedy2014, eykholt2018robust, carlini2018audio, ebrahimi2018adversarial}. However, the objective function designs in these prior works are mainly for image classification~\cite{eykholt2018robust, brown2017advpatch} or object detection~\cite{chen2018shapeshifter, eykholt2018physical, zhao2018seeing} models, which thus aim at decreasing class or bounding box probabilities. However, as introduced in~\S\ref{sec:background_alc}, LD models output detected lane line curves, and thus to achieve our attack goal the objective function needs to aim at changing the \textit{shape} of such curves. This is substantially different from decreasing probability values, and thus none of these existing designs can directly apply.

Closer to our problem, prior works that attack end-to-end autonomous driving models~\cite{pei2017deepxplore, tian2018deeptest, chernikova2019self, zhou2018deepbillboard} directly design their objective function to change the final steering angle decisions. However, as described in~\S\ref{sec:background_alc}, state-of-the-art LD models do not directly output steering angle decisions. Instead, they output lane line curves and rely on the lateral control step to compute the final steering angle decisions. However, many steps in the lateral control module, e.g., the desired driving patch calculation and the MPC framework, are generally not differentiable to the LD model input (i.e., camera frames), which makes it difficult to effectively optimize.

\vspace{-0.01in}
\nsection{Dirty Road Patch Attack Design}
\vspace{-0.05in}
\label{sec:design}
In this paper, we are the first to systematically address the design challenges above by designing a novel physical-world attack method on ALC, called \textit{Dirty Road Patch (DRP) attack}.

\vspace{-0.04in}
\nsubsection{Design Overview}
\vspace{-0.04in}
\label{sec:design_drp_overview}

\begin{figure}[t]
    \centering
    \includegraphics[width=\columnwidth]{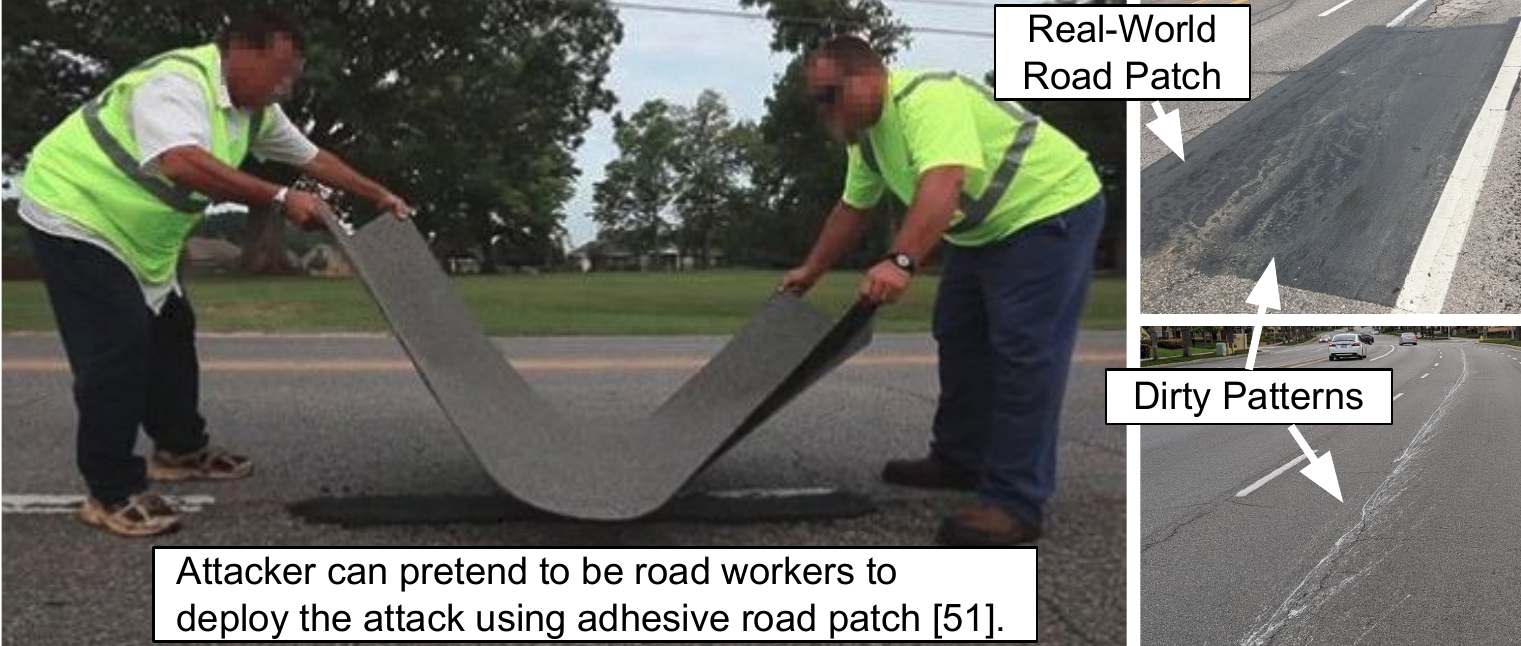}
    \vspace{-0.3in}
    \caption{Illustration of our novel and domain-specific attack vector: Dirty Road Patch (DRP).}
    \label{fig:threat_model}
    \vspace{-0.28in}
\end{figure}

To address the 3 design challenges in~\S\ref{sec:problem_formulation_challenges}, our DRP attack method has the following novel design components:

\textbf{Dirty road patch: Domain-specific \& stealthy physical-world attack vector.} To address challenge \textbf{C1}, we are the first to identify \textit{dirty road patch} as an attack vector in physical-world adversarial attacks. This design has 2 unique advantages. First, road patches can appear to be legitimately deployed on traffic lanes in the physical world, e.g., for fixing road cracks. Today, deploying them is made easy with adhesive designs~\cite{adhesive_patch} as shown in Fig.~\ref{fig:threat_model}. The attacker can thus take time to prepare the attack in house by carefully printing the malicious input perturbations on top of such adhesive road patches, and then pretend to be road workers like those in Fig.~\ref{fig:threat_model} to quickly deploy it when the target road is the most vacant, e.g., in late night, to avoid drawing too much attention.

Second, since it is common for real-world roads to have dirt or white stains such as those in Fig.~\ref{fig:threat_model}, using similar dirty patterns as the input perturbations can allow the malicious road patch to appear more normal and thus stealthier. To mimic the normal dirty patterns, our design only allows color perturbations on the gray scale, i.e., black-and-white. To avoid changing the lane information as discussed in~\S\ref{sec:problem_formulation_challenges}, in our design we (1) require the original lane lines to appear exactly the same way on the malicious patch, if covered by the patch, and (2) restrict the brightness of the perturbations to be strictly lower than that of the original lane lines. To further improve stealthiness, we also design parameters to adjust the perturbation size and pattern, which are detailed in~\S\ref{sec:design_opt_stealthy}.

So far, none of the popular production ALC systems today such as Tesla, GM, etc.~\cite{tesla2020support, gm_cadillac_ct6 ,hondasensing,toyotasafetysense,volvoxc90manual,kiamanual,fordescapemanual,nissanroguesports,hyundaisonatamanual} identify roads with such dirty road patches as driving scenarios that they do not handle, which can thus further benefit the attack stealthiness.

\textbf{Motion model based input generation.} 
To address the strong inter-dependencies among the camera frames (\textbf{C2}), we need to dynamically update the content of later camera frames according to the vehicle actuation decisions applied at earlier ones in the attack generation process. Since adversarial attack generation typically takes thousands of optimization iterations~\cite{carlini2017towards, madry2017towards}, it is practically highly difficult, if not impossible, to drive real vehicles on the target road to obtain such dynamic frame update in every optimization iteration. Another idea is to use vehicle simulators~\cite{lgsvl, dosovitskiy2017carla}, but it requires the attacker to first create a high-definition 3D scene of the target road in the real world, which requires a significant amount of hardware resource and engineering efforts. Also, launching a vehicle simulator in each optimization iteration can greatly harm the attack generation speed.

To efficiently and effectively address this challenge, we combine \textit{vehicle motion model}~\cite{bicyclemodel} and \textit{perspective transformation}~\cite{hartley2003perspective, tanaka2011perspective} to dynamically synthesize camera frame updates according to a driving trajectory simulated in a lightweight way. This method is inspired by Google Street View~\cite{anguelov2010google} that synthesizes 360$^{\circ}$ views from a limited number of photos utilizing perspective transformation. Our method only requires one trace of the ALC system inputs (i.e., camera frames) from the target road without attack, which can be easily obtained by the attacker (\S\ref{sec:problem_formulation_threat_model}).

\textbf{Optimization-based DRP generation.} To systematically generate effective malicious patches, we adopt an optimization-based approach similar to prior works~\cite{Szegedy2014, eykholt2018robust}. To address challenge \textbf{C3}, we design a novel lane-bending objective function as a differentiable surrogate that aims at changing the derivatives of the desired driving path before the lateral control module, which is equivalent to change the steering angle decisions at the lateral control design level. Besides this, we also have other domain-specific designs in the optimization problem formulation, e.g., for a differentiable construction of the curve fitting process, malicious road patch robustness, stealthiness, and physical-world realizability.

\begin{figure}[t!]
\centering
\includegraphics[width=\columnwidth]{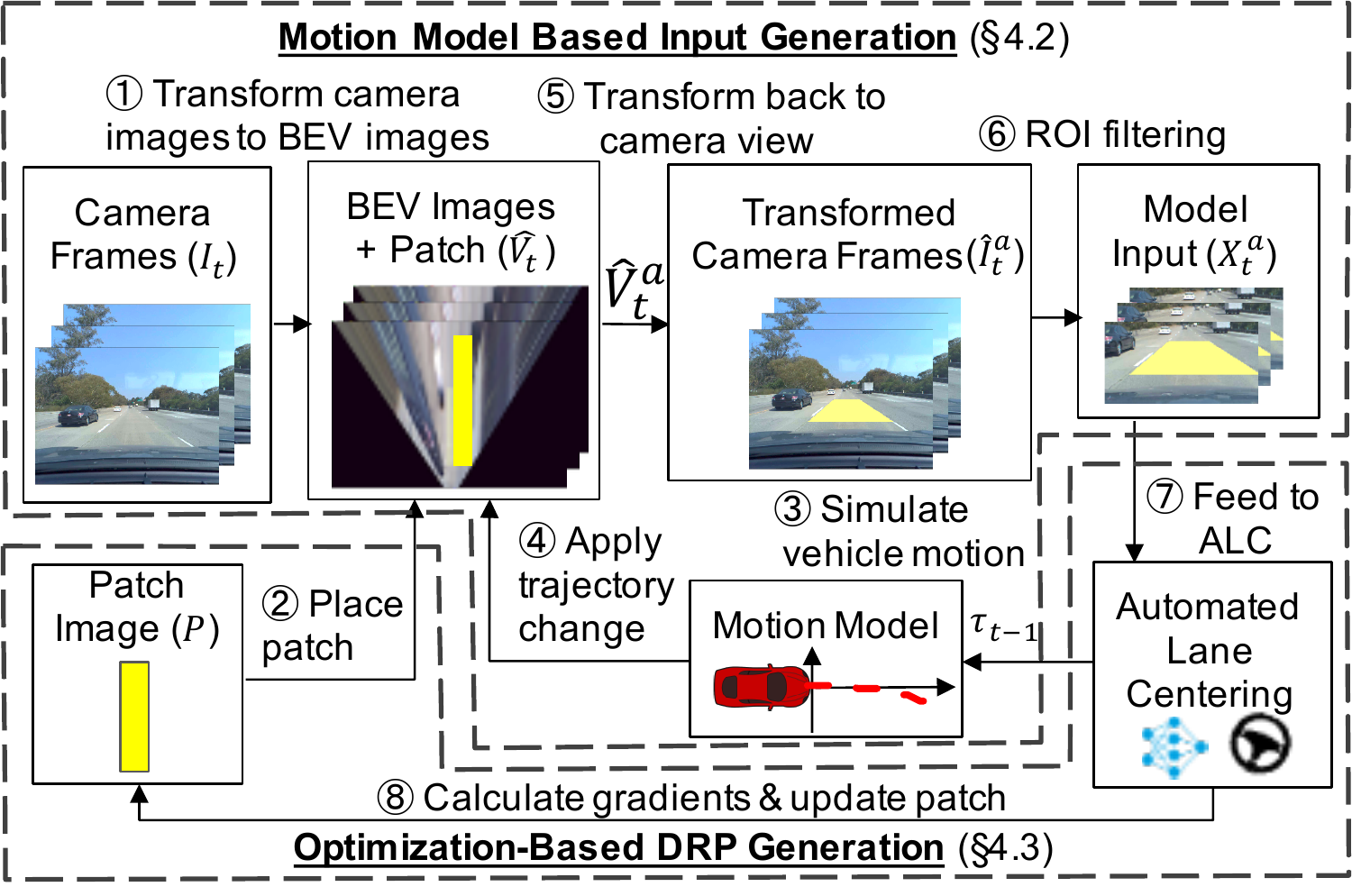}
\vspace{-0.3in}
\caption{Overview of our DRP (Dirty Road Patch) attack method. ROI: Region of Interest; BEV: Bird's Eye View.}
\label{fig:drp_attack_overview}
\vspace{-0.25in}
\end{figure}

Fig.~\ref{fig:drp_attack_overview} shows an overview of the malicious road patch generation process, which is detailed in the following sections.

\nsubsection{Motion Model based Input Generation} \label{sec:design_motion_model}

In Fig.~\ref{fig:drp_attack_overview}, step \circled{1}--\circled{7} belong to the motion model based input generation component. As described earlier in~\S\ref{sec:design_drp_overview}, the input to this component is a trace of ALC system inputs such as camera frames from driving on the target road without attack. In \circled{1}, we apply \textit{perspective transformation}, a widely-used computer vision technique that can project an image view from a 3D coordinate system to a 2D plane~\cite{hartley2003perspective, tanaka2011perspective}. Specifically, we apply it to the original camera frames from the driver's view to obtain their Bird's Eye View (BEV) images. This transformation is highly beneficial since it makes our later patch placement and attack-influenced camera frame updates much more natural and thus convenient. We denote this as $V_t := {\rm BEV}(I_t)$, where $I_t$ and $V_t$ are the original camera input and its BEV view respectively at frame $t$. This process is inversible, i.e.,  we can also obtain $I_t$ with ${\rm BEV}^{-1}(V_t)$.

Next, in \circled{2}, we obtain the generated malicious road patch image $P$ from the optimization-based DRP generation step (\S\ref{sec:design_opt_patch_gen}) and place it on $V_t$ to obtain the BEV image with the patch, denoted as $\widehat{V}_t := \Lambda(V_t, P)$. To achieve consistent patch placements in the world coordinate across frames, we calculate the \textit{pixel-meter relationship}, i.e., the number of pixels per meter, in BEV images based on the driving trace of the target road. With this, we can place the patch in each frame precisely based on the driving trajectory changes across frames.

Next, we compute the vehicle moving trajectory changes caused by the placed malicious road patch, and reflect such changes in the camera frames. We represent the vehicle moving trajectory as a sequence of vehicle states $S_t := [x_t, y_t, \beta_t, v_t], (t = 1, ..., T)$, where $x_t, y_t, \beta_t, v_t$ are the vehicle's 2D position, heading angle, and speed at frame $t$, and $T$ is the total number of frames in the driving trace. Thus, the trajectory change at frame $t$ is $\delta_t := S^{a}_t - S^{o}_t$, where $S^{a}_t$ and $S^{o}_t$ are vehicle states with and without attack respectively.

To calculate $\delta_{t}$ caused by the attack effect at the frame $t-1$, we need to know the attack-influenced vehicle state $S^{a}_{t}$. To achieve that, we use a \textit{vehicle motion model} to simulate the vehicle state $S^{a}_{t}$ by feeding the steering angle decision $\tau_{t-1}$ from the lateral control step in the ALC system (\S\ref{sec:background_alc}) given the attacked frame at $t-1$ and the previous vehicle state $S^{a}_{t-1}$, denoted as $S^{a}_{t} := {\rm MM}(S^{a}_{t-1}, \tau_{t-1})$. A vehicle motion model is a set of parameterized mathematical equations representing the vehicle dynamics and can be used to simulate its driving trajectory given the speed and actuation commands. In this process, we set the vehicle speed as the speed limit of the target road as described in our threat model (\S\ref{sec:problem_formulation_threat_model}). In our design, we adopt the kinematic bicycle model~\cite{kong2015kinematic}, which is the most widely-used motion model for vehicles~\cite{coursera, kong2015kinematic,watzenig2016automated}.

\begin{figure*}[t!]
    \begin{minipage}[t]{0.31\linewidth}
        \centering
        \includegraphics[width=\columnwidth]{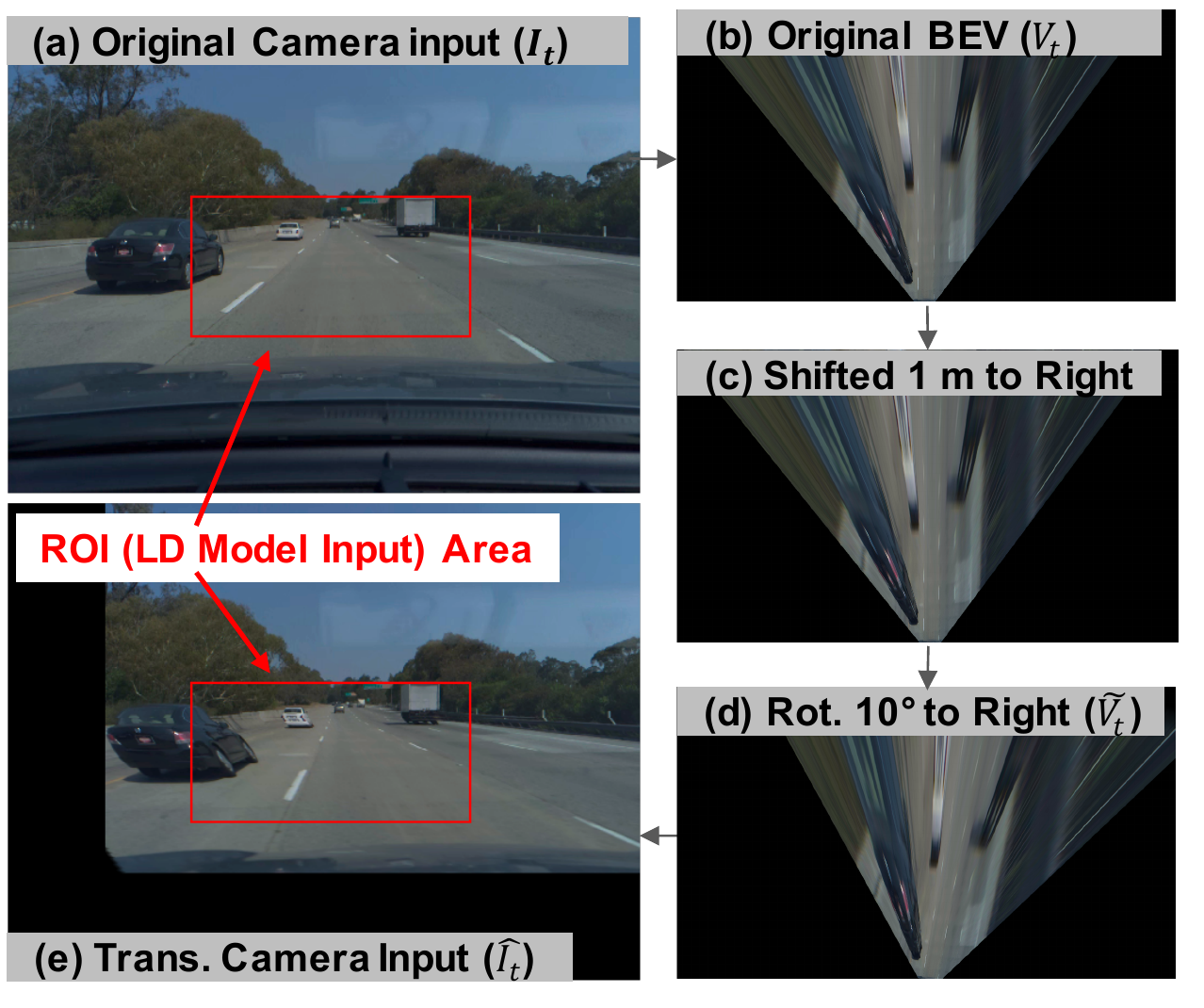}
        \vspace{-0.32in}
        \caption{Motion model based input generation from original camera input. 
        }
        \label{fig:input_generation}
    \end{minipage}
    \hspace{0.01\linewidth}
    \begin{minipage}[t]{0.42\linewidth }
        \centering
        \includegraphics[width=\columnwidth]{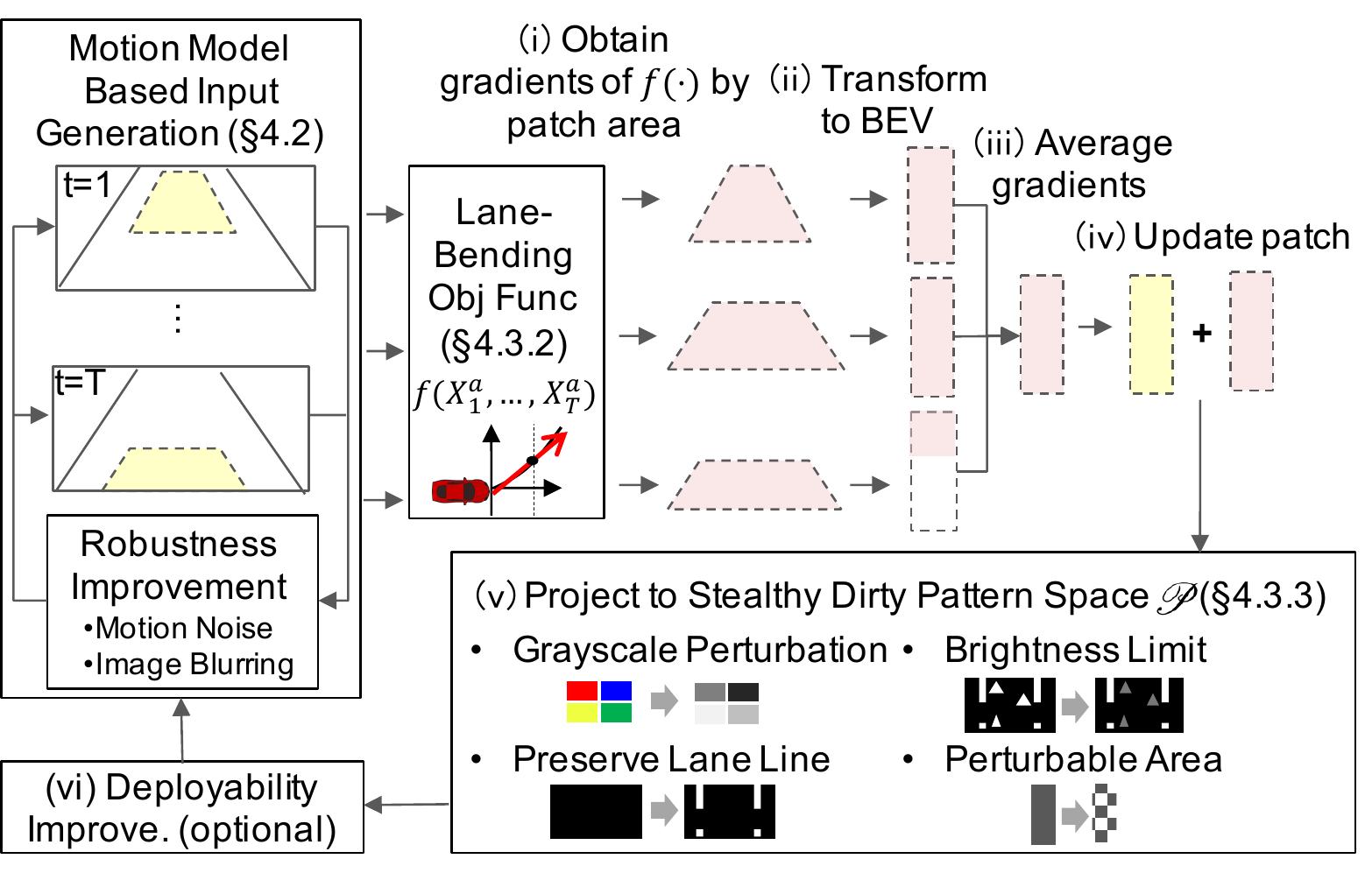}
        \vspace{-0.32in}
        \caption{Iterative optimization process design for our optimization-based DRP generation.
        }
        \label{fig:grad_agg}
    \end{minipage}
    \hspace{0.01\linewidth}
    \begin{minipage}[t]{0.22\linewidth}
        \vspace{-1.8in}
        \centering
        \includegraphics[width=0.8\columnwidth]{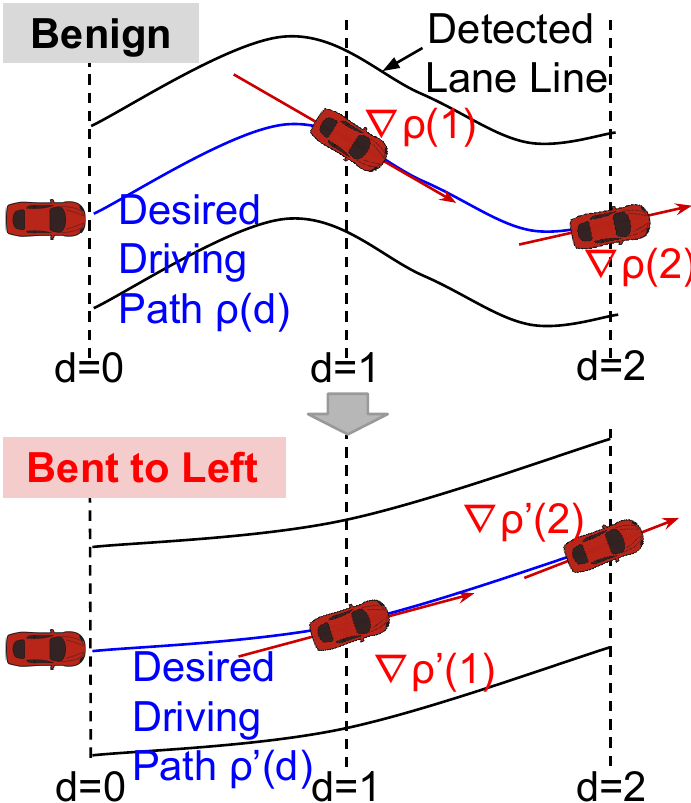}
        \vspace{-0.14in}
        \caption{``Lane bending'' effect of our objective function by maximizing $\nabla \rho(d)$ at each curve point. %
        }
        \label{fig:object_function}
    \end{minipage}
\vspace{-0.2in}
\end{figure*}

With $\delta_{t}$, in \circled{4} we then apply affine transformations on the BEV image $\widehat{V}_{t}$ to obtain the attack-influenced one $\widehat{V}^a_{t}$, denoted as $\widehat{V}^a_{t} := T(\widehat{V}_{t}, \delta_{t})$. Fig.~\ref{fig:input_generation} shows an example of the shifting and rotation $T(\cdot)$ in the BEV, which synthesizes a camera frame with the vehicle position shifted by 1 meter and rotated by 10$^{\circ}$ to the right. Although it causes some distortion and missing areas on the edge, the ROI area (red rectangle), i.e., the LD model input, is still complete and thus sufficient for our purpose. Since the ROI area is typically focused on the center and much smaller than the raw camera frame (\S\ref{sec:background_alc}), our method can successfully synthesize multiple complete LD model inputs from only 1 ALC system input trace.

Next, in \circled{5}, we obtain the attack-influenced camera frame at the driver's view $\widehat{I}^a_{t}$, i.e., the direct input to ALC, by projecting $\widehat{V}^a_{t}$ back using $\widehat{I}^a_{t} := {\rm BEV^{-1}}(\widehat{V}^a_{t})$. Next, in \circled{6}, the ROI filtering is used to extract the model input $X^a_t := {\rm ROI}(\widehat{I}^a_t)$. $X^a_t$ and vehicle state $S^{a}_t$ are then fed to ALC system in \circled{7} to obtain the steering angle decision $\tau_t$, denoted as $\tau_t := {\rm ALC}(X^a_t, S^{a}_t)$. Step \circled{3}--\circled{7} are then iteratively applied to obtain $\widehat{I}^a_{t+1}, \widehat{I}^a_{t+2}, ...$ one after one until all the original frames are updated to reflect the moving trajectory changes caused by $P$. These updated attack-influenced inputs are then fed to the optimization-based DRP generation component, which is detailed next.

\nsubsection{Optimization-Based DRP Generation} \label{sec:design_opt_patch_gen}

In Fig.~\ref{fig:drp_attack_overview}, step \circled{8} belongs to the optimization-based road path generation component. In this step, we design a domain-specific optimization process on the target ALC system to systematically generate the malicious dirty road patch $P$.

\textbf{DRP attack optimization problem formulation.} We formulate the attack as the following optimization problem:
\vspace{-0.1in}
\small
\begin{align}
    \min \ \ 
    & \mathscr{L} \label{math:obj_l}\\[-5pt]
    {\rm s.t.} \ \
    & X^a_t = {\rm ROI}({\rm BEV}^{-1}(T(\Lambda(V_t, P), S^{a}_t-S^{o}_t))) \ \  (t = 1,...,T)\label{math:input_gen}\\[-3pt]
    & \tau^a_t = {\rm ALC}(X^a_t, S^{a}_t) \hspace{10em}  (t = 1,...,T) \label{math:alc}\\[-3pt]
    & S^{a}_{t+1} = {\rm MM}(S^{a}_t, \tau^a_t) + \epsilon_t \hspace{6.2em} (t = 1,...,T-1) \label{math:mm}\\[-3pt]
    & S^{a}_1 = S^{o}_1 \label{math:init}\\[-3pt]
    & P = {\rm BLUR} ({\rm FILL}(B) + \Delta)  \label{math:patch_decomp}\\[-3pt]
    & \Delta \in \mathscr{P} \label{math:stealthy_space}
    \vspace{-0.3in}
\end{align}
\vspace{-0.28in}
\normalsize

\noindent{}where the $\mathscr{L}$ in Eq.~\ref{math:obj_l} is an objective function that aims at deviating the victim out of the current lane boundaries as fast as possible (detailed in~\S\ref{sec:design_lane_bending}). Eq.~\ref{math:input_gen}--\ref{math:init} have been described in~\S\ref{sec:design_motion_model}. In Eq.~\ref{math:patch_decomp}, the patch image $P \in \mathbb{R}^{H \times W \times C}$ consists of a base color $B \in \mathbb{R}^{C}$ and the perturbation $\Delta \in \mathbb{R}^{H \times W \times C}$, where $W, H$, and $C$ are the patch image width, height, and the number of color channels respectively. We select an asphalt-like color as the base color $B$ since the image is designed to mimic a road patch. Function FILL$:\mathbb{R}^C \rightarrow \mathbb{R}^{H \times W \times C}$ fills $B$ to the entire patch image. Since we aim at generating perturbations that mimic the normal dirty patterns on roads, we restrict $\Delta$ to be within a stealthy road pattern space $\mathscr{P}$, which is detailed in~\S\ref{sec:design_opt_stealthy}. We also include a noise term $\epsilon_t$ in Eq.~\ref{math:mm} and an image blurring function ${\rm BLUR}(\cdot)$ in Eq.~\ref{math:patch_decomp} to improve the patch robustness to vehicle motion model inaccuracies and camera image blurring, which are detailed in~\S\ref{sec:design_opt_robust}.

\nsubsubsection{Optimization Process Overview}
\label{sec:design_opt_overview}

Fig.~\ref{fig:grad_agg} shows an overview of our iterative optimization process design. Given an initial patch image $P$, we obtain the model input $X^a_{1},...,X^a_T$ from the motion model based input generation process. In step (i), we calculate the gradients of the objective function with respect to $X^a_{1},...,X^a_T$, and only keep the gradients corresponding to the patch areas. In step (ii), these gradients are projected into the BEV space. In step (iii), we calculate the average BEV-space gradients weighted by their corresponding patch area sizes in the model inputs. \diff{This step involves an approximation of the gradient of $\text{BEV}^{-1}(\cdot)$, which are detailed in Appendix~\ref{appendix:design_details}.} Next, in step (iv), we update the current patch with Adam~\cite{kingma2014adam} using the averaged gradient as the gradient of the patch image. In step (v), we then project the updated patch into the stealthy road pattern space $\mathscr{P}$. This updated patch image is then fed back to the motion model based input generation module, where we also add robustness improvement such as motion noises and image blurring. We terminate this process when the attack-introduced lateral deviations obtained from the motion model  are large enough.

\nsubsubsection{Lane-Bending Objective Function Design} \label{sec:design_lane_bending}
As discussed in~\S\ref{sec:design_drp_overview}, directly using steering angle decisions as $\mathscr{L}$ makes the objective function non-differentiable to $X^a_{1},...,X^a_T$. To address this, we design a novel lane-bending objective function $f(\cdot)$ as a differentiable surrogate function. In this design, our key insight is that at the design level, the lateral control step aims at making steering angle decisions that follows a \textit{desired driving path} in the middle of the detected left and right lane line curves from the lane detection step (\S\ref{sec:background_alc}). Thus, changing the steering angle decisions is equivalent to changing the derivatives of (or ``bending'') such desired driving path curve. This allows us to design $f(\cdot)$ as:

\small
\vspace{-0.25in}
\begin{align}
    &f(X^a_1,...,X^a_T) = 
    \sum_{t=1}^{T}\sum_{d \in D_t} \nabla \rho_{t}(d; { \{X^a_j| j \leq t\}, \theta}) + \lambda ||\Omega_t(X^a_t)||_p
    \label{math:objective}
\end{align}
\vspace{-0.16in}
\normalsize

\noindent{}where $\rho_t(d)$ is a parametric curve whose parameters are decided by (1) both the current and previous model inputs $ \{X^a_j| j \leq t\}$ due to frame inter-dependencies (\S\ref{sec:problem_formulation_challenges}), and (2) the LD DNN parameters $\theta$. $D_t$ is a set of curve point index $d=0, 1, 2, ...$ for the desired driving path curve at frame $t$.
$\lambda$ is the weight of the $p$-norm regularization term, designed for stealthiness (\S\ref{sec:design_opt_stealthy}). We then can define $\mathscr{L}$ in Eq.~\ref{math:obj_l} as $f(\cdot)$ and $-f(\cdot)$ when attacking to the left and right. Fig.~\ref{fig:object_function} illustrates this surrogate function when attacking to the left. As shown, by maximizing $\nabla \rho_{t}(d)$ at each curve point in Eq.~\ref{math:objective}, we can achieve a ``lane bending'' effect to the desired driving path curve. Since the direct LD output is lane line points (\S\ref{sec:background_alc}) but $\rho_{t}(\cdot)$ require lane line curves, we further perform a differentiable construction of curve fitting process (Appendix~\ref{appendix:design_details}).

\vspace{-0.01in}
\nsubsubsection{Designs for Dirty Patch Stealthiness}
\vspace{-0.04in}
\label{sec:design_opt_stealthy}

To mimic real-world dirty patterns like in Fig.~\ref{fig:threat_model}, we have 4 stealthiness designs in  stealthy road pattern space $\mathscr{P}$ in Eq.~\ref{math:stealthy_space}:

\textbf{Grayscale perturbation.} Real-world dirty patterns on the road are usually created by dust or white stains (Fig.~\ref{fig:threat_model}), and thus most commonly just appear white. Thus, we cannot allow perturbations with arbitrary colors like prior works~\cite{zhao2018seeing}.
Thus, our design restricts our perturbation $\Delta$ in the grayscale (i.e., black-and-white) by only allowing increase the Y channel in the YCbCr color space~\cite{hamilton2004jpeg}, denoted as $\Delta_{Y} \geq 0$.

\textbf{Preserving original lane line information.} We preserve the original lane line information by drawing the same lane lines as the original ones on the patch (if covered by the patch). Note that without this our attack can be easier to succeed, but as discussed in~\S\ref{sec:problem_formulation_challenges}, it is much more preferred to preserve such information so that the attack deployment can more easily appear as legitimate road work activities and the deployed patch is less likely to be legitimately removed.

\textbf{Brightness limits.} While the dirty patterns are restricted to grayscale, they are still the darker, the stealthier. Also, to best preserve the original lane information, the brightness of the dirty patterns should not be more than the original lane lines. Thus, we (1) add the $p$-norm regularization term in Eq.~\ref{math:objective} to suppress the amount of $\Delta_{Y}$, and (2) restrict $B_{Y} + \Delta_{Y} < {\rm LaneLine}_{Y}$, where $B_{Y}$ and ${\rm LaneLine}_Y$ are Y channel values for the base color and original lane line color respectively.

\textbf{Perturbation area restriction.} Besides brightness, also the fewer patch areas are perturbed, the stealthier. Thus, we define Perturbable Area Ratio (PAR) as the percentage of pixels on $P$ that can be perturbed. Thus, when PAR=30\%, 70\% pixels on $P$ will only have the base color $B$.

\nsubsubsection{Designs for Improving Attack Robustness, Deployability, and Physical-World Realizability}
\label{sec:design_opt_robust}
\label{sec:multiple_patch}
\label{sec:design_opt_realize}

We also have domain-specific designs for improving (1) \textbf{attack robustness}, which addresses the driving trajectory/angle deviations and camera sensing inaccuracies in real-world attacks; (2) \textbf{attack deployability}, which designs an optional \textit{multi-piece patch attack} mode that allows deploying DRP attack with multiple small and quickly-deployable road patch pieces; and (3) \textbf{physical-world realizability}, which addresses the color and pattern distortions due to physical-world factors such as lighting condition, printer color accuracy, and camera color sensing capability. Details are in Appendix~\ref{appendix:rbst_deploy_real}.

\nsection{Attack Methodology Evaluation}
\vspace{-0.05in}
\label{sec:eval}

In this section, we evaluate the effectiveness, robustness, generality, and realizability of our DRP attack methodology.

\textbf{Targeted ALC system.} In our evaluation, we perform experiments on the production ALC system in \op{}~\cite{openpilot}, which follows the state-of-the-art DNN-based ALC system design (\S\ref{sec:background_alc}). \op{} is an open-source production Level-2 driving automation system that can be easily installed in over 80 popular vehicle models (e.g., Toyota, Cadillac, etc.) by mounting a dashcam. We select \op{} due to its (1) \textit{representativeness}, since it is reported to have close performance to Tesla Autopilot and GM Super Cruise and better than many others~\cite{vanderwerp2020opteslagm, openpilotreview, hall2020ophandson}, (2) \textit{practicality}, from the large quantity and diversity of vehicle models it can support~\cite{openpilot}, and (3) \textit{ease to experiment with}, since it is the only production ALC system that is open sourced. In this paper, we mainly evaluate on the lane detection model in \op{} v0.7.0, which is released in Dec. 2019. \diff{More details of the OpenPilot ALC system are in Appendix~\ref{appendix:openpilot_details}.}

\textbf{Evaluation dataset.} We perform experiments using the comma2k19 dataset~\cite{comma2k19}, which contains over 33 hours driving traces between California's San Jose and San Francisco in a Toyota RAV4 2017 driven by human drivers. These traces are collected using the official \op{} dashcam device, called EON.
From this dataset, we manually look for short free-flow driving periods to make road patch placement convenient. In total, we obtain 40 eligible short driving clips, 10 seconds each, with half of them on the highway, and half on local roads. For each driving clip, we consider two attack scenarios: attack to the left, and to the right. Thus, in total we evaluate 80 different attack scenarios.

\begin{figure*}[t]
    \begin{minipage}{.78\linewidth}
        \centering
        \includegraphics[width=\linewidth]{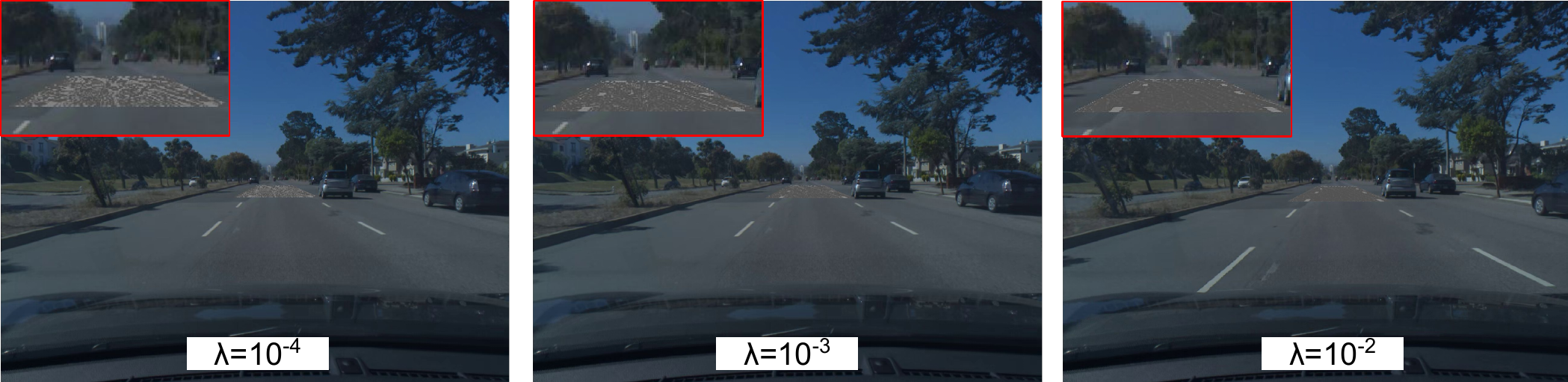}
        \vspace{-0.3in}
        \caption{Driver's view at 2.5 sec (average driver reaction time to road hazards~\cite{cali_driver_reaction_time}) before our attack succeeds under different stealthiness levels in local road scenarios. Inset figures are the zoomed-in views of the malicious road patches. Larger images for both local and highway are in Fig.~\ref{fig:poc_large_picture}.} 
        \label{fig:attack_poc}
    \end{minipage}\hspace{0.05in}
    \begin{minipage}{.2\linewidth}
      \vspace*{\fill}
      \centering
      \vspace{-0.05in}
      \includegraphics[width=\linewidth]{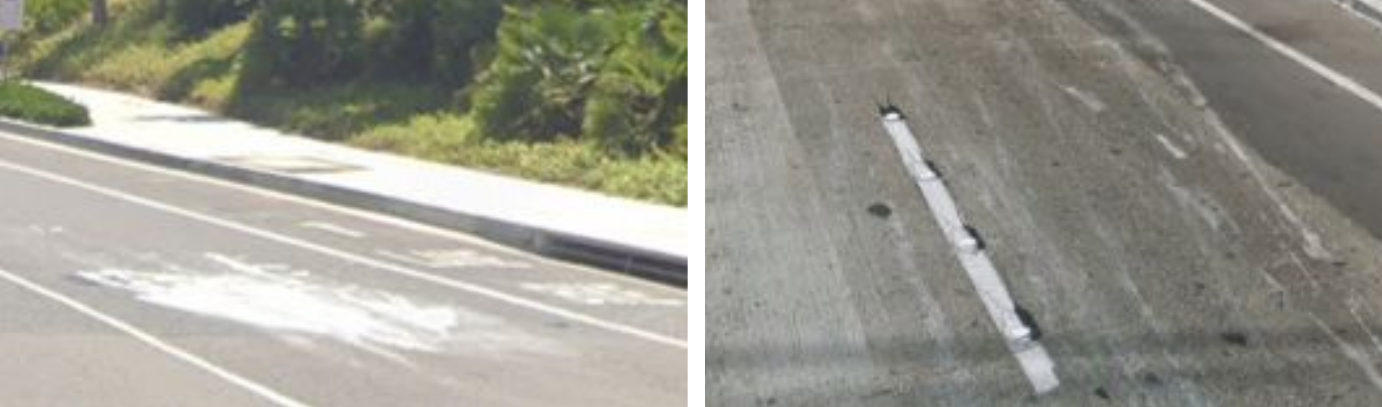}
      \vspace{-0.3in}
      \caption{Real-world dirty road patterns.}
      \label{fig:attack_poc_2.5}\par\vfill
      \includegraphics[width=\linewidth]{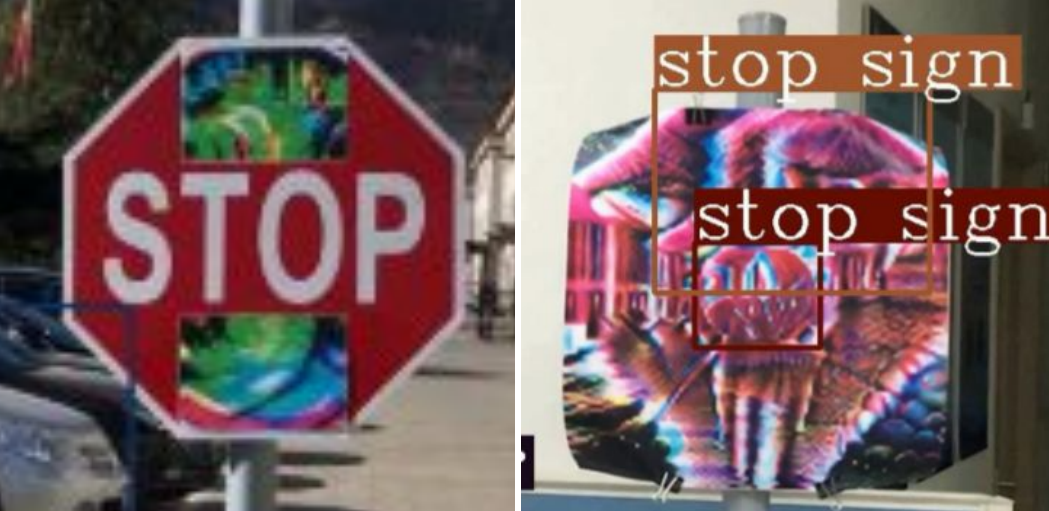}
      \vspace{-0.3in}
      \caption{Stop sign hiding and appearing attacks~\cite{zhao2018seeing}.}
      \label{fig:other_attacks}
    \end{minipage}
    \vspace{-0.2in}
\end{figure*}

\nsubsection{Attack Effectiveness} \label{sec:eval_results}

\textbf{Evaluation methodology and metrics.} We evaluate the attack effectiveness using the evaluation dataset described above. For each attack scenario, we generate an attack road patch, and use the motion model based input generation method in~\S\ref{sec:design_motion_model} to simulate the vehicle driving trajectory influenced by the malicious road patch. To judge the attack success, we use the attack goal defined in~\S\ref{sec:problem_formulation_goals} and concrete metrics listed in Table~\ref{tbl:required_dev_time}, i.e., achieving over 0.735m and 0.285m lateral deviations on highway and local road scenarios respectively within the average driver reaction, 2.5 sec. We measure the achieved deviation by calculating the lateral distances at each time point between the vehicle trajectories with and without the attack, and use the earliest time point to reach the required deviation to calculate the success time.

Since ALC systems assume a human driver who is prepared to take over, it is better if the malicious road patch can also look stealthy enough at 2.5 sec (driver reaction time) before the attack succeeds so that the driver won't be alerted by its looking and decide to take over. Thus, in this section, we also study the stealthiness of the generated road patches. Specifically, we quantify their perturbation degrees using the average pixel value changes from the original road surface in $\mathcal{L}_{1}, \mathcal{L}_{2}$ and $\mathcal{L}_{\inf}$ distances~\cite{ling2019deepsec, athalye2018obfuscated} and also a user study.

\textbf{Experimental setup.} For each scenario in the evaluation dataset, we manually mark the road patch placement area in the BEV view of each camera frame based on the lane width and shape. To achieve consistent road patch placements in the world coordinate across a sequence of frames, we calculate the number of pixels per meter in the BEV images and adjust the patch position in each frame precisely based on the driving trajectory changes across consecutive frames. The road patch sizes we use are 5.4 m wide, and 24--36 m long to ensure at least a few seconds of visible time at high speed. The patches are placed 7 m far from the victim at the starting frame. For stealthiness levels, we evaluate the $\mathcal{L}_2$ regularisation coefficient $\lambda=10^{-2}, 10^{-3}$, and $10^{-4}$, with PAR set to 50\%. According to Eq.~\ref{math:objective}, larger $\lambda$ value means more suppression of the perturbation, and thus should lead to a higher stealthiness level.
For the motion model, we directly use the vehicle parameters (e.g., wheelbase) of Toyota RAV4 2017, the vehicle model that collects the traces in our dataset.

\begin{table}[t]
\centering
\caption{Attack success rate and time under different stealthiness levels. Larger $\lambda$ means stealthier. Average success time is calculated only among the successful cases. Pixel $\mathcal{L}_{1}$, $\mathcal{L}_{2}$, and $\mathcal{L}_{inf}$ are the average pixel value changes from the original road surface in the RGB space and normalized to $[0, 1]$.}
\vspace{-0.12in}
\label{tbl:effectiveness}
 \footnotesize
\renewcommand{\arraystretch}{0.9}
\begin{tabular}{cccccc}
\toprule
\begin{tabular}[c]{@{}c@{}}Stealth. \\ Level $\lambda$\end{tabular} &
\begin{tabular}[c]{@{}c@{}}Succ.\\ Rate\end{tabular} & 
\begin{tabular}[c]{@{}c@{}}Succ.\\ Time (s)\end{tabular} &
\begin{tabular}[c]{@{}c@{}}Pixel \\$\mathcal{L}_{1}$ \end{tabular} &
\begin{tabular}[c]{@{}c@{}}Pixel \\$\mathcal{L}_{2}$ \end{tabular} &
\begin{tabular}[c]{@{}c@{}}Pixel \\$\mathcal{L}_{inf}$ \end{tabular} \\ \hline
$10^{-2}$ & 97.5\% & 0.903 & 0.018 & 0.045 & 0.201 \\
$10^{-3}$ & 100\% & 0.887 & 0.033 & 0.066 & 0.200 \\
$10^{-4}$ & 100\% & 0.886 & 0.071 & 0.109 & 0.200\\ \toprule
\end{tabular}
\vspace{-0.25in}
\end{table}

\textbf{Results.}
As shown in Table~\ref{tbl:effectiveness}, our attack has high effectiveness ($\ge$97.5\%) under all the 3 stealthiness levels. Fig.~\ref{fig:attack_poc} shows the malicious road patch appearances at different stealthiness levels from the driver's view at 2.5 seconds before our attack succeeds. %
As shown, even for the lowest stealthiness level ($\lambda=10^{-4}$) in our experiment, the perturbations are still smaller than some real-world dirty patterns such as the left one in Fig.~\ref{fig:attack_poc_2.5}. In addition, the perturbations for all these 3 stealthiness levels are a lot less intrusive than those in previous physical-world adversarial attacks in the image space~\cite{zhao2018seeing}, e.g., in Fig.~\ref{fig:other_attacks}. 
Among the successful cases, the average success time is all under 0.91 sec, which is substantially lower than 2.5 sec, the required success time. This means that even for a fully attentive human driver who is always able to take over as soon as the attack starts to take effect, the average reaction time is still far from enough to prevent the damage. A more detailed result discussion is in Appendix~\ref{appendix:attack_effectiveness}.

\textbf{Stealthiness user study.}
To more rigorously evaluate the attack stealthiness, we conduct a user study with 100 participants, and find that (1) even for the lowest stealthiness level at $\lambda=10^{-4}$, only \textit{less than 25\%} of the participants decide to take over the driving before the attack starts to take effect. This suggests that the majority of human drivers today do not treat dirty road patches as road conditions where ALC systems cannot handle; and (2) at 2.5 seconds before the attack succeeds, the attack patches with $\lambda=10^{-2}$ and $10^{-3}$ appear to be \textit{as innocent as normal clean road patches to human drivers}, with only less than 15\% participants deciding to take over. More detailed results and discussion are in Appendix~\ref{appendix:user_study}.

From these results, the stealthiness level with $\lambda=10^{-3}$ strikes an ideal balance between attack effectiveness and stealthiness: it does not increase driver suspicion compared to even a benign clean road patch at 2.5 seconds before our attack succeeds, while having no sacrifice of attack effectiveness as shown in Table~\ref{tbl:effectiveness}. We thus use it as the default stealthiness configuration in our following experiments.

\nsubsection{Comparison with Baseline Attacks}
\label{sec:eval_baseline}

\textbf{Evaluation methodology.} To understand the benefits of our current design choices over possible alternatives, we evaluate against 2 baseline attack methods: (1) \textit{single-frame EoT attack}, which still uses our lane-bending objective function but optimizes for the EoT (Expectation over Transformation) of the patch view (e.g., different positions/angles) in a single camera frame, and (2) \textit{drawing-lane-line attack}, which directly draws straight solid white lane line instead of placing dirty road patches. EoT is a popular design in prior works to improve attack robustness across sequential frames~\cite{athalye2018synthesizing, brown2017advpatch}. Thus, comparing with such a baseline attack can evaluate the benefit of our motion model based input generation design (\S\ref{sec:design_motion_model}) in addressing the challenge of frame inter-dependencies due to attack-influenced vehicle actuation (C2 in~\S\ref{sec:problem_formulation_challenges}).

The drawing-lane-line attack is designed to evaluate the type of ALC attack vector identified in the prior work by Tencent~\cite{tencent2019}, which uses straightly-aligned white stickers to fool Tesla Autopilot on road regions \textit{without lane lines}. In our case, we perform evaluations in road regions \textit{with lane lines}, and use a more powerful form of it (directly drawing solid lane lines) to understand the upper-bound attack capability of this style of perturbation for ALC systems.

\textbf{Experimental setup.} For single-frame EoT attack, we apply random transformations of the patch in BEV via (1) lateral and longitudinal position shifting. We apply up to $\pm0.735 \text{m}$ and $\pm0.285 \text{m}$ for highway and local respectively, which are their maximum in-lane lateral shifting from the lane center; and (2) viewing angle changes. we apply up to $\pm5.8^{\circ}$ changes, the largest average angle deviations under possible real-world trajectory variations (Table~\ref{tbl:eval_robustness} in Appendix~\ref{appendix:eval_robustness}).
For each scenario, we repeat the experiments for each frame with a complete patch view (usually the first 4 frames), and take the most successful one to obtain the upper-bound effectiveness. Other settings are the same as the DRP attack, e.g., $\lambda=10^{-3}$.

For the drawing-lane-line attack, we use the same perturbation area (i.e., the patch area) as the others for a fair comparison. Specifically, we sample points every 20cm at the top and bottom patch edges respectively, and form possible attacking lane lines by connecting a point at the top with one at the bottom. We exhaustively try all possible top and bottom point combinations and take the most successful one. The attacking lane lines are 10cm wide (a typical lane marking width~\cite{lane_marking_width}) with the same white color as the original lane lines.

\textbf{Results.} Table~\ref{tbl:baselines} shows the results under different patch area lengths. As shown, 
the DRP attack always has the highest attack success rate than these two baselines (with a $\ge$46\% margin).
When the patch area length is shorter and thus the perturbation capability is more limited, such advantage becomes larger; when the length is 12m, the success rates of single-frame EoT attack and the drawing-lane-line attack drops to 0\% and 2.5\%, while that for DRP is still 66\%. This shows that our motion model based input generation can indeed benefit attack effectiveness, as it can more accurately synthesize subsequent frame content based on attack-influenced vehicle actuation, instead of the blind synthesis in EoT. Also note that the single-frame EoT attack still uses our domain-specific lane-bending objective function design. The drawing-lane-line attack only has 2.5\% success rate when the length is 12m; the length used in the Tencent work is actually even shorter ($<$5m)~\cite{tencent2019}. This shows that in the road regions \textit{with lane lines}, simply adding lane-line-style perturbations, especially a short one, can barely affect production ALC systems. Instead, an attack vector with larger perturbation area, e.g., in DRP attack, may be necessary.

\begin{table}[t]
\centering
\footnotesize
\caption{\diff{Attack success rates of the DRP attack and 2 baseline attacks under different patch area lengths.}}
\vspace{-0.1in}
\begin{tabular}{ccccc}
\toprule
             & \multicolumn{4}{c}{Patch Area Length}      \\ \cline{2-5} 
Attack       & 12m     & 18m     & 24m     & 36m     \\ \hline
DRP   & 66.25\% & 82.50\% & 90.75\% & 100\%   \\
Single-frame EoT & 0.00\%  & 8.75\% & 21.25\% & 50.00\% \\
Drawing-lane-line & 2.50\%  & 13.75\% & 31.25\% & 53.75\% \\
\toprule
\end{tabular}
\label{tbl:baselines}
\vspace{-0.2in}
\end{table}

\nsubsection{Attack Robustness, Generality, and Deployability Evaluations} \label{sec:eval_robustness}

\textbf{Robustness to run-time driving trajectory and angle deviations.} As described in~\S\ref{sec:design_opt_robust}, the run-time victim driving trajectories and angles will be different from the motion model predicted ones in attack generation time due to run-time driving dynamics. To evaluate attack robustness against such deviations, we use (1) 4 levels of vehicle position shifting at each vehicle control step in attack evaluation time, and (2) 27 vehicle starting positions to create a wide range of approaching angles and distances to the patch, e.g., from (almost) the leftmost to the rightmost position in the lane. Our attack is shown to maintains a high effectiveness ($\ge$ 95\% success rate) even when the vehicle positions at the attack evaluation time has 1m shifting on average from those at the attack generation time at each control step. Details are in Appendix~\ref{appendix:eval_robustness}.

\label{sec:eval_generality}

\textbf{Attack generality evaluation.} To evaluate the generality of our attack against LD models of different designs, ideally we hope to evaluate on LD models from other production ALC besides \op{}, e.g., from Tesla Autopilot. However, \op{} is the only one that is currently open sourced. Fortunately, we find that the LD models in some older versions of \op{} actually have different DNN designs, which thus can also serve for our purpose. 
We evaluate on 3 versions of LD models with large DNN architecture differences (detailed in Appendix~\ref{appendix:dnn_design_diff}), and find that our attack is able to achieve $\ge$90\% success rates against all 3 LD models, with an average attack transferability of 63\%. More details are in Appendix~\ref{appendix:dnn_design_diff}.

\label{sec:deployability}

\textbf{Attack deployability evaluation.} We evaluate the attack deployability by estimating the required efforts to deploy the attack road patch. We perform experiments using our multi-piece patch attack mode design (\S\ref{sec:multiple_patch}), and find that the attack success rate can be as high as \textit{93.8\% with only 8 pieces of quickly-deployable road patches}, each requiring only 5-10 sec for 2 people to deploy based on videos of adhesive patch deployment~\cite{adhesive_video}. More details are in Appendix~\ref{appendix:deployability}.

\nsubsection{Physical-World Realizability Evaluation} \label{sec:eval_realizability}

While we have shown high attack effectiveness, robustness, and generality on real-world driving traces, the experiments are performed by synthesizing the patch appearances digitally, which is thus still different from the patch appearances in the physical world. As discussed in~\S\ref{sec:design_opt_realize}, there are 3 main practical factors that can affect the attack effectiveness in physical world: (1) the lighting condition, (2) printer color accuracy, and (3) camera sensing capability. Thus, in this section we perform experiments to understand the physical-world attack realizability against these 3 main practical factors.

\textbf{Evaluation methodology: miniature-scale experiments.} To perform the DRP attack, a real-world attacker can pretend to be road workers and place the malicious road patch on public roads. However, due to the access limit to private testing facilities, we cannot do so ethically and legally on public roads with a real vehicle.
Thus, we try our best to perform such evaluation by designing a \textit{miniature-scale experiment}, where the road and the malicious road patch are first physically printed out on papers and placed according to the physical-world attack settings but in miniature scale. Then the real ALC system camera device is used to get camera inputs from such a miniature-scale physical-world setting. Such miniature-scale evaluation methodology can capture all the 3 main practical factors in the physical-world attack setting, and thus can sufficiently serve for the purpose of this evaluation.

\begin{figure*}[t!]
    \begin{minipage}{0.270\linewidth}
        \centering
        \includegraphics[width=\linewidth]{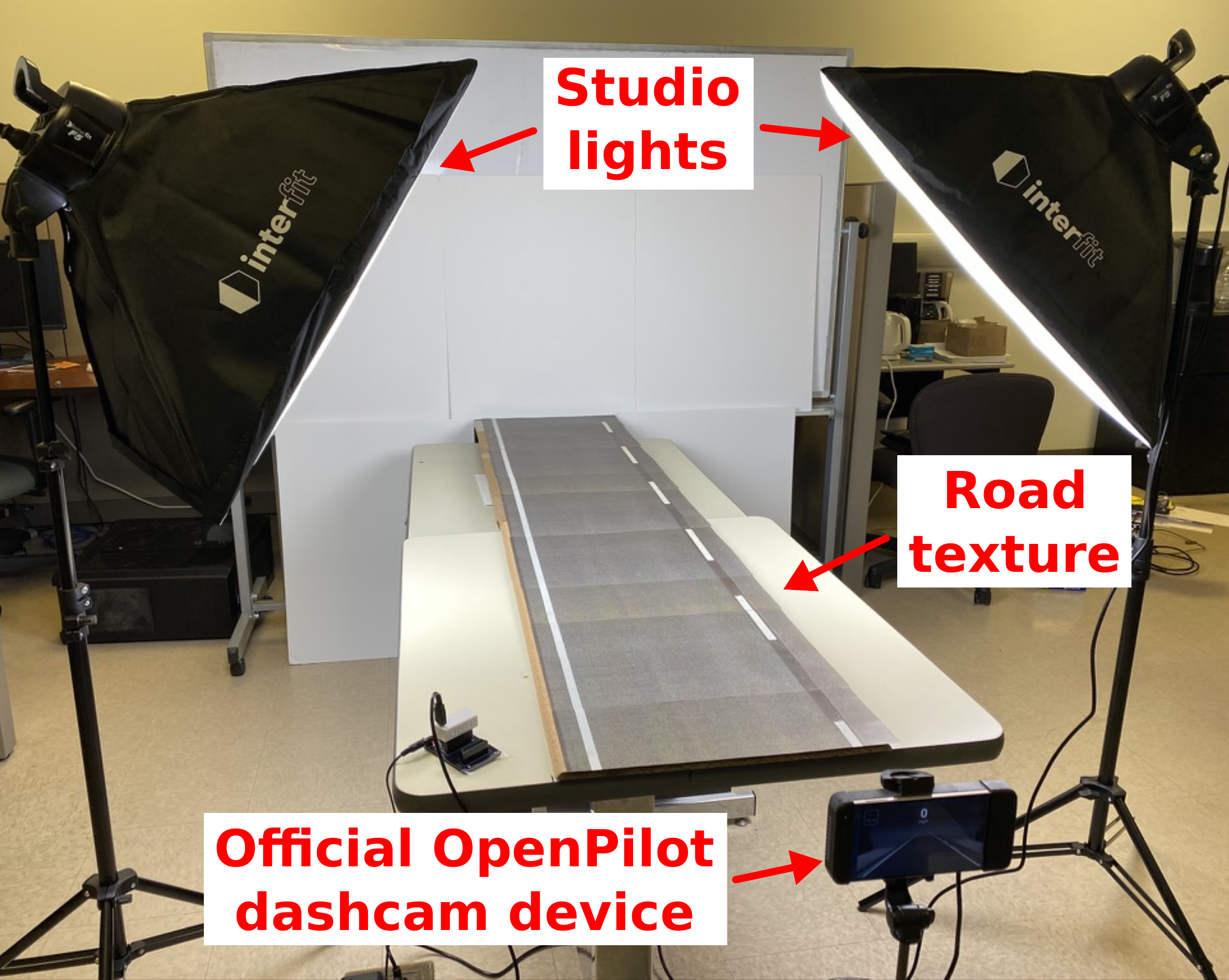}
        \vspace{-0.3in}
        \caption{Miniature-scale experiment setup. Road texture/patch are printed on ledger-size papers.
        }
        \label{fig:small_scale_expr_setup}
    \end{minipage}
    \hspace{0.005\linewidth} %
    \begin{minipage}{0.345\linewidth}
        \centering
        \includegraphics[width=\columnwidth]{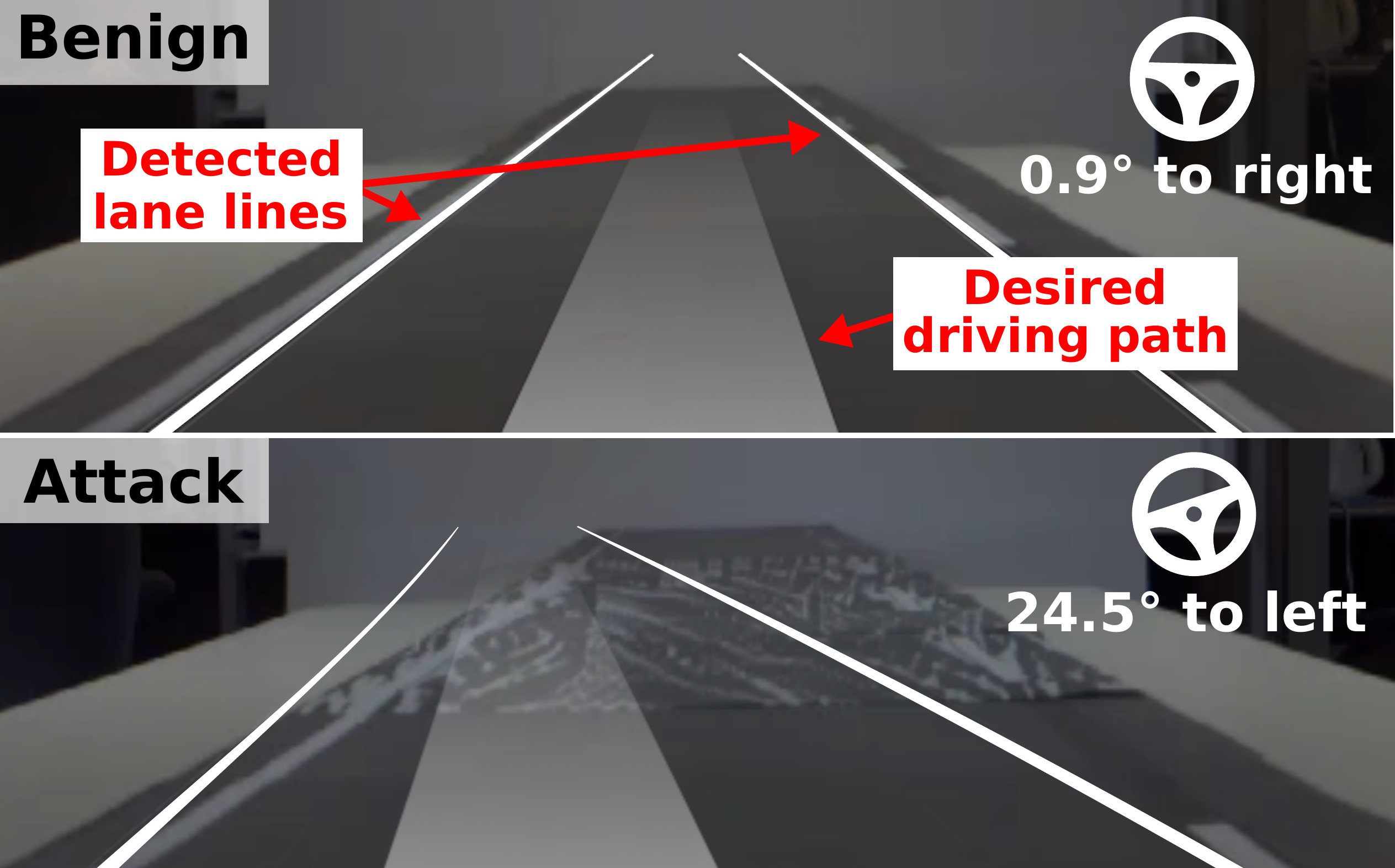}
        \vspace{-0.3in}
        \caption{Lane detection
        and steering angle decisions in benign and attacked scenarios in the miniature-scale experiment.
        }
        \label{fig:small_scale_results}
    \end{minipage}
    \hspace{0.005\linewidth} %
    \begin{minipage}{0.345\linewidth}
        \centering
        \includegraphics[width=\columnwidth]{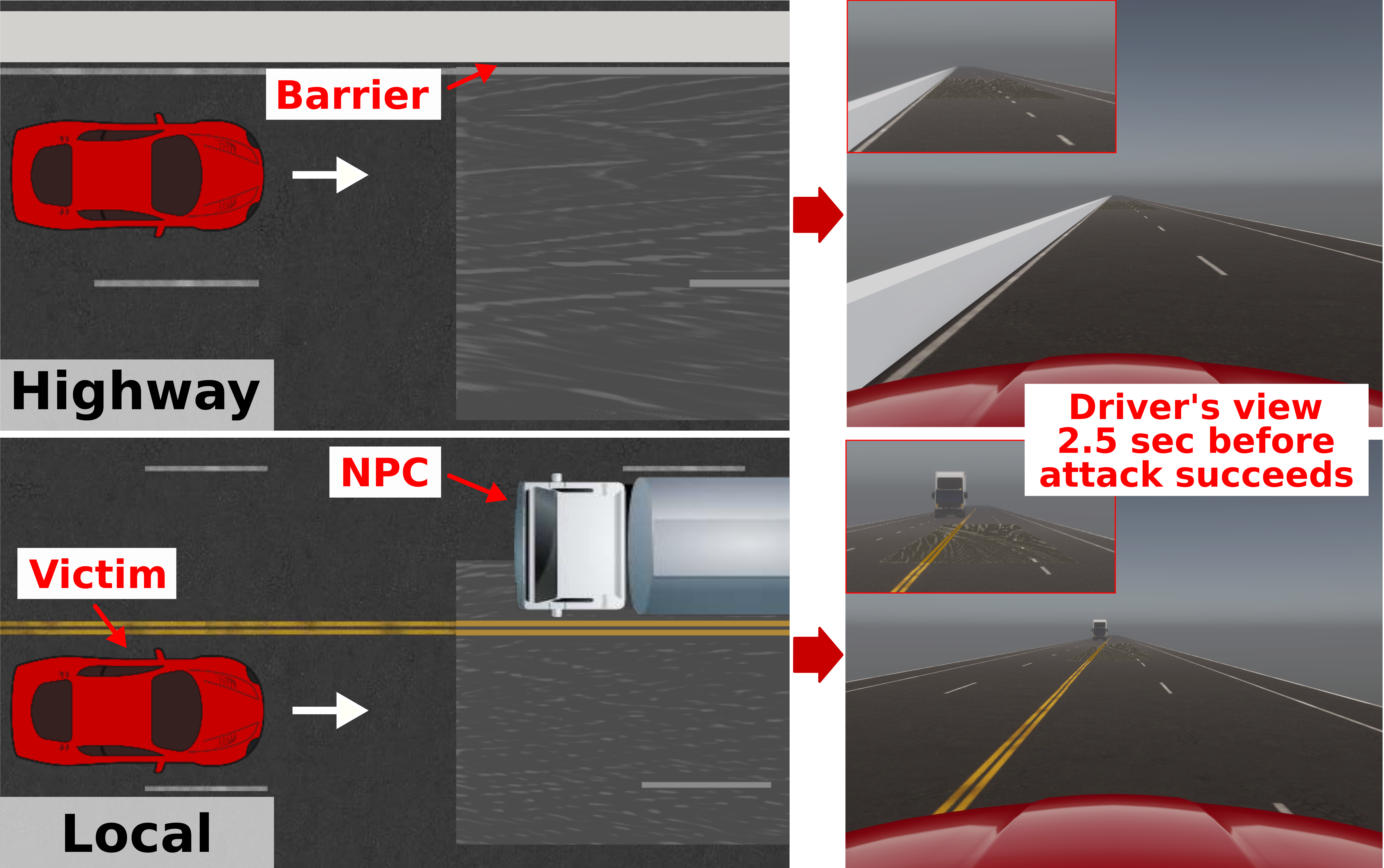}
        \vspace{-0.3in}
        \caption{\diffst{End-to-end}\diff{Software-in-the-loop} simulation scenarios and driver's view 2.5 sec before attack succeeds. Larger images are in Fig.~\ref{fig:sim_starts_large_picture}.}
        \label{fig:sim_roads}
    \end{minipage}
\vspace{-0.25in}
\end{figure*}

\textbf{Experimental setup.} As shown in Fig.~\ref{fig:small_scale_expr_setup}, we create a miniature-scale road by printing a real-world high-resolution BEV road texture on multiple ledger-size papers and concatenating them together to form a long straight road. In the attack evaluation time, we create the miniature-scale malicious road patch using the same method, and place it on top of the miniature-scale road following our DRP attack design. 
The patch is printed with a commodity printer: RICOH MP C6004ex Color Laser Printer.
We mount EON, the official \op{} dashcam device, on a tripod and face it to the miniature-scale road. The road size, road patch size, and the EON mounting position are carefully calculated to represent \op{} installed on a Toyota RAV4 driving on a standard 3.6-meter wide highway road at 1:12 scale. We also create different lighting conditions with two studio lights. 
The patch size is set to represent a 4.8me wide and 12m long one in the real world scale.
The other settings are the same as in~\S\ref{sec:eval_robustness}. %

\textbf{Evaluation metric.} Since the camera is mounted in a static position, we evaluate the attack effectiveness directly using the steering angle decision at the frame level instead of the \diffst{end-to-end }lateral deviation used in previous sections. This is equivalent from the attack effectiveness point of view since the large \diffst{end-to-end} lateral deviation is essentially created by a sequence of large steering angle decisions at the frame level. Specifically, we first find the camera frame that has the same relative position between the camera and the patch as that in the miniature-scale experimental setup. Then we compare its \textit{designed} steering angle at the attack generation time and its \textit{observed} steering angle that the ALC system in \op{} intends to apply to the vehicle in the miniature-scale experiment. Thus, the more similar these two steering angles are, the higher realizability our attack has in the physical world.

\textbf{Results.} Fig.~\ref{fig:small_scale_results} shows a visualization of the lane detection results of the benign and attacked scenarios in the miniature-scale experiment using the \op{}'s official visualization tool.
As shown, in the benign scenario, both detected lane lines align accurately with the actual lane lines, and the desired driving path is straight as expected. However, when the malicious road patch is placed, it bends the detected lane lines significantly to the left and causes the desired driving path to be curving to the left, which is exactly the designed attack effect of our lane-bending objective function (\S\ref{sec:design_lane_bending}). 
In this case, the designed steering angle is 23.4$^{\circ}$ to the left at the digital attack generation time, and the observed one in the physical miniature-scale experiment is 24.5$^{\circ}$ to the left, which only differs by 4.7\%. In contrast, in the benign scenario the observed steering angle for the same frame is 0.9$^{\circ}$ to the right. 

\textbf{Robustness under different lighting conditions.} 
We repeat this experiment under 12 lighting conditions ranging from 15 lux (corresponding to sunset/sunrise) to 1210 lux (corresponding to midday of overcast days). The results show that the same attack patch above is able to \textit{maintain a desired steering angle of 20-24$^{\circ}$ to the left under all 12 lighting conditions}, which are all significantly different from the benign scenario (0.9$^{\circ}$ to the right). More details are in Appendix~\ref{appendix:lighting}.

\diff{
\textbf{Robustness to different viewing angles.} We evaluate the attack robustness from 45 different viewing angles created by different distances to the patch and lateral offsets to the lane center. Our results show that our attack always achieves over 23.4$^{\circ}$ to the left from all viewing angles. We record videos in which we dynamically change viewing angles in a wide range while showing real-time lane detection results under attack, available at \textbf{\url{https://sites.google.com/view/cav-sec/drp-attack/}}.}

\nsection{
\texorpdfstring{\diffst{End-to-End Attack Evaluation}\diff{Software-in-the-Loop Simulation}}{Software-in-the-Loop Simulation}
} \label{sec:end_to_end}
To understand the \diffst{end-to-end }safety impact, we perform \diff{software-in-the-loop} evaluation of our attack on LGSVL, a production-grade autonomous driving simulator~\cite{lgsvl}. We overcame several engineering challenges in enabling this setup (Appendix~\ref{appendix:lgsvl_op_bridging}) and open-sourced via our website~\cite{drp_attack_page}.

\textbf{\diffst{End-to-end}\diff{Evaluation} scenarios.} We construct \diffst{end-to-end }2 attack scenarios for highway and local road settings respectively, as shown in Fig.~\ref{fig:sim_roads}. For the former, we place a concrete barrier on the left, and for the latter, we place a truck driving on an opposite direction lane. The attack goals are to hit the concrete barrier or the truck. 
Detailed setup are in Table~\ref{tbl:sim_configs} in Appendix.

\begin{figure}[t]
  \centering
  \includegraphics[width=\linewidth]{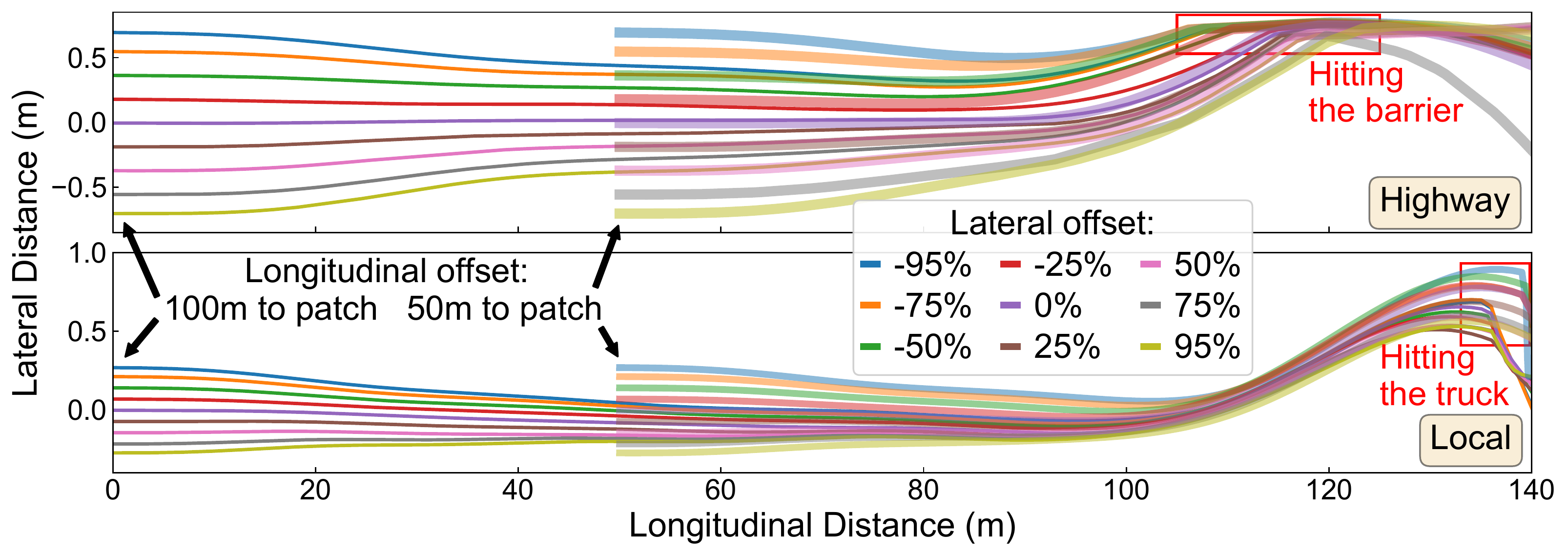}
  \vspace{-0.3in}
  \caption{\diff{Victim driving trajectories in the software-in-the-loop evaluation from 18 different starting positions for highway and local road scenarios. Lateral offset values are percentages of the maximum in-lane lateral shifting from lane center; negative and positive signs mean left and right shifting.}}
  \label{fig:simulation_trjector}
  \vspace{-0.15in}
\end{figure}

\textbf{Experimental setup and evaluation metrics.} We perform evaluation on \op{} v0.6.6 with the Toyota RAV4 parameters. We follow the methodology in~\S\ref{sec:design_opt_realize} to obtain and apply the color mapping in our simulation environment. The patch size is 5.4m wide and 70m long, and we place it in the simulation environment by importing the generated patch image into Unity. The other parameters are the same as~\S\ref{sec:eval_generality}. \diff{To evaluate the attack effectiveness from different victim approaching angles, for each scenario we evaluate the same patch from 18 different starting positions, created from the combinations of 2 longitudinal distances to the patch (50 and 100 m) and 9 lateral offsets (from -95\% to 95\%) as shown in Fig.~\ref{fig:simulation_trjector}. The patch is visible at all these starting positions. We repeat 10 times for each starting position in each scenario.}

\textbf{Results and video demos.} \diff{Our attack achieves 100\% success rates from all 18 starting positions in both highway and local road scenarios (more details in Table~\ref{tbl:sim_configs} in Appendix). Fig.~\ref{fig:simulation_trjector} shows the averaged vehicle trajectories from each starting positions. As shown, the vehicle always first drives toward the lane center since the ALC system tries to correct the initial lateral deviations. After that, the patch starts to take effect, and causes the vehicle to deviate to the left significantly and hit the barrier or truck.}
We record demo videos at \textbf{\url{https://sites.google.com/view/cav-sec/drp-attack/}}. In the highway scenario, after the victim hits the concrete barrier, it bounces away quickly due to the abrupt collision. For local road, the victim crashes to the front of the truck, causing both the victim and truck to stop. This suggests that the safety impacts of our attack can be severe. Snapshots of the demo when the attack goals are achieved are in Fig.~\ref{fig:sim_snapshots} in Appendix.

\vspace{-0.02in}
\nsection{Safety Impact on Real Vehicle} \label{sec:real_vehicle}
\vspace{-0.05in}
While the simulation-based evaluation above has shown severe safety impacts, it does not simulate other driver assistance features that are commonly used with ALC at the same time in real-world driving, for example Lane Departure Warning (LDW), Adaptive Cruise Control (ACC), Forward Collision Warning (FCW), and Automatic Emergency Braking (AEB). This makes it unclear whether the safety damages shown in~\S\ref{sec:end_to_end} are still possible when these features are used, especially the safety-protection ones such as AEB. In this section, we thus use a real vehicle to more directly understand this.

\begin{figure}[t]
    \centering
    \includegraphics[width=\linewidth]{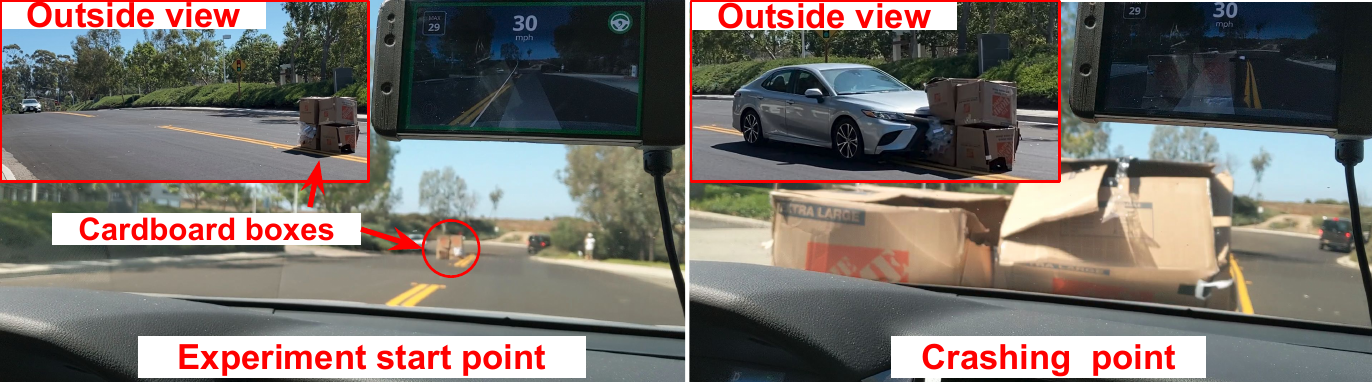}
    \vspace{-0.25in}
    \caption{Safety impact evaluation for our attack on a Toyota 2019 Camry with OpenPilot engaged. Even with other driver assistance features such as Automatic Emergency Braking (AEB), our attack still causes collisions in all the 10 trials. 
    } 
    \label{fig:realtest}
    \vspace{-0.2in}
\end{figure}

\textbf{Evaluation methodology.} We install OpenPilot on a Toyota 2019 Camry, in which case OpenPilot provides ALC, LDW, and ACC, and the Camry's stock features provide AEB and FCW~\cite{openpilot}. We then use this real-world driving setup to perform experiments on a rarely-used dead-end road, which has a double-yellow line in the middle and can only be used for U-turn. The driver's view of this road is shown on the left of Fig.~\ref{fig:realtest}. In our miniature-scale experiment in~\S\ref{sec:eval_realizability}, the attack realizability from the physically-printed patch to the LD model output has already been validated under 12 different lighting conditions. Thus, in this experiment we evaluate the safety impact by directly injecting an attack trace at the LD model output level \diff{(detailed in Appendix~\ref{appendix:openpilot_details})}. This can also avoid blocking the road for sticking patches to the ground and cleaning them up, which may affect other vehicles.

To create safety-critical driving scenarios, we place \diff{cardboard} boxes adjacent to but outside of the current lane as shown in Fig.~\ref{fig:realtest}, which can mimic road barriers and obstacles in opposite direction as in~\S\ref{sec:end_to_end} while not causing damages to the vehicle and driver safety. Similar setup is also used in today's vehicle crash tests~\cite{toyota_aeb_demo, honda_aeb_demo, iihs_aeb_test, aeb_target}. To ensure that we do not affect other vehicles, we place the \diff{cardboard} boxes only when the entry point of this dead-end road has no other driving vehicles in sight, and quickly remove them right after our vehicle passes them as required by the road code of conduct~\cite{road_objects, road_objects2, road_objects_idaho}.

\textbf{Experiment setup.} We perform experiments in day time with and without attack, each 10 times. The driving speed is kept at $\sim$28 mph \diff{($\sim$45 km/h)}, the min speed for engaging OpenPilot on our Camry. The injected attack trace is from our simulation environment (\S\ref{sec:end_to_end}) at the same driving speed.

\textbf{Results}. Our experiment results show that our attack causes the vehicle to hit the \diff{cardboard} boxes in all the 10 attack trials (100\% collision rate), including 5 front and 5 side collisions. The collision variations are caused by randomness in the dynamic vehicle control and the timing differences in OpenPilot engaging and attack launching. In contrast, in the trials without attack, OpenPilot can always drive correctly and does not hit or even touch the objects in any of the 10 trials. 

These results thus show that driver assistance features such as LDW, ACC, FCW, and AEB are not able to effectively prevent the safety damages caused by our attack on ALC. We examine the attack process and find that LDW is not triggered since it relies on the same lane detection module as ALC and thus are affected simultaneously by our attack. ACC does not take any action since it does not detect a front vehicle to follow and adjust speed in these experiments. FCW is triggered 5 times out of the 10 collisions, but it is only a warning and thus cannot prevent the collision by itself. Moreover, in our experiments FCW is triggered only \textit{0.46 sec} before the collision on average, which is far too short to allow human drivers to react considering the 2.5-second average driver reaction time to road hazard (\S\ref{sec:problem_formulation}).

In our Camry model, FCW and AEB are turned on together as a bundled safety feature~\cite{toyotacamry2019}. However, while we have observed some triggering of FCW, we were not able to observe any triggering of AEB among the 10 attack trials, leading to a \textit{100\% false negative rate}. We check the vehicle manual~\cite{toyotacamry2019} and find that this may be because the AEB feature (called \textit{pre-collision braking} for Toyota) is used very conservatively: it is triggered only when the possibility of a collision is \textit{extremely high}. This observation is also consistent with the previously-reported high failure rate (60\%) for AEB features on popular car models today~\cite{aeb-fail}. Such conservative use of AEB can reduce false alarms and thus avoid mistaken sudden emergency brakes in normal driving, but also makes it difficult to effectively preventing the safety damages caused by our attack --- in our experiments, it was not able to prevent any of the 10 collisions. The video recordings for these real-vehicle experiments are available at \textbf{\url{https://sites.google.com/view/cav-sec/drp-attack/}}.

\nsection{Limitations and Defense Discussion}

\vspace{0.15in}

\nsubsection{Limitations of Our Study} \label{sec:limitations}

\textbf{\hspace*{1em} Attack deployability.} As evaluated in~\S\ref{sec:deployability}, our attack can achieve a high success rate (93.8\%) with only 8 pieces of quickly-deployable road patches, each requiring only 5-10 sec to deploy for 2 people. To further increase stealthiness, the attacker can pretend to be road workers like in Fig.~\ref{fig:threat_model} to avoid suspicion, and pick a deployment time when the target road is the most vacant, e.g., at late night. Nevertheless, lower deployment efforts is always more preferred for attackers to reduce risks. One potential direction to further improve this is to explore other common road surface patterns besides dirty patterns, which we leave as future work.

\textbf{Generality evaluation.} Although we have shown high attack generality against LD models with different designs (\S\ref{sec:eval_generality}), all our evaluations are performed on only one production ALC in \op{}. Thus, it is still unclear whether other popular ALC, e.g., Tesla Autopilot and GM Cruise, are vulnerable to our attack. Unfortunately, to the best of our knowledge, the \op{} ALC is the only production one that is open sourced. 
Due to the same reason, we are also unable to evaluate the transfer attacks from OpenPilot to these other popular ALC systems.
Nevertheless, since the \op{} ALC is representative at both design and implementation levels (\S\ref{sec:eval}), we think our current discovery and results can still generally benefit the understanding of the security of production ALC today. Also, since DNNs are generally vulnerable to adversarial attacks~\cite{Szegedy2014, goodfellow2014explaining, carlini2017towards,
kurakin2016adversarial_b, sharif2016accessorize, athalye2018synthesizing, brown2017advpatch, chen2018shapeshifter, eykholt2018robust, eykholt2018physical, jack2018caraml, zhao2018seeing}, if these other ALC systems also adopt the state-of-the-art DNN-based design, at least at design level they are also vulnerable to our attack.

\diff{
\textbf{End-to-end evaluation in real world.} In this work, we evaluate our attack against various possible real-world factors such as lighting conditions, patch viewing angles, victim approaching angles/distances, printer color accuracy, and camera sensing capability (\S\ref{sec:eval_robustness},~\S\ref{sec:eval_realizability}), and also evaluate the safety impact using software-in-the-loop simulation (\S\ref{sec:end_to_end}) and attack trace injection in a real vehicle (\S\ref{sec:real_vehicle}). However, these setups still have a gap to real-world attacks as we did not perform direct end-to-end attack evaluation with real vehicles in the physical world. Such a limitation is caused by safety issues (vehicle-enforced minimum OpenPilot engagement speed at 28 mph, or 45 km/h) and access limits to private testing facilities (for patch placement). In the future, we hope to overcome this by finding ways to lower the minimum engagement speed and obtain access to private testing facilities.

}

\nsubsection{Defense Discussion} \label{sec:defenses}
\vspace{0.15in}
\nsubsubsection{Machine Learning Model Level Defenses} \label{sec:ml_defenses}

\begin{figure*}[t!]
  \centering
  \includegraphics[width=\linewidth]{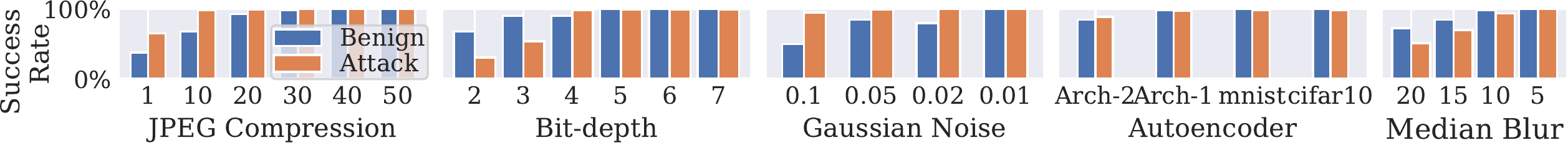}
  \vspace{-0.25in}
\caption{Evaluation results for 5 directly-applicable DNN model level defense methods. \textit{Attack}: Attack success rate. \textit{Benign}: Percentage of scenarios where the ALC can still behave correctly (i.e., not driving out of current lane) with defense applied.}
\label{fig:defenses}
  \vspace{-0.2in}
\end{figure*}

In the recent arms race between adversarial machine learning attacks and defenses, numerous defense/mitigation techniques have been proposed~\cite{liao2018defense, xie2019feature, xu2017feature, guo2017countering, madry2017towards, raghunathan2018certified, lecuyer2019certified, cohen2019certified }.
However, so far none of them studied LD models.
As a best effort to understand the effectiveness of existing defenses on our attack,
we perform evaluation on 5 popular defense methods that only require model input transformation without re-training: JPEG compression~\cite{dziugaite2016study}, bit-depth reduction~\cite{xu2017feature}, adding Gaussian noise~\cite{pmlr-v89-zhang19b}, median blurring~\cite{xu2017feature}, and autoencoder reformation~\cite{meng2017magnet}, since they are directly applicable to LD models. Descriptions and our configurations of these methods are in Appendix~\ref{appendix:defense}. Our experiments use the same dataset and success metrics as in~\S\ref{sec:eval}. Meanwhile, we also evaluate a \textit{benign-case success rate}, defined as the percentage of scenarios where the ALC can still behave correctly (i.e., not driving out of current lane) when the defense method is applied.

Fig.~\ref{fig:defenses} shows the evaluation results. As shown, for each defense method we also vary the parameters to explore the trade-off between attack success rate and benign-case success rate. As shown, while all methods can effectively decrease the attack success rate with certain parameter configurations, the benign-case success rates are also decreased at the same time. In particular, when the benign-case success rates are still kept at 100\%, the attack success rates are still 99 to 100\% for all methods. This shows that \textit{none of these methods can effectively defend against our attack without harming ALC performance in normal driving scenarios}. This might be because these defenses are mainly for disrupting digital-space human-imperceptible perturbations, and thus are less effective for physical-world realizable attacks with human-perceptible (but seemingly-benign) perturbations.

These results show that directly-applicable defense methods cannot easily defeat our attack. Thus, it is necessary to explore (1) novel adaptions of more advanced defenses such as adversarial training to LD, or (2) new defenses specific to LD and our problem setting, which we leave as future work.

\nsubsubsection{Sensor/Data Fusion Based Defenses} Besides securing LD models, another direction is to fuse camera-based lane detection with other independent sensor/data sources such as LiDAR and High Definition (HD) map~\cite{hdmaps}. For example, LiDAR can capture the tiny laser reflection differences for lane line markings, and thus is possible to perform lane detection~\cite{bai2018deep}. However, while LiDARs are commonly used in high-level (e.g., Level-4) AD systems such as Google Waymo~\cite{waymoone}
that provide self-driving taxi/truck, so far they are not generally used in production low-level (e.g., Level-2) AD such as ALC, e.g., Tesla, GM Cadillac, Toyota RAV4, etc.~\cite{tesla2020autopilot, gm_cadillac_ct6, hondasensing, toyotasafetysense}. This is mainly because LiDAR is quite costly for vehicle models sold to individuals (typically $\ge$\$4,000 each for AD~\cite{velodyne_cut_price}). For example, Elon Musk, the co-founder of Tesla, claims that LiDARs are ``\textit{expensive sensors that are unnecessary (for autonomous vehicles)}''~\cite{burns2019lidar}.

Another possible fusion source is lane information from a pre-built HD map of the targeted road, which can be used to cross-check with the run-time detected lane lines to detect our attack. However, this requires ALC providers to collect and maintain accurate lane line information for each road, which can be time consuming, costly, and also hard to scale. To the best of our knowledge, ALC systems in production Level-2 AD systems today do not use HD maps in general. For instance, Tesla explicitly claims that it does not use HD map for Autopilot driving since it is a \textit{``non-scalable approach''}~\cite{teslamap}.

Nevertheless, considering that Level-4 AD systems today are able to build and heavily utilize HD maps~\cite{waymo_hdmap, apollo2016hdmap}, we think leveraging HD maps is still a more feasible solution than requiring production Level-2 vehicle models to install LiDARs. If such a map can be available, a follow-up research question is how to effectively detect our attack without raising too many false alarms, since mismatched lane information can also occur in benign cases due to (1) vehicle position and heading angle inaccuracies when localized on the HD map, e.g., due to sensor noises in GPS and IMU~\cite{wan2018robust, levinson2007map}, and (2) normal-case LD model inaccuracies.

\nsection{Related Work}

\textbf{Autonomous Driving (AD) system security.} For AD systems, there are mainly two types of security research: \textit{sensor security} and \textit{autonomy software security}.
For \textit{sensor security}, prior works studied spoofing/jamming on camera~\cite{petit2015remote, yan2016can, nassi2020phantom}, LiDAR~\cite{petit2015remote, shin2017illusion, cao2019adversarial}, RADAR~\cite{yan2016can}, ultrasonic~\cite{yan2016can}, and IMU~\cite{tu2018injected, trippel2017walnut}.
For \textit{autonomy software security}, prior works have studied the security of object detection~\cite{eykholt2018physical, chen2018shapeshifter, zhao2018seeing, cao2019adversarial, povolny2020adas} and tracking~\cite{jia2019fooling},
localization~\cite{junjie2020fusionripper},
traffic light detection~\cite{kanglan2021fooling},
and end-to-end AD models~\cite{pei2017deepxplore, tian2018deeptest, chernikova2019self, zhou2018deepbillboard}. Our work studies autonomy software security in production ALC. The only prior effort is from Tencent~\cite{tencent2019}, but it neither attacks the designed operational domain for ALC (i.e., roads with lane lines), nor generates perturbations systematically by addressing the design challenges in \S\ref{sec:problem_formulation_challenges}.

\textbf{\hspace*{1em}Physical-world adversarial attacks.} Multiple prior works have explored image-space adversarial attacks in the physical world~\cite{kurakin2016adversarial_b, sharif2016accessorize, athalye2018synthesizing, brown2017advpatch, chen2018shapeshifter, eykholt2018physical, jack2018caraml, zhao2018seeing, li2019adversarial}.
In particular, various techniques have been designed to improve the physical-world robustness, e.g., non-printability score~\cite{sharif2016accessorize, chen2018shapeshifter, eykholt2018physical, jack2018caraml}, low-saturation colors~\cite{zhao2018seeing}, and EoT~\cite{athalye2018synthesizing, brown2017advpatch, chen2018shapeshifter, eykholt2018physical, zhao2018seeing}. In comparison, prior efforts concentrate on image classification and object detection, while we are the first to systematically design physical-world adversarial attacks on ALC, which require to address various new and unique design challenges (\S\ref{sec:problem_formulation_challenges}).

\nsection{Conclusion}

In this work, we are the first to systematically study the security of DNN-based ALC in its designed operational domains under physical-world adversarial attacks. With a novel attack vector, dirty road patch, we perform optimization-based attack generation with novel input generation and objective function designs. Evaluation on a production ALC using real-world traces shows that our attack has over 95\% success rates with success time substantially lower than average driver reaction time, and also has high robustness, generality, physical-world realizability, and stealthiness. We further conduct experiments using both simulation and a real vehicle, and find that our attack can cause a 100\% collision rate in different scenarios. We also evaluate and discuss possible defenses. Considering the popularity of ALC and the safety impacts shown in this paper, we hope that our findings and insights can bring community attention and inspire follow-up research.

\bibliographystyle{ieeetr}
{
\footnotesize
\bibliography{main}
}

\appendix
\section*{Appendix}

\section{Required Deviations and Success Time} \label{appendix:deviation_calculation}

\begin{figure}[t]
\centering
\includegraphics[width=.6\columnwidth]{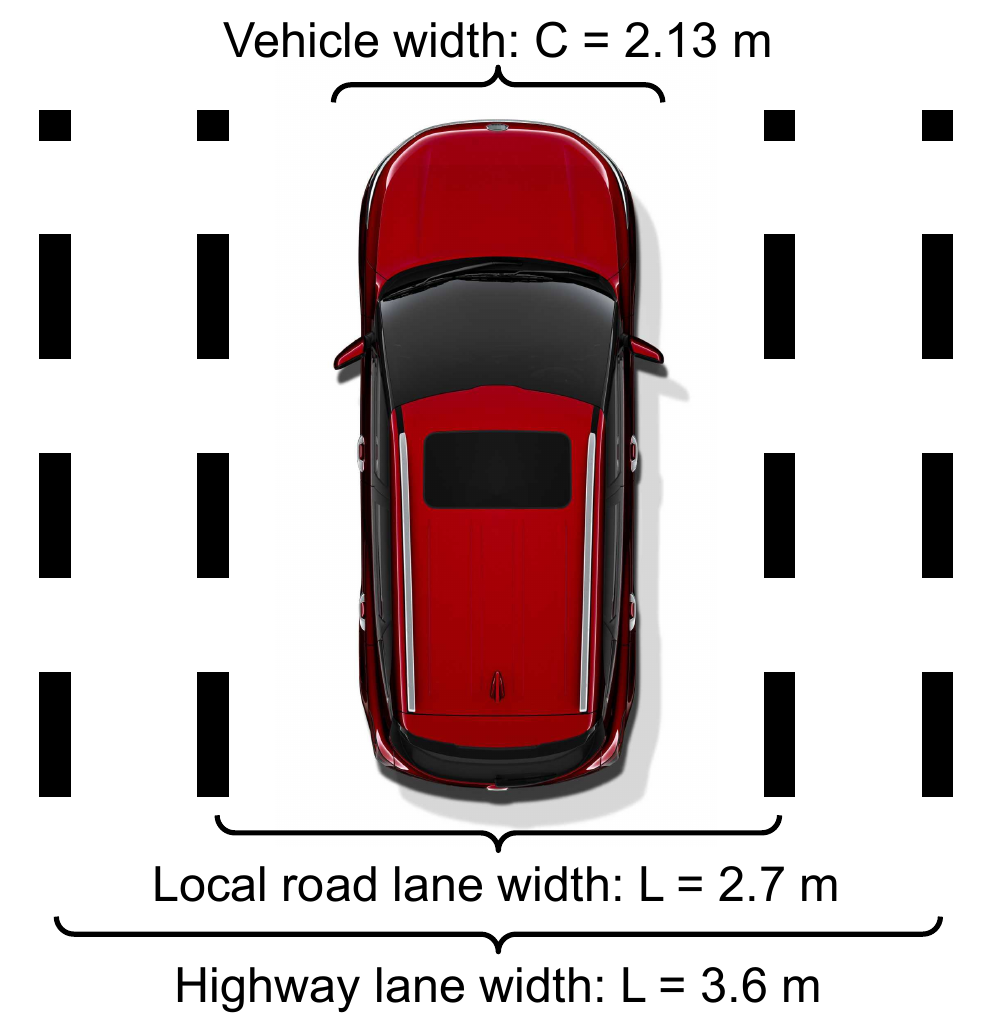}
\caption{Vehicle and lane widths used in this paper.}
\label{fig:car_lane_widths}
\vspace{-0.15in}
\end{figure}

\textbf{Required deviations.} The required deviations for the highway and local roads are calculated based on Toyota RAV4 width (including mirrors) and standard lane widths in the U.S.~\cite{lanewidth} as shown in Fig.~\ref{fig:car_lane_widths}.
We use Toyota RAV4 since it is the reference vehicle used by the \op{} team when collecting the comma2k19 data set~\cite{comma2k19}. For the lane widths, we refer to the design guidelines~\cite{lanewidth} published by the U.S. Department of Transportation Federal Highway Administration.
The required deviations to touch the lane line are calculated using $\frac{L-C}{2} = 0.735 m$~(highway) and $0.285 m$~(local), where $L$ is the lane width and $C$ is the vehicle width.

\textbf{Required success time.} Since ALC systems assume a fully attentive human driver who is prepared to take over at any moment~\cite{tesla2020support, sae2018}, the required deviation above needs to be achieved fast enough so that the human driver cannot react in time to take over and steer back. Thus, when we define the attack goal, we require not only the required deviation above, but also an attack success time that is smaller than the average driver reaction time to road hazards.
We select the average driver reaction time based on different government-issued transportation policy guidelines~\cite{cali_driver_reaction_time, uk_driver_reaction_time, uk_highway_code, nsc_driver_reaction_time}. In particular, in the California Department of Motor Vehicles Commercial Driver Handbook Section 2.6.1~\cite{cali_driver_reaction_time}, it describes (1) a 1.75 seconds \textit{average perception time}, i.e., the time from the time the driver's eyes see a hazard until the driver's brain recognizes it, and (2) a 0.75 to 1 seconds \textit{average reaction time}, i.e., the time from the driver's brain recognizing the hazard to physically take actions. Thus, in total it's \textbf{2.5 to 2.75 seconds} from the driver's eyes seeing a hazard to physically take actions. The UK ``Highway Code Book'' and ``Code of Practice for Operational Use of Road Policing Enforcement Technology'' use 3 seconds for driver reaction time~\cite{uk_driver_reaction_time, uk_highway_code}. National Safety Council also adopts a 3-second driver reaction time to calculate the minimum spacing between vehicles~\cite{nsc_driver_reaction_time}. Among them, we select the \textbf{smallest} one, i.e., 2.5 seconds from the California Department of Motor Vehicles~\cite{cali_driver_reaction_time}, as the required success time in this paper to avoid possible overestimation of the attack effectiveness in our evaluation.

Note that the driver reaction time above is commonly referring to the reaction time to apply the brake, instead of steering. In our paper, we use such reaction time to apply the brake as the reaction time to take over the steering wheel when the ALC systems are in control of the steering wheel. This is because in traditional driving, the driver is \textit{actively} steering the vehicle but \textit{passively} applying the brake. However, when the ALC system is controlling the steering, the human driver is \textit{passively} steering the vehicle, i.e., her hands are not actively controlling the steering wheel. Thus, the reaction time to take over the steering wheel during passive steering is analogous to that to apply the brake during passive braking.

In fact, the actual average driver reaction time when the ALC system is taking control is likely to be much higher than the 2.5 seconds measured in traditional driving, due to the reliance of human drivers on such convenient driving automation technology today. A recent study performed a simulation-based user study on Tesla Autopilot, and found that 40\% drivers fail to react in time to avoid a crash happening \textit{6.2 seconds} after the Autopilot fails to operate~\cite{loeb2019age}. In the real world, it is found multiple times that Tesla drivers fall asleep with Autopilot controlling the vehicle in high speed~\cite{kolodny2019watch}. Thus, the required success time of 2.5 seconds used in this paper is a relatively conservative estimation, and thus the attack effectiveness reported in our evaluations is likely only a \textbf{lower bound} of the actual effectiveness of our attack in the real world.

\section{Detailed Steps in the Attack Design}
\label{appendix:design_details}

\textbf{Differentiable construction of curve fitting (\S\ref{sec:design_lane_bending}).} Since the direct LD model output is the detected left and right lane line points (\S\ref{sec:background_alc}), a curve fitting step is further required to calculate $\rho_{t}(d; { \{X^a_j| j \leq t\}})$ in Eq.~\ref{math:objective} from the lane line points. This step also needs to be differentiable to allow the entire $f(\cdot)$ differentiable with respect to $\{X^a_j| j \leq t\}$. Thus, we further perform a differentiable construction of the curve fitting process as follows. We use $P_{l}, P_{r} \in \mathbb{R}^{|D_t|}$ to represent the left and right lane line points respectively, where their indexes represent $x$-axis (longitudinal coordinate) and their values represent the $y$-axis (lateral coordinate). We first fit the lane line points into polynomial curves in a least-square manner with $\xi_{l} = (V^{T}V)^{-1}V^{T} P_{l}$ and $\xi_{r} = (V^{T}V)^{-1}V^{T} P_{r}$, where $\xi_{l}, \xi_{r} \in \mathbb{R}^{n + 1}$ are the coefficients of the $n$-degree polynomial functions of the left and right lane lines respectively, and $V \in \mathbb{R}^{|D_t| \times n+1}$ is a Vandermonde matrix. Then we calculate the desired driving path coefficients $\xi_{d}$ by averaging those for the left and right lane lines: $\xi_{d} = \frac{1}{2} (\xi_{l} +  \xi_{r})$. As all operations above are written in closed form, the desired driving path polynomial $\rho_t(d) = [1, d, d^2, ..., d^{n}]\xi_d$ is differentiable by each lane line point.

\diff{
\textbf{Gradient aggregation in BEV space (\S\ref{sec:design_opt_overview}).}
In the gradient averaging step in Fig.~\ref{fig:grad_agg} (step iii), we project the gradients w.r.t $X^a_{1},...,X^a_T$ into the BEV space, and then calculate the average value of them weighted by their corresponding visible patch area sizes in the model inputs. The weight is the number of pixels of the patch in the model input and normalized over all frames, i.e., the sum of the weights equals 1. This weighted averaging is designed to prevent the averaged gradient from being dominated by the earlier frames, where the patch is far and small but the whole patch is visible. 

Note that this procedure does not produce the true gradient on
$\text{BEV}^{-1}(X^a_{i})$. 
Instead, this is approximation of this true gradient to avoid engineering efforts in deriving the differentiation of  $\text{BEV}^{-1}(\cdot)$ code in OpenCV. This also allows us to control the aggregation weights more flexibly. $\text{BEV}(\cdot)$ consists of matrix-vector multiplication and scaling. This approximation works since in our case $\text{BEV}(\cdot)$ and $\text{BEV}^{-1}(\cdot)$ are close to a linear transformation as the scaling is not substantially different across consecutive frames.
}

\section{Designs for Attack Robustness, Deployability, and Physical-World Realizability}
\label{appendix:rbst_deploy_real}
\label{appendix:design_opt_robust}
\label{appendix:multiple_patch}
\label{appendix:design_opt_realize}

\textbf{Attack robustness improvement.} In our attack generation process, the synthesized attack-influenced camera image frames can differ from the actual ones at the attack evaluation time due to (1) victim driving trajectory and angle deviations from the motion model predicted ones in the attack generation time, due to run-time driving dynamics from the complex and dynamic interactions among the road, tires, and the vehicle internal components in the real world, and (2) camera sensing inaccuracies such as blurring due to imperfect autofocus, especially when the vehicle is in motion. Thus, to improve the robustness of the generated patch in the real world, we add noises $\epsilon_t$ to the vehicle state $S^{a}_t$ during the motion model based input generation process. More specifically, we apply $\epsilon_t$ following a normal distribution $\mathcal{N}(0, \alpha)$ to both of the $MM(\cdot)$ outputs used in our input generation: the lateral position $y_t$, and the heading angle $\beta_t$. Meanwhile, we apply Gaussian blur to both the patch and gradients during the optimization process to address potential image blurring.

\begin{figure}[t]
    \centering
    \includegraphics[width=0.9\linewidth]{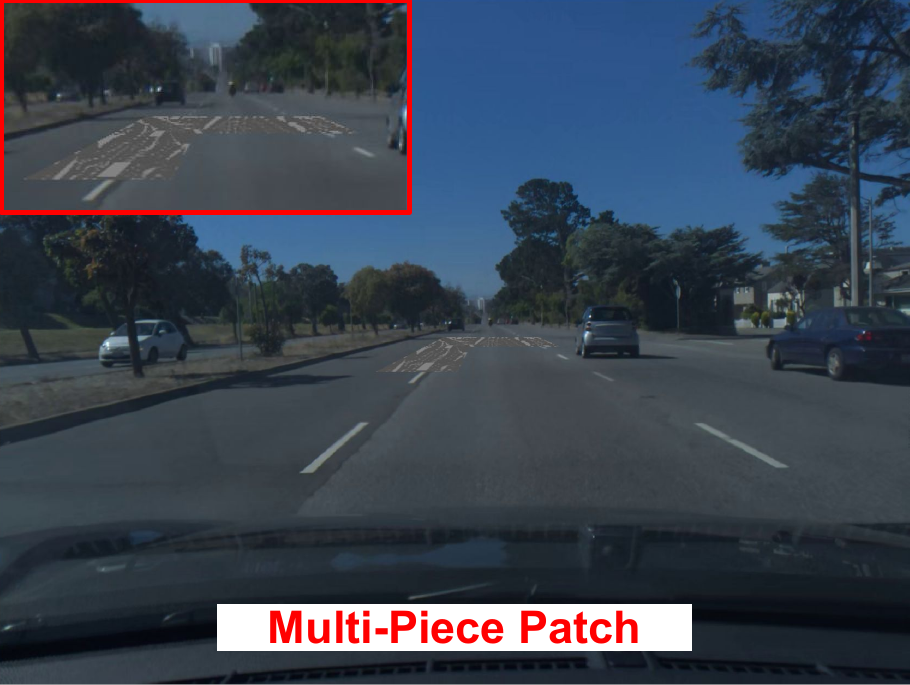}
    \vspace{-0.1in}
    \caption{Driver's view in multi-piece patch attack mode (\S\ref{sec:deployability}) at 2.5 seconds before the attack succeeds. The attack in the figure has 8 pieces of 1.8m$\times{}$7.2m road patches, each requiring only 5-10 sec to deploy for 2 people.} 
    \label{fig:poc_multiple}
    \vspace{-0.1in}
\end{figure}

\textbf{Attack deployability improvement.} To deploy the attack, the direct approach is to print the malicious dirty patterns on one single road patch and deploy it at once. However, if the required patch size is large, it may not be easy to be quickly deployed at once, which may increase the risk of being noticed by police officers or road guards. To improve this, we design an optional \textit{multi-piece patch attack} mode, which allows the attackers to deploy the DRP attack with multiple small pieces of road patches. In this mode, the attacker can specify (1) the size of a small piece that they can quickly deploy at once, denoted as $size_p$, and (2) the total number of such pieces based on their affordable deployment efforts and risks, denoted as $N_p$. Fig.~\ref{fig:poc_multiple} shows an example multi-piece patch attack from the driver's view, which includes 8 pieces of quickly-deployable small road patches. With this, the attacker can deploy one small piece at a time to avoid drawing too much attention, and can also parallelize the deployment of different pieces to further accelerate the process. 

To support this mode, we partition the original patch placement area (yellow area in Fig.~\ref{fig:drp_attack_overview}) into a grid, with each grid cell no larger than $size_p$. Next, in the optimization process shown in Fig.~\ref{fig:grad_agg}, we add a step right before the patch placement in the motion model based input generation to greedily select $N_p$ cells with the largest perturbation amounts (e.g., $L_1$ distances), and only perform patch placement on these cells. Here, our intuition is that the cells with larger perturbation amounts should be more important for the attack to succeed.

\textbf{Designs for physical-world realizability.} After the malicious road patch is generated digitally, the color and pattern of its perturbations can be perceived differently by camera devices in the physical world due to 3 main practical factors: (1) the lighting condition, (2) printer color accuracy, and (3) camera sensing capability, e.g., color accuracy and image clarity. Such differences can thus degrade the attack effectiveness when applying our attack in practice.

To address this, we create a color mapping between the intended colors and the camera-perceived colors after the intended ones are physically printed and captured by the camera device of the targeted ALC system. In our attack context, the attacker can do the same on the target road. Since our malicious road patches are designed to be only in grayscale for stealthiness (\S\ref{sec:design_opt_stealthy}), we constrain the color mapping in the gray scale instead of full color. Then after the patch is generated digitally, we perform a post-processing step that uses the color mapping to transform the color of each pixel to the color that can be mapped to the intended one after being printed.

\section{Detailed Discussion of Attack Effectiveness and Stealthiness Evaluation} \label{appendix:attack_effectiveness}

This section provides a more detailed discussions about the results of the effectiveness and stealthiness evaluation on 80 attack scenarios from real-world driving traces in \S\ref{sec:eval_results}.

\textbf{Effectiveness.} Table~\ref{tbl:effectiveness} shows the results. As shown, our attack is highly effective under all the 3 stealthiness levels: the success rates are 100\% for all scenarios when $\lambda=10^{-4}$ and $10^{-3}$, and are only slightly lower,
but still over 97.5\% when
$\lambda=10^{-2}$, the highest stealthiness level in our experiment.
Among the successful cases, the average success time is all under 0.91 sec, which is substantially lower than 2.5 sec, the required success time. This means that even for a fully attentive human driver who is always able to take over as soon as the attack starts to take effect, the average reaction time is still far from enough to prevent the damage.

\textbf{Stealthiness.} The 3 rightmost columns in Table~\ref{tbl:effectiveness} show the perturbation amounts of our attack in L-norm pixel distances. As shown, $\mathcal{L}_{1}$ and $\mathcal{L}_{2}$ distances are at most 0.071 and 0.109 respectively, which means that the perturbations are on average only 10.9\% brighter than the original surface even at the lowest stealthiness level in our experiment. Fig.~\ref{fig:attack_poc} shows the malicious road patch appearances at different stealthiness levels from the driver's view at 2.5 seconds before our attack succeeds. %
As shown, even for the lowest stealthiness level ($\lambda=10^{-4}$) in our experiment, the perturbations are still smaller than some real-world dirty patterns such as the left one in Fig.~\ref{fig:attack_poc_2.5}. In addition, we find that the perturbations for all these 3 stealthiness levels are already much less intrusive than those in previous physical-world adversarial machine learning attacks in the image space~\cite{zhao2018seeing}, e.g., in Fig.~\ref{fig:other_attacks}.

\section{Attack Stealthiness User Study} \label{appendix:user_study}

In this section, we conduct a user study to more directly evaluate the stealthiness of the DRP attack. We have gone through the IRB process and our study is determined as in the IRB Exempt category since it does not involve the collection of any Personally Identifiable Information (PII) or target any sensitive population.

\textbf{Evaluation methodology.} We use the generated attacks on real-world driving traces in~\S\ref{sec:eval_results} to perform the user study. For an attack scenario, we ask the participants to imagine that they are driving with the ALC system taking control, and then show a sequence of image frames with the malicious road patch from the driver's view at 3, 2.5, 2, 1.5, and 1 second(s) before the attack succeeds. Here, 1 second before the attack succeeds is right before the attack starts to take effect. For each image frame, we ask whether they will decide to take over the driving to avoid danger or potential safety risks. These questions are also asked for the image frames with a benign road patch that only has the base color without the malicious dirty patterns as a control group.

Since our attack is designed for drivers who are in favor of using ALC system in normal cases, the same set of questions are asked at the beginning for the original image frames without attack, and we only accept a participant if she does not choose to take over the driving for these cases. This process also helps filter out ill-behaved participants who just provide random answers. Since DRP is a new form of attack vectors on the road, we do not tell the participants that the study is related to security attacks. Instead, we only tell them that our focus is on surveying driver's decisions under ALC systems for different road surface patterns such as road patches and scratches. At the beginning of the study, we also provide an introduction of ALC systems with demo videos to ensure that the participants fully understand what driving technology we are surveying about. To understand the distribution of the participant background, we also ask demographic information and background information related to driving and ALC usage. None of the questions in our study involve PII or target any sensitive population; our study is thus determined as in the IRB Exempt category.

\begin{figure}[t]
\centering
\includegraphics[width=\columnwidth]{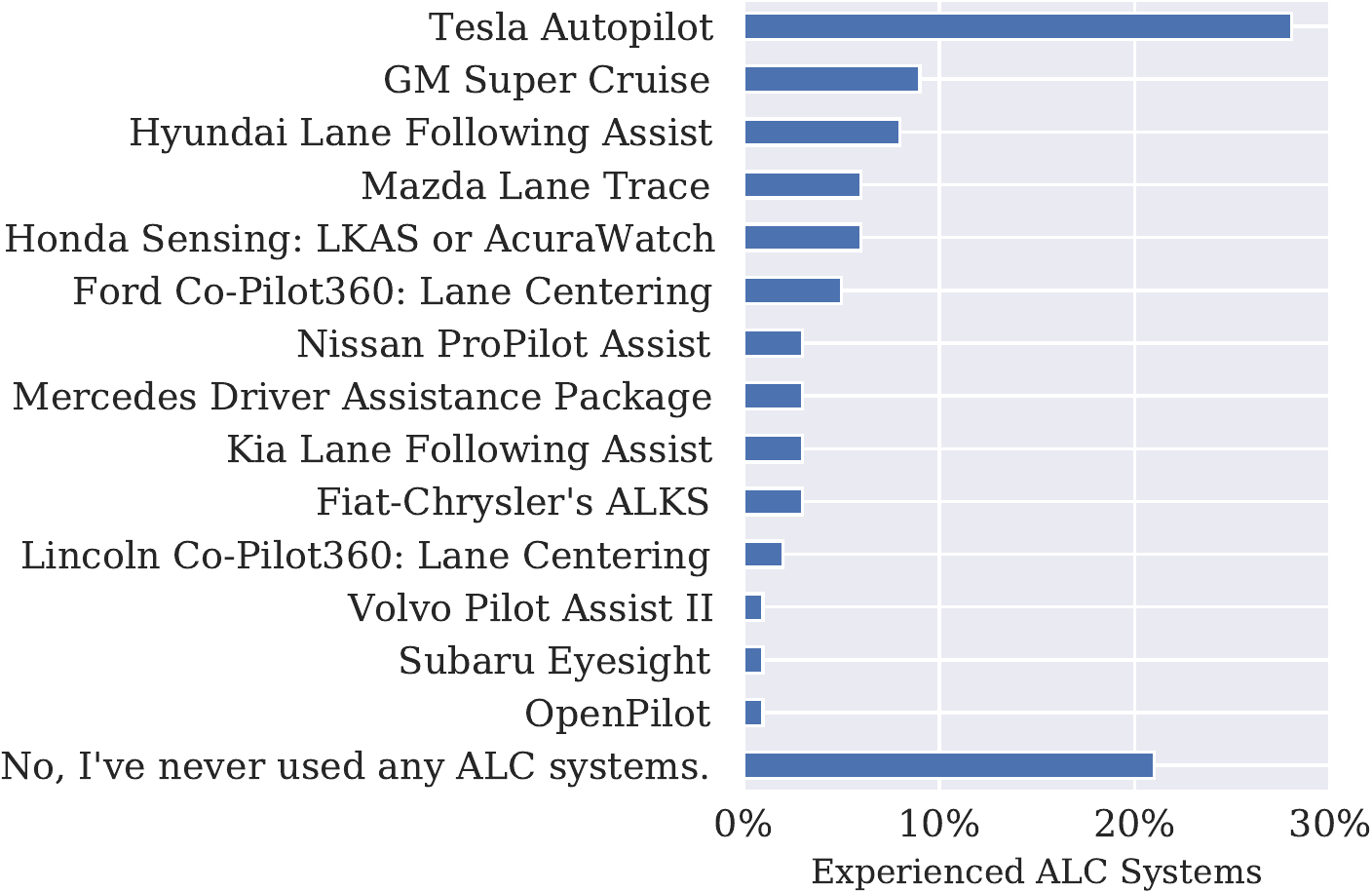}
\includegraphics[width=\columnwidth]{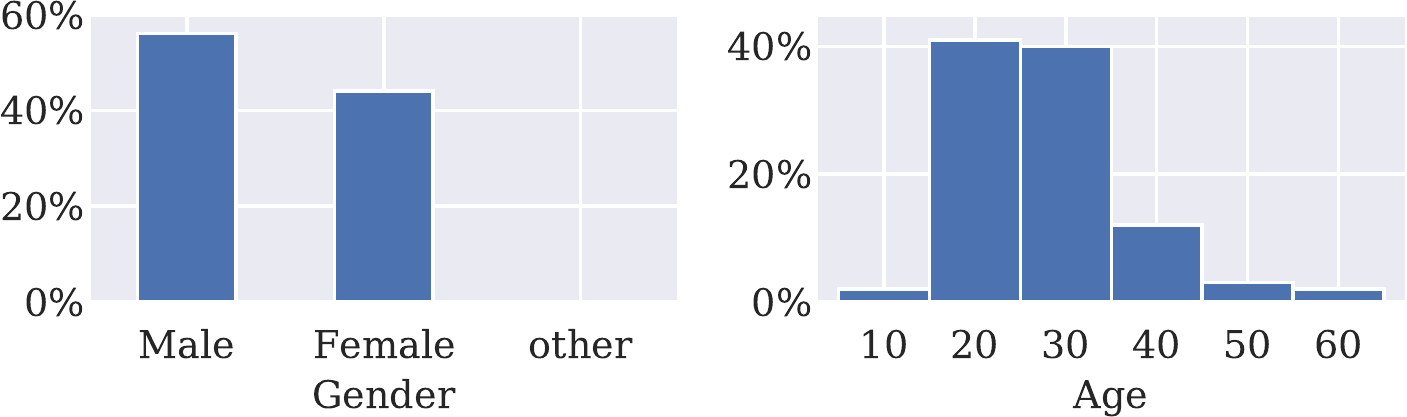}
\vspace{-0.25in}
\caption{Statistics of the ALC system experience and demographic information in the attack stealthiness user study.}
\label{fig:user_study_demo}
\end{figure}

\textbf{Evaluation setup.} We use Amazon Mechanical Turk~\cite{mturk} to perform this study, and in total collected 100 participants. All of them have driving experience, which is confirmed by asking them the age when first licensed and the weekly driving mileage. A local-road driving trace is used in this study, and for the scenarios with attack, we evaluate 3 stealthiness levels as in~\S\ref{sec:eval_results} (i.e., $\lambda=10^{-2}, 10^{-3}, 10^{-4}$). The survey is available at~\cite{user_test_link}. Among the 100 participants, 56\% are male and 44\% are female. The average age is 32.3 years old. 79\% have experienced at least one ALC system, among which Tesla Autopilot has the largest share (28\%). Statistics of ALC experiment and demographic information are shown in Fig.~\ref{fig:user_study_demo}.

\begin{figure}[t]
\centering
\includegraphics[width=0.8\columnwidth]{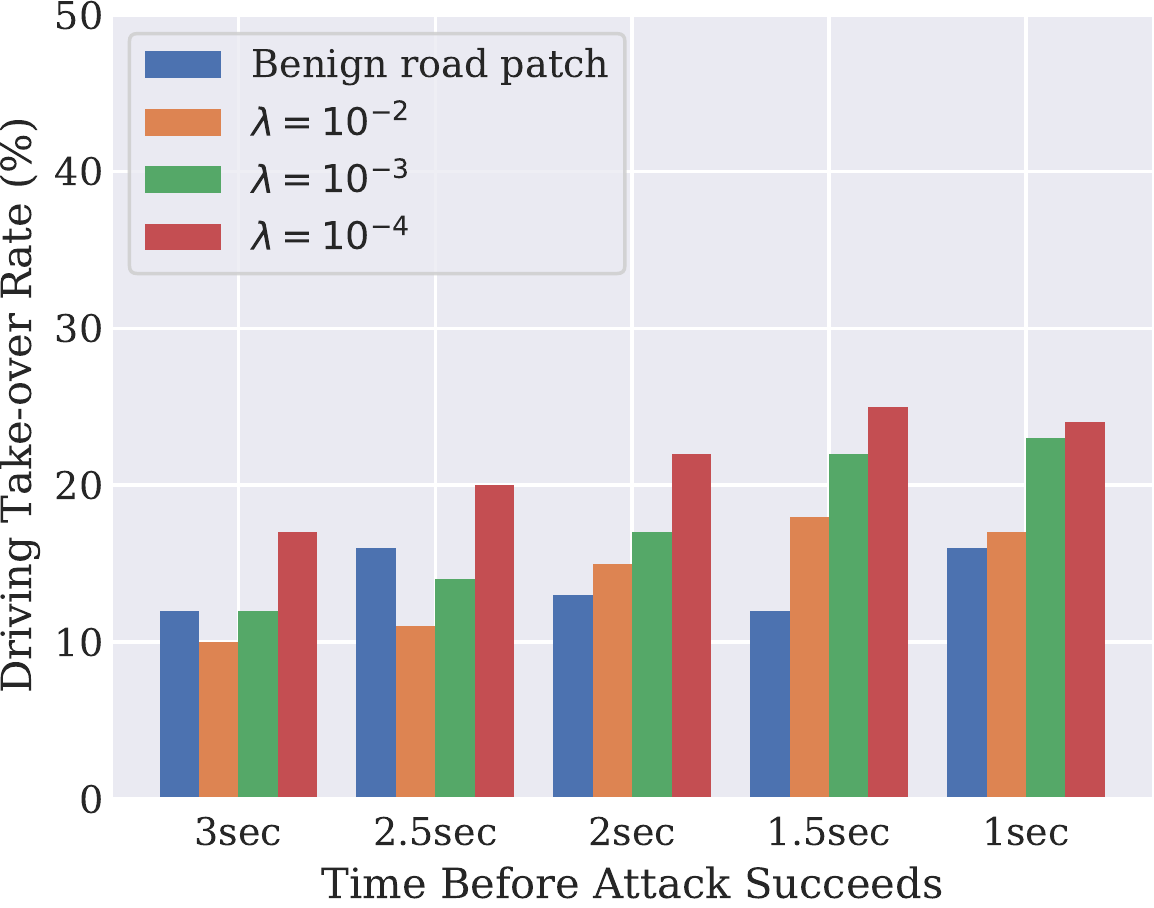}
\vspace{-0.1in}
\caption{Results of the attack stealthiness user study. Driving take-over rate is the percentage of participants who choose to take over the driving at a particular time point before the attack succeeds.}
\label{fig:user_study}
\end{figure}

\begin{figure*}[p]
\centering
\includegraphics[width=2\columnwidth]{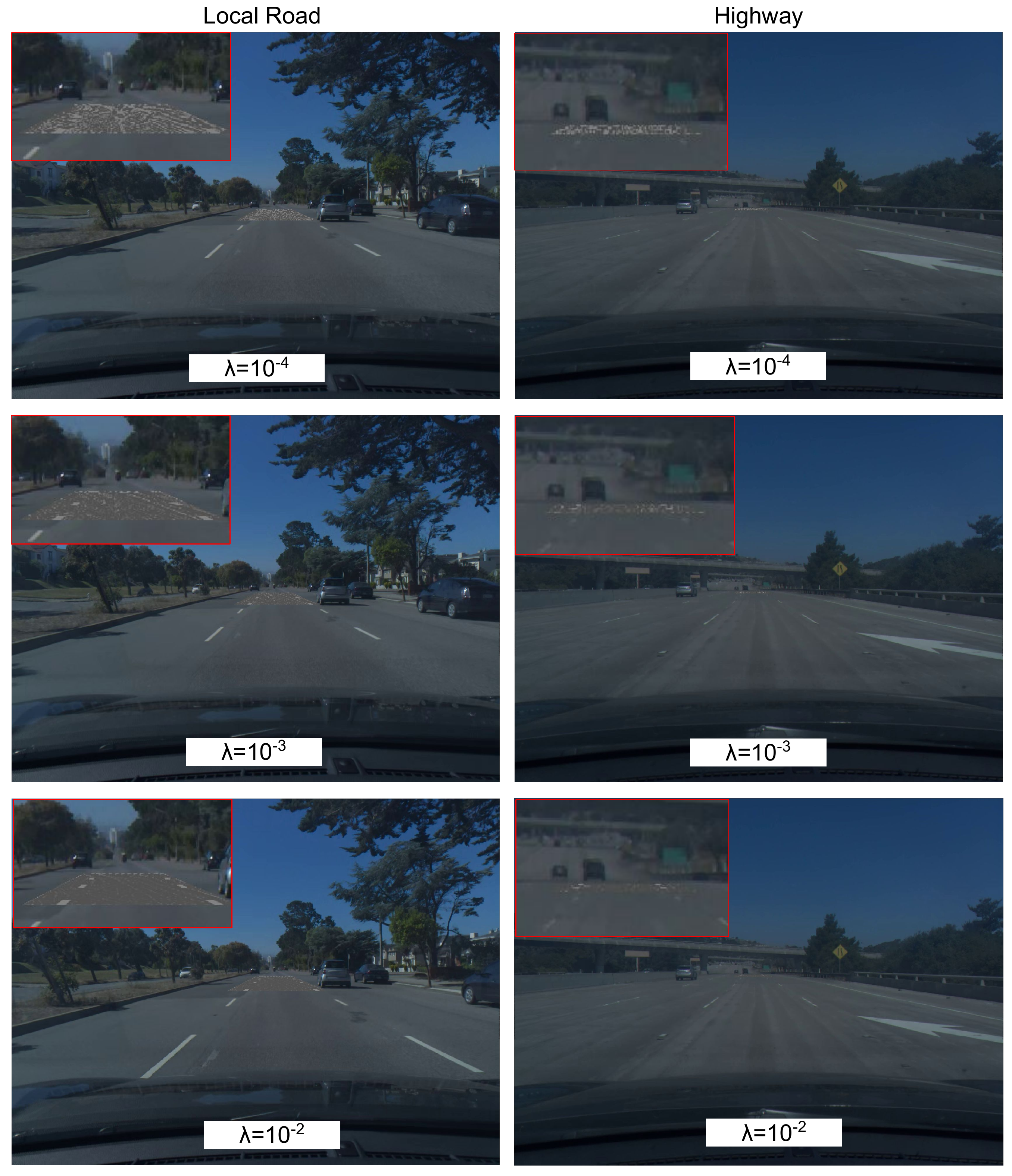}  %
\caption{Large images of the driver's view images in Fig.~\ref{fig:attack_poc}, which correspond to the driver's view 2.5 sec (average driver reaction time~\cite{cali_driver_reaction_time}) before the attack succeeds under different stealthiness levels in our real world trace based attack effectiveness evaluation in~\S\ref{sec:eval_results}. The inset figures are the zoomed-in views of the malicious road patches.}
\label{fig:poc_large_picture}
\end{figure*}

\textbf{Results.} Fig.~\ref{fig:user_study} shows the study results. As shown, the closer it is to the attack success time, the more participants choose to take over the driving in the attacked scenarios since the dirty patterns become increasingly larger and clearer. Among the 3 stealthiness levels, the driver decisions are consistent with our design: the lowest stealthiness level ($\lambda=10^{-4}$) has the highest take-over rate, while the highest level ($\lambda=10^{-2}$) has the lowest. In particular, we find that even for the lowest stealthiness level ($\lambda=10^{-4}$), only \textit{less than 25\%} of the participants decide to take over before the attack starts to take effect. As shown in Fig.~\ref{fig:attack_poc} and Fig.~\ref{fig:poc_large_picture}, at this stealthiness level the white dirty patterns are quite dense and prominent. Thus, these results suggest that the majority of human drivers today do not treat dirty road patches as road conditions where ALC systems cannot handle.

As introduced in~\S\ref{sec:problem_formulation_goals}, 2.5 seconds is commonly used as the average driver reaction time to road hazard. Thus, at 2.5 seconds or more before the attack succeeds, the human driver still has a chance to take over the driving to prevent the damage in common cases, as long as she can realize that it is a road hazard. However, our results show that only less than 20\% of the participants decide to take over at 2.5 and 3 seconds before our attack succeeds even for the lowest stealthiness level. In particular, when the stealthiness levels are $\lambda=10^{-2}$ and $\lambda=10^{-3}$, the take-over rates at these 2 time points are similar to the rates for the benign road patch with only the base color. This suggests that at the time when there is still a chance to prevent the damage in common cases, our attack patches at $\lambda=10^{-2}$ and $10^{-3}$ \textit{appear to be as innocent as normal clean road patches to human drivers}. In these cases, the take-over rates are only \textit{less than 15\%}, which are from participants who will take over even for normal clean road patches. Note that the take-over rates in practice are likely to be lower than this since (1) this study is performed for a local-road scenario, while the road patches in highway scenarios are much farther and thus much less noticeable as shown in Fig.~\ref{fig:attack_poc} and Fig.~\ref{fig:poc_large_picture}, and (2) the road patches in this study are digitally synthesized into the image frames, which may appear less natural and thus may more easily alert the participants.

\textbf{Stealthiness from pedestrian view.} In local road scenarios, the stealthiness from the pedestrian's view is also an aspect worth considering, as pedestrians may report anomalies if our attack patch looks too suspicious. Our user study includes the driver’s view at 1 second before the attack succeeds, which is ~7 meters to the driver’s eyes so similar to the distance from the pedestrian on local roads. However, only <25\% of the participants choose to take over driving, meaning that >75\% do not think our attack patch at this distance looks suspicious enough to affect driving. This may be because the general public today does not know that dirty road patches can be a road hazard. We hope that our paper can expose this and thus help raise such awareness.

\diff{
\section{Robustness to Run-Time Driving Trajectory and Angle Deviations}
\label{appendix:eval_robustness}

\textbf{Experimental setup.} To model the run-time victim driving trajectory and angle deviations from the motion model predicted ones, we introduce vehicle position shifting at each vehicle control step following different motion model prediction error levels. According to previous studies, the kinematic bicycle model used in our attack has around 0.08-meter errors on average with a standard deviation of around 0.02 meters when the control frequency is 10 Hz (i.e., control applied every 100 ms)~\cite{kong2015kinematic}. Since the control frequency in \op{} is 100 Hz, we estimate the errors in one control step (applied every 10 ms) as having 0.008 meters mean with 0.002 meters standard deviation. Based on this, we evaluate the attack effectiveness with motion model errors randomly sampled from the uniform distribution $U(-\Delta, \Delta)$ with $\Delta=0.012, 0.024, 0.048$ and $0.096$, which correspond to 2$\times$, 4$\times$, 8$\times$, and 16$\times$ standard deviations respectively. These errors are added to the longitudinal and lateral position changes independently at each control step when we use the motion model to obtain the driving trajectory of the victim at the attack evaluation time.

Besides adding vehicle position shifting at each control step, we also vary vehicle starting positions in the current lane to create a wide range of approaching angles and distances to the patch. Specifically, considering the vehicle drives around 1 meter in a frame, we evaluate 3 longitudinal offsets, 0m, 0.5m, and 1 m, which are the forward shifting distances from the default starting position in~\S\ref{sec:eval_results} (i.e., at the lane center and 7 m to the patch). For example, 1m longitudinal offset means 6 m to the patch. For the lateral direction, we evaluate 9 offsets from the lane center as shown in Table~\ref{tbl:trace_different_starting_pos}, which are represented as the percentages of the maximum in-lane lateral shifting from the lane center. Negative and positive signs mean left and right shifting. For example, 95\% lateral offset means almost the rightmost possible position within the current lane. In total, these longitudinal and lateral offsets create 27 different approaching angles and distances to the patch. In this experiment, we also add a motion model error of $\Delta=0.024$ (4-sigma) at each control step to incorporate run-time victim driving trajectory and angle deviations as described above. The evaluation dataset and metrics are the same as those in~\S\ref{sec:eval_results}. The experiments use the same malicious road patches generated in~\S\ref{sec:eval_results} with $\lambda=10^{-3}$ and PAR=50\%, the default stealthiness level as discussed in~\S\ref{sec:eval_results}.

\textbf{Results.} Table~\ref{tbl:eval_robustness} shows the attack success rates and time of the same patches generated in~\S\ref{sec:eval_results} under different levels of driving trajectory and angle deviations at the attack evaluation time. As shown, at the 4 motion model error levels, the driving trajectory changes, measured by RMSE (Root-Mean-Squared Error) between the attack evaluation time trajectory and the attack generation time trajectory, vary significantly from 33.6cm to even 1m on average at each control step. This creates $2.0^{\circ}$ to $5.8^{\circ}$ changes of the viewing angles to the patch center at each control step on average. Note that the largest motion model error level here ($\Delta=0.096$) is already 16$\times$ standard deviation (or 16-sigma) and thus very unlikely to happen in real world. In these cases, we find that the attack patches can still have at least 95\% success rates and at most 0.869 sec success time, which thus shows a high robustness to run-time driving trajectory and angle deviations.

Table~\ref{tbl:trace_different_starting_pos} shows the attack success rates from the 27 different starting positions. As shown, the attack success rate  monotonically decreases when the starting position has larger lateral offsets from the lane center. In contrast, although the lateral offset range ($<$0.735m) is smaller than the longitudinal offset range (1m), the attack success rate are almost unaffected by the longitudinal direction shifting. Overall, even for the leftmost and rightmost starting positions in the current lane, our attack still achieves at least 88\% attack success rates, which shows a high robustness to a wide range of victim approaching angles and distances. 
}

\begin{table}[t]
\caption{
\diff{
Evaluation results for attack robustness under different levels of run-time driving trajectory and angle deviations. $\Delta$ controls the deviations introduced to lateral and longitudinal driving directions independently following the uniform distribution $\mathcal{U}(-\Delta, \Delta)$. Avg. $\Delta\theta$ and RMSE are the average patch viewing angle and vehicle position differences between the attack evaluation time trajectory and the attack generation time trajectory at each control step. $\theta$ is calculated from the vehicle position to the center of the patch visible area.
}
}
\setlength\tabcolsep{3pt} %
\label{tbl:eval_robustness}
\footnotesize
\centering
\begin{tabular}{ccccc}
\toprule
\begin{tabular}[c]{@{}c@{}}Error Level\\  $\Delta$\end{tabular} & \begin{tabular}[c]{@{}c@{}}Avg.\\ $\Delta\theta$\end{tabular} & RMSE & Succ. Rate & Succ. Time (s) \\ \hline
0.012 & $2.0^{\circ}$ & 33.6 cm & 97.5\% & 0.869 \\
0.024 & $2.6^{\circ}$ & 39.6 cm & 97.5\% & 0.830 \\
0.048 & $3.7^{\circ}$ & 54.4 cm & 97.5\% & 0.795 \\
0.096 & $5.8^{\circ}$ & 100 cm & 95.0\% & 0.804 \\ \toprule
\end{tabular}
\end{table}

\begin{table}[t]
\caption{\diff{Attack success rates from 27 different starting positions. The lateral offset is the percentage of the maximum in-lane lateral shifting from the lane center; negative and positive signs mean left and right shifting. Longitudinal offset is the forward shifting distance towards the patch.}
}
\centering
\footnotesize
\setlength{\tabcolsep}{3.5pt}
\begin{tabular}{cccccccccc}
\toprule
\multirow{2}{*}{\begin{tabular}[c]{@{}c@{}}Long.\\ Offset\end{tabular}} & \multicolumn{9}{c}{Lateral Offset (negative/positive sign: left/right shifting)}                              \\ \cline{2-10} 
                                                                           & -95\% & -75\% & -50\% & -25\% & 0\%   & 25\% & 50\% & 75\% & 95\% \\ \hline
1m                                                                         & 93\%  & 98\%  & 99\%  & 99\%  & 100\% & 98\% & 95\% & 93\% & 89\%  \\
0.5m                                                                       & 93\%  & 98\%  & 99\%  & 99\%  & 100\% & 98\% & 95\% & 93\% & 89\%  \\
0m                                                                         & 94\%  & 98\%  & 99\%  & 99\%  & 98\%  & 98\% & 94\% & 93\% & 88\% \\
\toprule
\end{tabular}
\label{tbl:trace_different_starting_pos}
\end{table}

\section{Attack Generality Evaluation}
\label{appendix:dnn_design_diff}

To achieve \diffst{end-to-end}\diff{real-world} attack impact on ALC systems, our attack is designed around exploiting the LD model, the most critical element in ALC systems (\S\ref{sec:background}). Thus, in this section we evaluate the generality of our attack against LD models of different designs. Ideally we hope to evaluate on LD models from other production ALC systems besides \op{}, e.g., from Tesla Autopilot. However, to best of our knowledge, \op{} is the only one that is currently open sourced. Fortunately, we find that the LD models in some older versions of \op{} actually have different DNN designs (detailed below), which thus can also serve for our purpose.

\textbf{Experimental setup.} We select LD models from \op{} v0.7.0 (used in previous sections), v0.6.6, and v0.5.9 to perform generality evaluation considering their large DNN architecture differences with each other, which are summarized in Table~\ref{tab:eval_generality}. Among them, the v0.5.9 model has the largest DNN architecture differences to the other two. First, the v0.5.9 model has 1.6 million trainable parameters and 49 layers, which are substantially smaller than the other models: 58\% fewer weights and 58\% fewer layers than the v0.7.0 model, and as high as 63\% fewer weights and 60\% fewer layers than the v0.6.6 model. At the design level, after a sequence of the CNN residual blocks~\cite{he2016deep}, the v0.5.9 model uses a naive RNN layer~\cite{cho2014learning}, while the other two models use a more advanced RNN design called Gated Recurrent Unit (GRU). GRU is proposed to deal with long text in the natural language processing domain similar to LSTM~\cite{hochreiter1997lstm} because the naive RNN is known to focus too much on the current and recent few words~\cite{bengio1994learning}. Thus, the v0.6.6 and v0.7.0 models with such more advanced RNN design should be able to consider more historical states and thus produce more stable results during driving.

Between v0.6.6 model and the v0.7.0 model, the former has 4.2 million weights and 124 layers, which are 11\% more weights and 7\% more layers than the latter and thus the largest among the three. A key design-level difference between them is that the Batch Normalization (BN) layer~\cite{ioffe2015batchnorm} is used in the v0.6.6 model but later removed in the v0.7.0 model, potentially because BN layers can harm RNN performance~\cite{ba2016layer}. The evaluation dataset and metrics are the same as~\S\ref{sec:eval_results}. The experiments use $\lambda=10^{-3}$ and PAR=50\%, the default stealthiness level as discussed in~\S\ref{sec:eval_results}.

\textbf{Results.}
Table~\ref{tab:eval_generality} shows the results. As shown, our attack is able to achieve at least 90\% success rates against all three LD models of different DNN designs, which shows a high generality of the attack methodology. Among them, the v0.6.6 model seems to have the highest robustness against our attack, with 5-15\% lower success rates and also higher success time than those for the other models. Also, the $\mathcal{L}_1$ and $\mathcal{L}_2$ pixel distances for it are the smallest. Since its major design-level difference to the v0.7.0 model is BN layers, we suspect that such robustness increase might come from the use of BN layers. Interestingly, this is contradicting to latest researches that find BN layers can harm DNN model robustness~\cite{galloway2019batch}. We plan to systematically investigate this in the future. Between the v0.7.0 and v0.5.9 models, the attack effectiveness is very close. Thus, even though GRU is supposed to be more advanced and make the driving more stable, it does not show increased robustness to our attack.

\textbf{Transferability.} We also evaluate the transferability of the patch generated by one model to the other two models. The transfer success rate is 63\% on average, which is similar to the state-of-the-art rate for MNIST and CIFAR10~\cite{suya2020hybrid}.

\begin{table}[t]
\caption{Evaluation results for attack generality against LD (Lane Detection) models of different DNN designs in \op{}. Evaluation metrics are the same as those in Table~\ref{tbl:effectiveness}.}
\label{tab:eval_generality}
\centering
 \footnotesize
\setlength{\tabcolsep}{3.5pt}
\begin{tabular}{ccccccc}
\toprule
\begin{tabular}[c]{@{}c@{}}LD \\ Model \end{tabular} & 
\begin{tabular}[c]{@{}c@{}}DNN \\ Arch. \end{tabular}& 
\begin{tabular}[c]{@{}c@{}}Succ.\\ Rate\end{tabular} & 
\begin{tabular}[c]{@{}c@{}}Succ.\\ Time (s)\end{tabular} &
\begin{tabular}[c]{@{}c@{}}Pixel \\$\mathcal{L}_{1}$ \end{tabular} &
\begin{tabular}[c]{@{}c@{}}Pixel \\$\mathcal{L}_{2}$ \end{tabular} &
\begin{tabular}[c]{@{}c@{}}Pixel \\$\mathcal{L}_{inf}$ \end{tabular} \\ \hline
v0.7.0 & \begin{tabular}[c]{@{}l@{}}CNN+\\ GRU\end{tabular} & 100\% & 0.887 & 0.033 & 0.066 & 0.200 \\\hline
v0.6.6 & \begin{tabular}[c]{@{}c@{}} \ CNN+\\ BN+GRU\end{tabular} & 90\% & 0.940 & 0.034 & 0.069 & 0.201 \\\hline
v0.5.9 & \begin{tabular}[c]{@{}l@{}} \ CNN+\\ naive RNN\end{tabular} & 100\% & 0.919 & 0.038 & 0.077 & 0.201 \\ \toprule
\end{tabular}
\end{table}

\section{Attack Deployability Evaluation} \label{appendix:deployability}

In this section, we evaluate the attack deployability by estimating the required efforts to deploy the attack road patch.

\textbf{Evaluation methodology and setup.} First, we consider a road patch of size 1.6-1.8 m wide and 6-8 m long as a \textit{deployment unit} for 2 people at a time, based on normal arm span and available adhesive road patch lengths online~\cite{americanroadpatch}. Different from a legitimate road patch deployment, the attacker does not need to fill asphalt and stamp the road, which are the most time-consuming steps. Instead, she only needs to place the patch on the road surface, and this step only takes 5-10 seconds for the defined deployment unit above based on demo videos of the adhesive patch deployment~\cite{adhesive_video}. With this, we can then estimate the deployment effort for 2 people by calculating the number of deployment units for a given attack road patch. We start from an attack patch size of 5.4m$\times{}$36m (width$\times{}$length), the size used in previous sections, and then vary the allowed number of deployment units in the patch area using multi-piece patch attack mode (\S\ref{sec:multiple_patch}) to evaluate the trade-off between deployability and attack effectiveness. 

\textbf{Results.} Table~\ref{tab:multiple_patch_attack} shows the experiment results with 4 to 15 deployment units. To fully cover the 5.4m wide and 36m long patch placement area, 15 deployment units, each for a piece of 1.8m$\times{}$7.2m patch, are required for 2 people. When the allowed piece number decreases, the attack success rate decreases accordingly since the perturbable area size becomes smaller. Interestingly, the success rate is still as high as 93.8\% when only 8 pieces are allowed, which is able to significantly reduce the original deployment efforts by nearly 50\%. This suggests that it is not necessary to cover the entire attack deployable area to achieve a high success rate for the DRP attack, which thus concretely shows the benefit of our multi-piece patch attack mode design in improving deployability. Fig.~\ref{fig:poc_multiple} shows an example DRP attack with such 8 pieces of 1.8m$\times{}$7.2m road patches from the driver's view. To deploy 8 such pieces, the attacker only needs to find opportunities to block the road for 1-2 min in total. To maximize the stealthiness, such an effort can be spread over 8 different times, each for a single piece. Each time, it only requires a 5-10 seconds vacant period of the target road, which is common in the U.S. since the majority (e.g., 78\%~\cite{zhao2017trafficnet}) of the driving scenarios in the U.S. are \textit{free-flow} scenarios, where a vehicle has at least 5-9 seconds headway. In the late night, such opportunities can be even more frequent.

\begin{table}[t]
\caption{Evaluation results for attack deployability using multi-piece patch attack mode. Piece \#: Allowed number of 1.8m$\times$7.2m patches for the attack; 15 pieces cover the entire deployable area.}
\vspace{-0.1in}
\centering
\footnotesize
\label{tab:multiple_patch_attack}
\begin{tabular}{cccr}
\toprule
Piece \# & \begin{tabular}[c]{@{}c@{}}Succ. \\ rate\end{tabular} & \begin{tabular}[c]{@{}c@{}}Succ. \\ time (s) \end{tabular}&  \begin{tabular}[c]{@{}c@{}} Deploy. time \\ improvement\end{tabular} \\\hline
\begin{tabular}[c]{@{}c@{}}15 (full)\end{tabular}       & 100\%      & 0.89            & -          \\ \hline
12       & 98.8\%     & 0.94    &    -20\%    \\
8        & 93.8\%     & 1.28  &    -47\%    \\
6        & 76.3\%     & 1.63    &    -60\%    \\
\toprule
\end{tabular}
\vspace{-0.15in}
\end{table}

\section{Robustness To Different Lighting Conditions in Physical World} \label{appendix:lighting}

In the physical world, the lighting condition would change dramatically at different times of day. Thus, in this section, we further evaluate the robustness of our DRP attack to the variation of lighting conditions by configuring the lighting sources in our miniature-scale experiment setup (\S\ref{sec:eval_realizability}).

\textbf{Experimental setup.}
In our miniature-scale experiment setup (\S\ref{sec:eval_realizability} and Fig.~\ref{fig:small_scale_expr_setup}), there are 3 light sources: (1) 2 studio lights, (2) room light in the lab ceiling, and (3) window light from outdoor. We control the light intensity of each of them as follows. For (1), each studio light has 4 levels of light intensity with different numbers of light bulbs turned on or off. We denote them as ``All On'', ``3 On'', ``1 On'', and ``All Off''. For (2), the room light in the lab can be turned on or off. Thus, we denote them as ``On'' and ``Off''. For (3), our lab window is facing east, and we conduct all experiments between 2-3 pm during sunny summer days. To control the intensity of window light, we adjust the window blinds to ``Open'' or ``Closed''. By combining the light intensity control for the 3 light sources above, we create 12 different lighting conditions in total as listed in Table~\ref{tbl:lighting_conditions}. To understand the correspondence between the lighting conditions and the physical world scenarios, we measure the light intensity (in \textit{lux}) using the popular lighting condition measurement App \textit{Lux Light Meter Pro}~\cite{lux_measure_app}. The measurements are repeated 10 times for each lighting condition. We use the same physically-printed malicious dirty road patch as in~\S\ref{sec:eval_realizability} for all lighting conditions.

\begin{table}[tbp]
\caption{Attack effectiveness for the same physically-printed malicious road patch under 12 different lighting conditions in our miniature-scale experiment setup (\S\ref{sec:eval_realizability}), ranked by the light intensity (i.e., average lux).}
\label{tbl:lighting_conditions}
\vspace{-0.1in}
\footnotesize
\centering
\setlength{\tabcolsep}{2pt}
\begin{tabular}{@{}ccccccc@{}}
\toprule
\begin{tabular}[c]{@{}c@{}}Lighting\\ cond. \#\end{tabular} & \begin{tabular}[c]{@{}c@{}}Left studio\\ light\end{tabular} & \begin{tabular}[c]{@{}c@{}}Right studio\\ light\end{tabular} & \begin{tabular}[c]{@{}c@{}}Room\\ light\end{tabular} & \begin{tabular}[c]{@{}c@{}}Window\\ blinds\end{tabular} & \begin{tabular}[c]{@{}c@{}}Avg.\\ lux\end{tabular} & \begin{tabular}[c]{@{}c@{}}Desired steer angle\\ (under attack)\end{tabular} \\ \midrule
1 & All On & All On & On & Open & 1210 & 24.5$^{\circ}$ to left \\
2 & All Off & All On & On & Open & 1055 & 24.4$^{\circ}$ to left \\
3 & All On & All Off & On & Open & 885 & 24.5$^{\circ}$ to left \\
4 & 3 On & 3 On & On & Open & 883 & 23.9$^{\circ}$ to left \\
5 & All Off & 3 On & On & Open & 772 & 24.4$^{\circ}$ to left \\
6 & 3 On & All Off & On & Open & 612 & 24.5$^{\circ}$ to left \\
7 & 1 On & 1 On & On & Open & 503 & 24.5$^{\circ}$ to left \\
8 & All Off & 1 On & On & Open & 426 & 24.3$^{\circ}$ to left \\
9 & 1 On & All Off & On & Open & 411 & 24.3$^{\circ}$ to left \\
10 & All Off & All Off & On & Open & 252 & 24.4$^{\circ}$ to left \\
11 & All Off & All Off & Off & Open & 160 & 20.7$^{\circ}$ to left \\
12 & All Off & All Off & Off & Closed & 15 & 20.2$^{\circ}$ to left \\ \bottomrule
\end{tabular}
\end{table}

\begin{figure}[t]
\centering
\includegraphics[width=.95\linewidth]{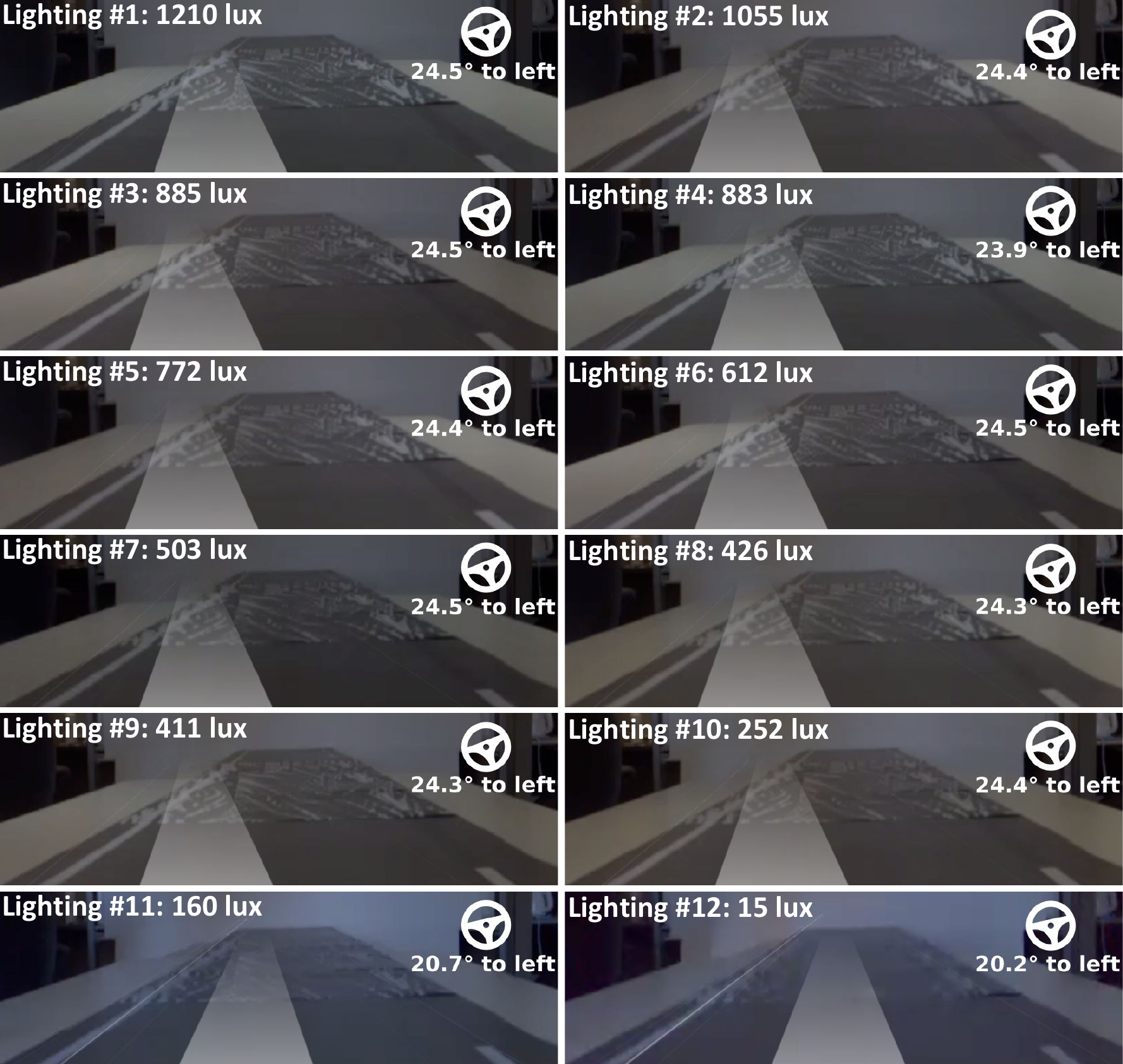}
\vspace{-0.1in}
\caption{Visualization of the desired steering angle and desired driving path under 12 different lighting conditions. Due to \op{} camera's automatic exposure adjustment feature, the brightness of these images appears relatively similar to each other except the two darkest ones at the bottom. However, such a feature also makes our attack more robust.}
\label{fig:different_lighting}
\end{figure}

\textbf{Results.} Table~\ref{tbl:lighting_conditions} shows the attack effectiveness under 12 different lighting conditions, in which the collected camera images are also visualized in Fig.~\ref{fig:different_lighting}. As shown, our patch can maintain a desired steering angle of 20-24$^{\circ}$ to the left under a wide range of light intensity (15-1210 lux, similar to lighting conditions ranging from sunset/sunrise of a fully overcast day to midday of an overcast day as shown in Table~\ref{tab:lighting_reference}~\cite{lighting}), which are all significantly different from the desired steering angle in the benign scenario (0.9$^{\circ}$ to the right). This thus shows high robustness of our attack patch to common lighting condition changes in the day time. 

Interestingly, despite the substantial lighting condition variation, we find that the camera images appear relatively similar in all lighting conditions except the two darkest ones as shown in Fig.~\ref{fig:different_lighting}. This might be because the \op{} camera has an \textit{auto exposure adjustment} feature, which can keep the brightness of the DNN model input to be relatively the same even under different lighting conditions in the real world. For vision applications, it is a common practice to enable auto exposure adjustment as it is critical to the model performance~\cite{yang2018challenges}. However, such a feature also makes our attack more robust.

\begin{table}[t]
\caption{The correspondence between light intensity and the physical world scenarios~\cite{lighting}.}
\vspace{-0.12in}
\label{tab:lighting_reference}
\centering
\small
\begin{tabular}{rl} \toprule
\multicolumn{1}{r}{Illuminance} & \multicolumn{1}{l}{Example}                                                            \\\hline
120,000 lux     & Brightest sunlight               \\
111,000 lux     & Bright sunlight                  \\
20,000 lux                      & \begin{tabular}[l]{@{}l@{}}Shade illuminated by \\ entire clear blue sky, midday\end{tabular} \\
1,000-2,000 lux & Typical overcast day, midday     \\
400 lux         & Sunrise or sunset on a clear day \\
0.25 lux        & A full Moon, clear night sky     \\
0.01 lux        & A quarter Moon, clear night sky  \\\toprule
\end{tabular}
\end{table}

\diff{
\section{Robustness to Different Viewing Angles in Physical World} \label{appendix:miniature_robustness}

\textbf{Experimental setup.} In this section, we evaluate the robustness of the physically-printed malicious road patch in our miniature-scale experiment setup (\S\ref{sec:eval_realizability}) to different patch viewing angles. Specifically, we move the official OpenPilot dashcam device in the setup (Fig.~\ref{fig:small_scale_expr_setup}) to create 45 different viewing angles in total using the combinations of 5 distances to the patch and 9 lateral offsets to the lane center. The 5 distances are 5, 6, 7, 8, and 9 m in the real-world scale (41.6, 50, 58.3, 66.7, and 75 cm in the miniature scale). The 9 lateral offsets as shown in Table~\ref{tbl:miniature_robustness} are represented as the percentages of the maximum in-lane lateral shifting from the lane center, where negative and positive signs mean left and right shifting. For example, 95\% lateral offset means almost the rightmost possible position within the current lane. Note that 7m and 0\% is the distance and lateral offset used at the attack generation time. The light intensity is 471 Lux in this experiment, a roughly medium one among the 12 in Table~\ref{tbl:lighting_conditions}.

\textbf{Results.}
Table~\ref{tbl:miniature_robustness} shows the desired steering angles from these 45 different viewing angles. As shown, the same physically-printed patch is able to maintain an attack impact of 23.4-26.0$^{\circ}$ to the left from all viewing angles, while those in the benign scenario is 0.9$^{\circ}$ to the right (\S\ref{sec:eval_realizability}). Such high robustness to viewing angles is likely due to our robustness improvement designs in~\S\ref{sec:design_opt_robust} and Appendix~\ref{appendix:design_opt_robust}. 
We also notice that our patch is slightly less effective at the top-right corner of Table~\ref{tbl:miniature_robustness}, e.g., 95\% lateral offset at 5m to the patch. Since 7m is the distance at the patch generation time and the attack goal is deviating to the left, at 5m to the patch the victim should not locate near the rightmost position in the lane due to the attack impact at 7m. Thus, such slightly reduced attack effectiveness is consistent with our attack design. Nevertheless, since we consider the run-time driving trajectory/angle deviations and camera sensing inaccuracies in the patch generation (\S\ref{sec:design_opt_robust} and Appendix~\ref{appendix:design_opt_robust}), our patch still maintains high effectiveness in such viewing angles.

\textbf{Video demo.} To illustrate such attack robustness, we record videos in which we dynamically move the official OpenPilot dashcam device around to create wide patch viewing angles and distances in the miniature-scale setup while showing the real-time lane detection results on the device screen. Video link is at \textbf{\url{https://sites.google.com/view/cav-sec/drp-attack}}. Representative screenshots from the videos are in Fig.~\ref{fig:small_scale_dynamic_moving}. As shown, the physically-printed attack patch can always cause a significant lane bending effect to the left compared to the benign case at similar viewing angles.

\begin{figure*}[tbp]
\centering
  \centering
  \includegraphics[width=\linewidth]{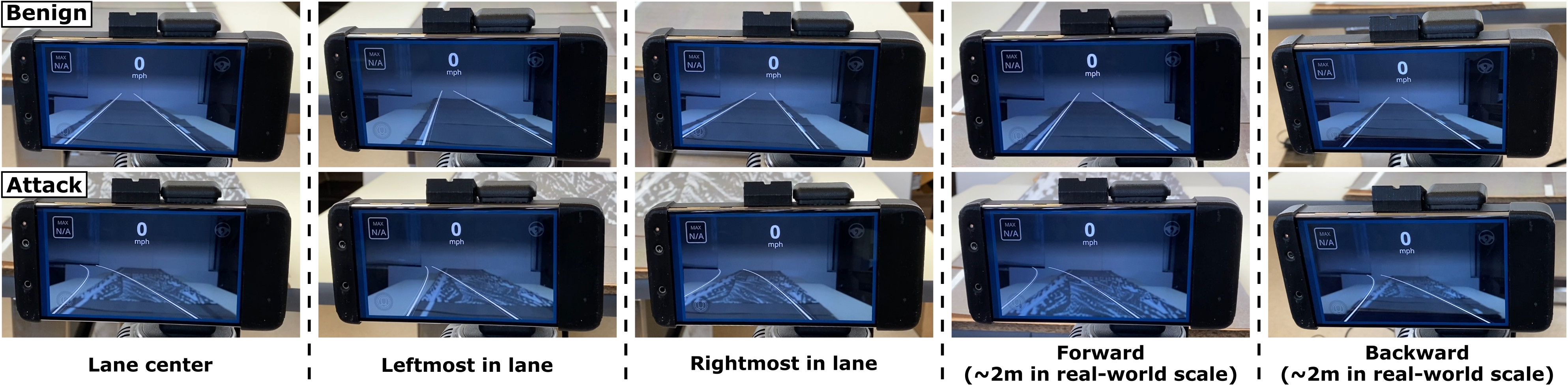}
\caption{\diff{Representative video snapshots of the real-time benign and attacked lane detection results on the official OpenPilot dashcam device screen when we dynamically move the device around in the miniature-scale setup (\S\ref{sec:eval_realizability}) to create wide patch viewing angle and distance changes. The videos are available at \textbf{\url{https://sites.google.com/view/cav-sec/drp-attack}}.}
}
\label{fig:small_scale_dynamic_moving}
\end{figure*}

\begin{table}[t]
\centering
\footnotesize
\setlength{\tabcolsep}{3pt}

\caption{\diff{Desired steering angles for the same physically-printed malicious road patch under 45 different positions in our miniature-scale experiment setup (\S\ref{sec:eval_realizability}). The distance to patch is shown in the real-world scale (12$\times$ miniature scale). The lateral offset is the percentage of the maximum in-lane lateral shifting from the lane center; negative and positive signs mean left and right shifting.}
}
\begin{tabular}{cccccccccc}
\toprule
\multicolumn{1}{l}{}                                 & \multicolumn{9}{c}{Lateral Offset (negative/positive sign: left/right shifting)}                            \\ \cline{2-10} 
\begin{tabular}[c]{@{}c@{}}Distance\\ to Patch \end{tabular} & -95\% & -75\% & -50\% & -25\% & 0\%  & 25\% & 50\% & 75\% & 95\% \\ \hline
5m                  & 24.5$^{\circ}$  & 25.0$^{\circ}$  & 25.5$^{\circ}$  & 25.6$^{\circ}$  & 26.0$^{\circ}$ & 25.6$^{\circ}$ & 24.1$^{\circ}$ & 23.2$^{\circ}$ & 23.5$^{\circ}$ \\
6m                  & 24.5$^{\circ}$  & 25.8$^{\circ}$  & 25.9$^{\circ}$  & 25.8$^{\circ}$  & 25.6$^{\circ}$ & 23.9$^{\circ}$ & 24.1$^{\circ}$ & 24.7$^{\circ}$ & 23.4$^{\circ}$ \\
7m                  & 24.4$^{\circ}$  & 25.5$^{\circ}$  & 25.8$^{\circ}$  & 25.6$^{\circ}$  & 25.6$^{\circ}$ & 26.1$^{\circ}$ & 25.9$^{\circ}$ & 25.9$^{\circ}$ & 26.0$^{\circ}$ \\
8m                  & 25.5$^{\circ}$  & 25.3$^{\circ}$  & 25.8$^{\circ}$  & 25.6$^{\circ}$  & 25.9$^{\circ}$ & 25.8$^{\circ}$ & 25.5$^{\circ}$ & 25.3$^{\circ}$ & 24.7$^{\circ}$ \\
9m                  & 25.0$^{\circ}$  & 25.5$^{\circ}$  & 24.9$^{\circ}$  & 25.4$^{\circ}$  & 25.9$^{\circ}$ & 25.6$^{\circ}$ & 25.8$^{\circ}$ & 24.8$^{\circ}$ & 24.7$^{\circ}$ \\
\toprule
\end{tabular}
\label{tbl:miniature_robustness}
\end{table}
}

\diff{
\section{Details of \op{} ALC system} \label{appendix:openpilot_details}
In this section, we describe the implementation details of the \op{} ALC system, which follows the typical modular ALC system design introduced in~\S\ref{fig:alc_overview}:

\paragraph{Lane Detection (LD).} 
The LD model used in OpenPilot uses recurrent DNN structures (e.g., RNN and GRU), which are more detailed in Appendix~\ref{appendix:dnn_design_diff} for 3 specific versions of it. 
In each frame, the recurrent model receives a front-camera input of 512 pixels wide by 256 pixels high and 512-dimensional recurrent features from the previous frame. The recurrent features are the output of a middle layer. The final output for ALC consists of information of 3 lines (left and right lines and driving path). Each line has coordinates of 192 points (1 m interval from the vehicle to driving direction), uncertainty scores of each coordinate, and a confidence score of its lane. Thus, there are $(192 \times 2 + 1) \times 3 = 1,155$ output values in total. The desired driving path is calculated by the weighted average of the driving path and the center line of the left and right lines weighted by the uncertainty and confidence scores. See OpenPilot code~\cite{openpilot} for more details.
Such recurrent structure is stateful: it allows leveraging the previous detection results to enhance the current frame detection since lane line shapes are typically not changed largely across consecutive frame. OpenPilot LD models output the detected lane line points of the left line, right line, and predicted driving path. Each line is fitted to the 3-degree polynomial, and the desired driving path is then calculated as the weighted average of the three lines with their confidence levels. OpenPilot LD operates at 20 Hz (every 50 ms).

In~\S\ref{sec:real_vehicle}, we inject the attack traces at the end of this step by modifying the ALC source code to replace the real-time LD model outputs with a sequence of attacked ones obtained from the software-in-the-loop simulation at the same driving speed (simulation environment described in~\S\ref{sec:end_to_end}). 

\paragraph{Lateral control.}
OpenPilot adopts Model Predictive Control (MPC)~\cite{MPC} to decide the desired steering angle, which will then be sent to the vehicle actuation step. The input of the MPC is the desired driving path, the current speed, and the current steering angle. This step works at the same frequency as LD, i.e., the desired steering angle is decided every 50 ms. The MPC is stateful: it reuses the solution of the previous frame as the initial solution for the current frame.

\paragraph{Vehicle actuation.} 
Based on the obtained desired steering angle from MPC, OpenPilot vehicle actuation decides the steering angle \textit{change} to actuate in the control step and sends actuation messages through CAN (Controller Area Network) bus. This thus makes the absolute value of the actuated steering angle stateful: the new actuated steering angle is built upon the previous one, by applying the angle \textit{change} actuations. OpenPilot actuation works at 100 Hz control frequency. The actuated steering angle change is \textit{up to 0.25$^\circ$ per control step} (every 10 ms). As described in~\S\ref{sec:background_alc}, such limit is typically imposed in production ALC systems due to the physical constraints of the mechanical control units and also for driving stability and safety~\cite{becker2018functional}.
OpenPilot is integrated to a vehicle by overriding the stock cruise control system. It thus is engaged to control the steering and throttle when the driver turns on the cruise control mode, and can work with stock safety features such as AEB and FCW~\cite{openpilot}.

\paragraph{Connection to the design challenge C2 (\S\ref{sec:problem_formulation_challenges}).} The design challenge C2 in \S\ref{sec:problem_formulation_challenges} (i.e., camera frame inter-dependency due to attack-influenced vehicle actuation) is mainly created by the \textit{actuated steering change limit} described above in the vehicle actuation step. Specifically, due to such limit (up to 0.25$^\circ$ per control step), the actuated steering angle can only be changed up to 1.25$^\circ$ for each LD frame (every 50 ms) even if the attacked desired steering angle from MPC is very large, e.g., 90$^\circ$. Such actuated steering angle change can only lead to up to \textit{0.3-millimeter} lateral deviations at 45 mph ($\sim$72 km/h), which means that a successful attack on only \textit{one single LD output frame} can hardly cause any damage in real-world driving. Thus, to achieve our attack deviation goal (\S\ref{sec:problem_formulation_goals}), the attack must be \textit{continuously effective on sequential camera frames} to accumulatively make the 1.25$^\circ$ actuated steering angle changes per attacked LD output in order to increasingly reach larger actuated steering angles and thus larger lateral deviations per control step. 

Note that although the recurrent DNN structure in the LD model and the MPC in the lateral control also have states, these alone cannot prevent achieving the attack deviation goal by successfully attacking a single LD output frame, as such a single-frame attack can be highly effective and cause the vehicle to actuate a very large steering angle (e.g., 90$^\circ$) in such a frame (if without actuated steering angle change limit). The need to maintain high attack effectiveness across \textit{sequential} camera frames is the foundational problem for challenge C2.

To achieve effective attacks cross sequential camera frames, previous designs for object classification or detection tasks popularly use EoT (Expectation over Transformation) of the patch view in a camera frame~\cite{athalye2018synthesizing, brown2017advpatch}. However, different from the object classification and detection models, the content of the sequential camera frames in ALC systems have inter-dependencies due to attack-influenced vehicle actuation as described in \S\ref{sec:problem_formulation_challenges} (i.e., a successful attack on earlier frames to the right will change the viewing angle towards the right of the road in later frames). Thus, in \S\ref{sec:design_motion_model} we design a novel motion model based input generation to systematically synthesis such frame content inter-dependencies. As evaluated in~\S\ref{sec:eval_baseline}, 
this design achieves higher attack success rate with $\ge$50\% margin than the EoT-based solution, likely because the latter is more blindly synthesizing possible patch views in subsequent frames without systematically considering the inter-dependency from vehicle actuation.

}

\section{LGSVL-\op{} Bridging} \label{appendix:lgsvl_op_bridging}

For safety and ethical considerations, the \diffst{end-to-end}\diff{software-in-the-loop evaluation} in~\S\ref{sec:end_to_end} is performed in a production-grade Autonomous Driving (AD) simulator called LGSVL~\cite{lgsvl}. LGSVL is an open-source Unity-based simulator designed specifically for evaluating production-level AD systems. It leverages Unity's built-in physics engine to accurately simulate the vehicle dynamics and tire-road interaction, and provide photo-realistic simulation of the driving environment. It has already been demonstrated~\cite{lgsvl_overview_ppt} to be able to support production-grade AD systems such as Baidu Apollo~\cite{apollo} and Autoware~\cite{autoware}.

Since the original LGSVL simulator does not support \op{}, for the \diffst{end-to-end}\diff{software-in-the-loop} evaluation in~\S\ref{sec:end_to_end} we have to implement the bridging between them (i.e., passing actuation decisions and sensor data to each other) on our own. Since \op{} is designed and optimized for a dedicated device (EON), in this process we overcome some engineering challenges detailed below. This is an engineering contribution of this paper, and thus upon paper acceptance, we will open source our LGSVL-\op{} bridge code to benefit future works in related research areas.

\textbf{Sensing/actuation channel interfacing.} Currently, LGSVL supports a ROS~\cite{quigley2009ros} based bridging to AD systems for receiving vehicle actuation commands and publishing simulated sensor data. However, since \op{} is built with a ZeroMQ~\cite{hintjens2013zeromq} based messaging framework and has no existing support for simulation, we develop the interface to achieve a \textit{frame-by-frame} simulation of \op{} driving in LGSVL. Such interface has 2 directions:
\begin{itemize}
    \item \textit{(1) Passing sensor data from LGSVL to \op{}}. \op{} is originally designed as a driving agent directly running on EON (the official \op{} dashcam). To support simulation, we modify \op{} such that it accepts camera frames from LGSVL instead of the native camera device driver on EON. After our modification, as soon as \op{} receives a camera frame, the lane detection, lateral control, and vehicle actuation logic in \op{} are triggered one by one as described in~\S\ref{sec:background_alc}. 
    \item \textit{(2) Passing actuation commands from \op{} to LGSVL}. LGSVL allows the driving agent to control the built-in vehicle both laterally (i.e., steering) and longitudinally (i.e., gas and brake). In our interface design, we use \op{} to control the lateral movement and use a simple PID controller to control the longitudinal movement to maintain a stable cruising speed.
\end{itemize}

\textbf{Control interruption mechanism modification.} Due to the overhead of the frame rendering and sending, the simulation speed cannot satisfy the real time execution requirement of the control loop, which operates at 100 Hz. Thus, without loss of functional correctness, we disable the timer interrupt in the original control loop and directly invoke 5 control iterations per frame since \op{}'s control loop is $5\times$ faster than the camera frame rate (20 Hz). To match the simulation speed with the control frequency, we apply the steering angle decision of each control iteration to the LGSVL vehicle for a simulation duration of 0.01 seconds.

\begin{table}[t]
\centering
\footnotesize
\caption{Simulation scenario configurations and evaluation results. Lane widths and vehicle speeds are based on standard/common ones in the U.S.~\cite{lanewidth}. Simulation results without attack are confirmed to have 0\% success rates with $\le$0.018 m (std: $\le$9e-4) average maximum deviations.}
\label{tbl:sim_configs}
\vspace{-0.15in}
\setlength{\tabcolsep}{3pt}
\renewcommand{\arraystretch}{0.4}
\begin{tabular}{@{}ccccccc@{}}
\toprule
\begin{tabular}[c]{@{}c@{}}Sim.\\ Scenario\end{tabular} & \begin{tabular}[c]{@{}c@{}}Lane\\ Width\end{tabular} & \begin{tabular}[c]{@{}c@{}}Veh.\\ Speed\end{tabular}      & Attack Goal                                                                        & \begin{tabular}[c]{@{}c@{}}Ave. Max\\ Dev. (std)\end{tabular}          & \begin{tabular}[c]{@{}c@{}}Succ.\\ Rate\end{tabular}      & \begin{tabular}[c]{@{}c@{}}Succ.\\ Time\end{tabular} \\ \midrule
Highway                                                 & 3.6 m                                                & \begin{tabular}[c]{@{}c@{}}65 mph\\ (29 m/s)\end{tabular} & \begin{tabular}[c]{@{}c@{}}Hit barrier\\on the left\end{tabular}     & \begin{tabular}[c]{@{}c@{}}0.76 m\\ (5e-3)\end{tabular} & \begin{tabular}[c]{@{}c@{}}100\%\\ (100/100)\end{tabular}   & 0.97 s                                                \\ \midrule
Local                                                   & 2.7 m                                                & \begin{tabular}[c]{@{}c@{}}45 mph\\ (20 m/s)\end{tabular} & \begin{tabular}[c]{@{}c@{}}Hit truck in the\\opposite lane\end{tabular} & \begin{tabular}[c]{@{}c@{}}0.55 m\\ (7e-2)\end{tabular} & \begin{tabular}[c]{@{}c@{}}100\%\\ (100/100)\end{tabular} & 1.36 s                                               \\ \bottomrule
\end{tabular}
\end{table}

\begin{figure}[t]
\centering
\begin{subfigure}{.5\columnwidth}
  \centering
  \includegraphics[width=.95\linewidth]{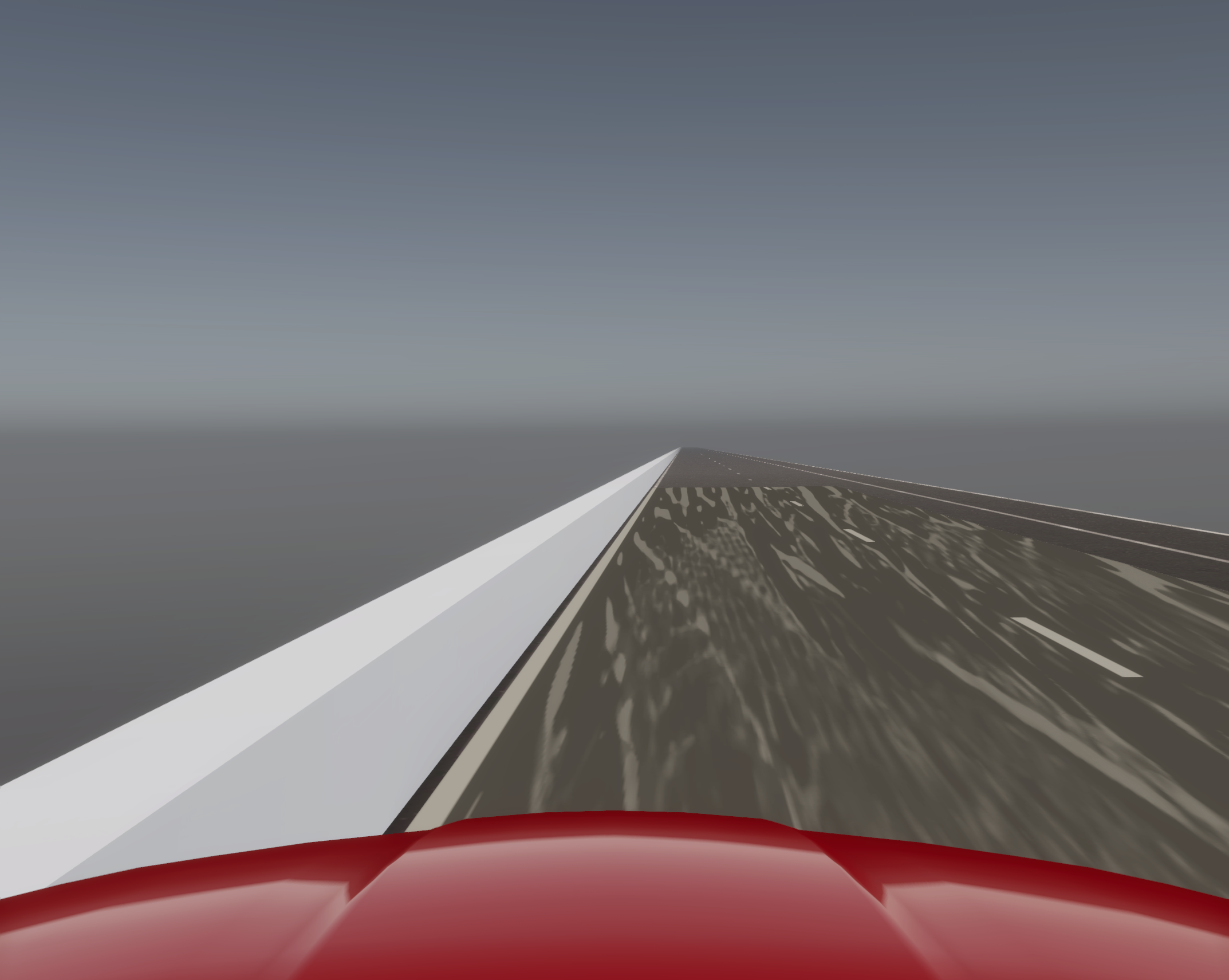}
  \caption{Highway: hit concrete barrier.}
  \label{fig:sim_snapshots_highway}
\end{subfigure}%
\begin{subfigure}{.5\columnwidth}
  \centering
  \includegraphics[width=.95\linewidth]{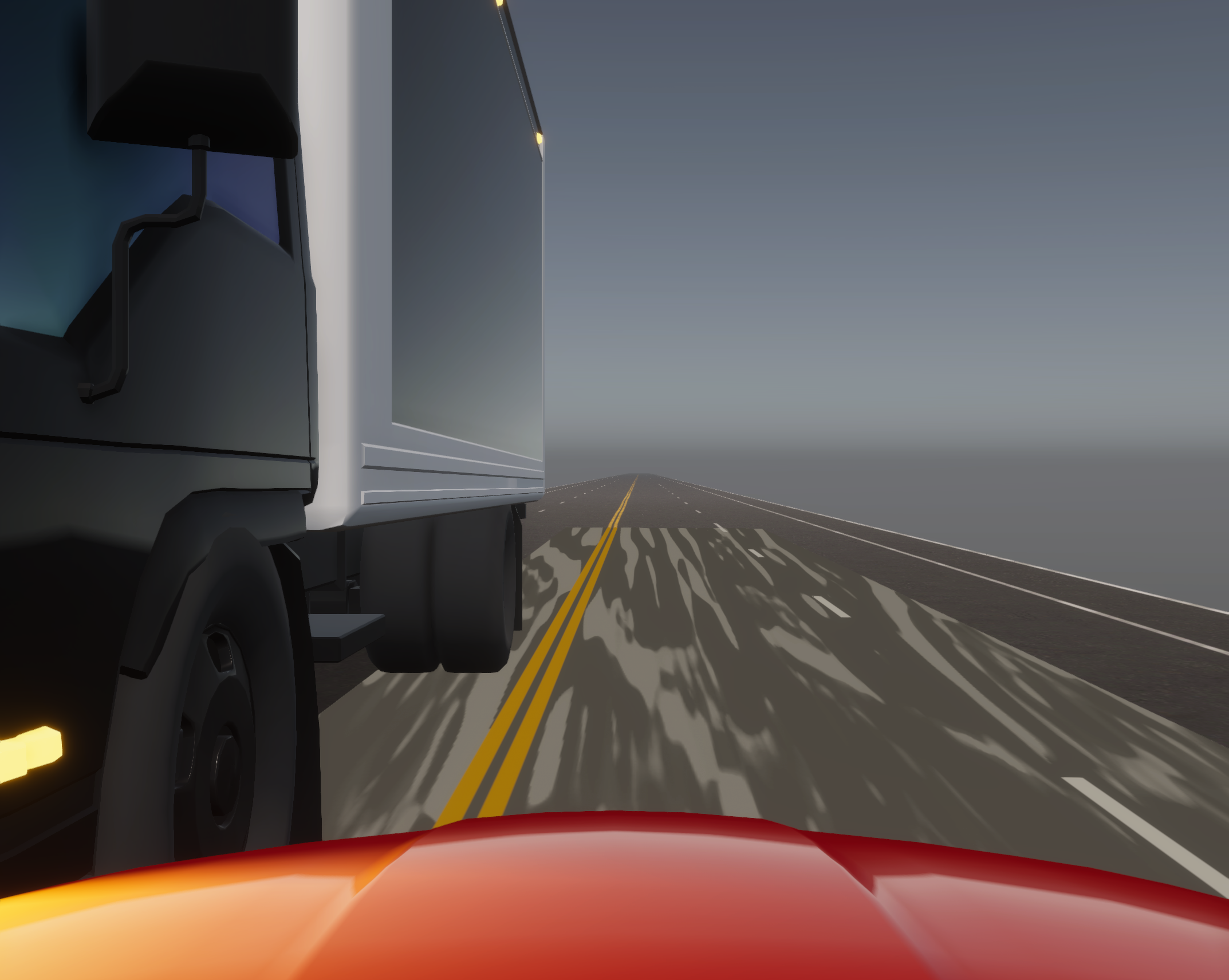}
  \caption{Local: crash into a truck.}
  \label{fig:sim_snapshots_local}
\end{subfigure}
\caption{Snapshots of the attack demo videos when the attack goals are achieved in the simulation for (a) the highway scenario, and (b) the local road scenario in the \diffst{end-to-end}\diff{software-in-the-loop} evaluation in~\S\ref{sec:end_to_end}.}
\label{fig:sim_snapshots}
\end{figure}

\begin{figure*}[p]
\centering
\begin{subfigure}{\linewidth}
  \centering
  \includegraphics[width=.65\columnwidth]{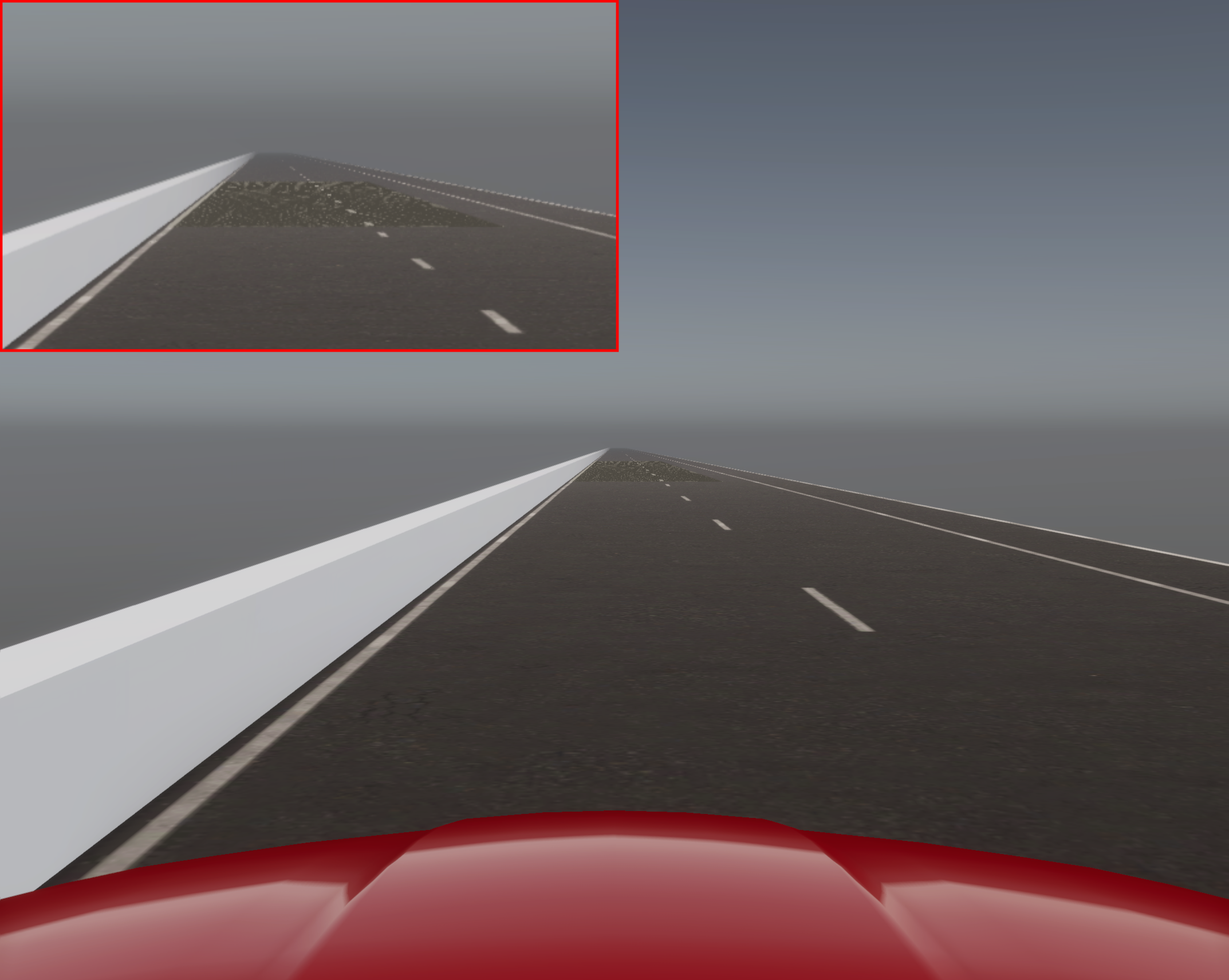} %
  \caption{Driver's view 2.5 sec before the attack succeeds in the highway scenario.}
  \vspace{0.2in}
\end{subfigure}
\begin{subfigure}{\linewidth}
  \centering
  \includegraphics[width=.65\columnwidth]{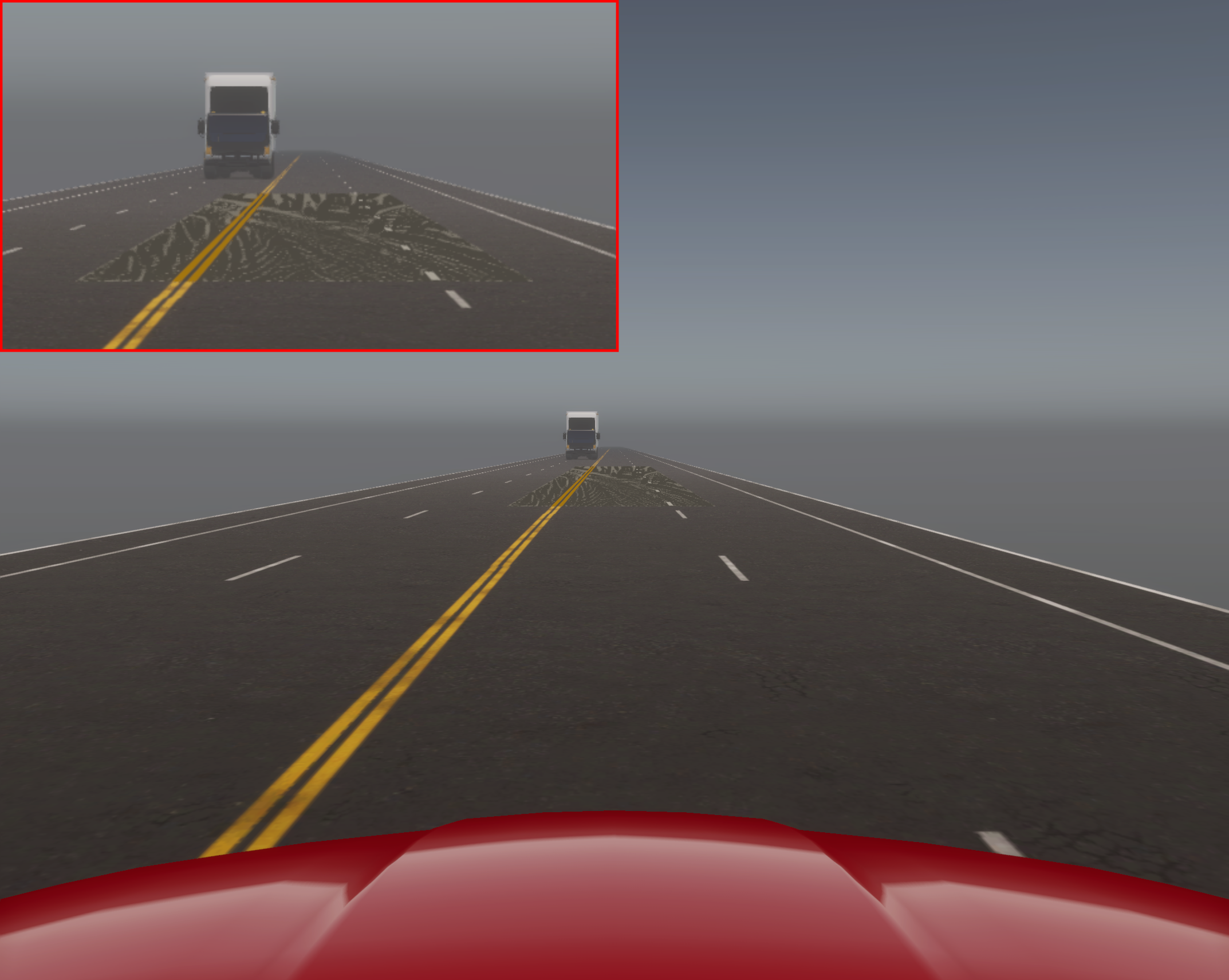} %
  \caption{Driver's view 2.5 sec before the attack succeeds in the local road scenario.}
\end{subfigure}
\caption{Large images of the driver's view images in Fig.~\ref{fig:sim_roads}, which corresponds to the driver's view 2.5 sec (average driver reaction time) before the attack succeeds in the simulation for (a) the highway scenario, and (b) the local road scenario in the \diffst{end-to-end}\diff{software-in-the-loop} evaluation in~\S\ref{sec:end_to_end}. The inset figures are the zoomed-in views of the malicious road patches.}
\label{fig:sim_starts_large_picture}
\end{figure*}

\section{Details of The DNN-Level Defense Methods Evaluated in~\S\ref{sec:ml_defenses}} \label{appendix:defense}
In~\S\ref{sec:ml_defenses}, we evaluate the effectiveness of 5 existing defense methods that are directly applicable to LD models. Brief descriptions and our configurations of these methods are:

\noindent\textbf{JPEG compression~\cite{dziugaite2016study}.} This defense applies lossy JPEG compression for the input image and feeds the generated JPEG image to DNN models. Since most adversarial perturbations are designed to be subtle pixel-level noises that human beings cannot perceive, this defense method expects that the lossy JPEG compression can disrupt such noise-level adversarial perturbations and thus defeat the adversarial attack. 

We use Python Image Library (PIL)~\cite{pillow} to apply image compression to a given input image, which has an argument called \textit{quality} to control the compression quality. Lower values of it mean higher compression rates. Our experiments use 6 values of this argument: 1, 10, 20, 30, 40, and 50, to explore the defense effectiveness at different compression rates.

\noindent\textbf{Bit-depth reduction~\cite{xu2017feature}.} In an RGB image, each pixel has 3 color channels (red, green, and blue) and each channel has an 8-bit depth (0-255). This defense reduces the bit depth of each color channel in the input image, by first projecting each color channel to a space with lower bit depth, and then projecting it back to the normal 8-bit depth space. This defense method expects this operation to disrupt the adversarial perturbations in the input image. In our experiments, we evaluated 6 different bit-depths, from 7-bit to 2-bit.

\noindent\textbf{Adding Gaussian noise~\cite{pmlr-v89-zhang19b}.}
This defense method adds Gaussian noises to the input image. As most adversarial perturbations are designed to be subtle pixel-level noises, this defense method expects to disrupt the adversarial noise by adding white noises. We add Gaussian noise to the input images in the YUV space normalized in [-1, 1]. The noise is sampled from $\mathcal{N}(0, \sigma)$ for each color channel in each pixel independently. In our experiments, we vary the standard deviation $\sigma$ from 0.1 to 0.001.

\noindent\textbf{Median blurring~\cite{xu2017feature}.}  We use the median filter implemented in \textit{SciPy}~\cite{2020SciPy-NMeth} as in~\cite{xu2017feature}. 
Median blurring is one of the local smoothing methods that use the nearby pixels to smooth each pixel. We define a kernel size and take a median around each pixel with the kernel size. This defense expects to smooth and thus eliminate the adversarial perturbations. In our experiments, we evaluate 4 different kernel sizes: 5$\times$5, 10$\times$10, 15$\times$15, and 20$\times$20.

\noindent\textbf{Autoencoder reformation~\cite{meng2017magnet}.} 
Autoencoder reformation is proposed as a part of the MagNet defense method~\cite{meng2017magnet}. First, we train an autoencoder that projects the input image to the latent space and also projects it back to the original space.
The projection onto the latent space is expected to work as a project to the data space where the DNN model is trained on and thus can handle correctly. The reformation process is expected to disrupt the adversarial perturbations that do not exist in the training data space of a given DNN model.

We evaluate 4 different autoencoder architectures: \textit{mnist}, \textit{cifar10}, \textit{Arch-1}, and \textit{Arch-2}, which are listed in the x-axis of Fig.~\ref{fig:defenses}. The former two, \textit{mnist} and \textit{cifar10}, are the architectures used in the MagNet paper~\cite{meng2017magnet} for the MNIST and CIFAR10 datasets respectively. Since the input image size of the OpenPilot LD model is much larger than the images in MNIST and CIFAR10, we also tune two other architectures, \textit{Arch-1} and \textit{Arch-2}. \textit{Arch-2} has more pooling layers than the other autoencoders so that it is compressed the most in the latent space. \textit{Arch-1} has less pooling layers than \textit{Arch-2} but still has more pooling layers than \textit{mnist} and \textit{cifar10}. In other word, if we rank these 4 autoencoder architectures based on the latent space dimension size, the ordering from the smallest to the largest is: \textit{Arch-2}, \textit{Arch-1}, \textit{mnist}, and \textit{cifar10}. All autoencoders are trained with real-world driving images in comma2k19~\cite{comma2k19}, the same dataset we use in our attack evaluation (\S\ref{sec:eval}). To avoid evaluation on the training data (i.e., to avoid in-sample evaluation), we exclude the 80 scenarios for the attack and benign success evaluation from the training data for the autoencoders.

\end{document}